\documentclass[fleqn,usenatbib]{mnras}

\usepackage{newtxtext,newtxmath}

\usepackage[T1]{fontenc}

\DeclareRobustCommand{\VAN}[3]{#2}
\let\VANthebibliography\thebibliography
\def\thebibliography{\DeclareRobustCommand{\VAN}[3]{##3}\VANthebibliography}

\usepackage{graphicx}	
\usepackage{amsmath}	
\usepackage{mathtools}
\usepackage{subcaption}
\usepackage{chngcntr}

\newcommand{\kms}{$\rm km\,s^{-1}$}
\newcommand{\unsim}{\mathord{\sim}}
\newcommand{\chisqr}{$\chi^2_r$}
\newcommand{\alm}{\alpha_{\ell,m}}
\newcommand{\blm}{\beta_{\ell,m}}
\newcommand{\glm}{\gamma_{\ell,m}}
\newcommand{\ylm}{Y_{\ell,m}}
\newcommand{\plm}{P_{\ell,m}}
\newcommand{\xlm}{X_{\ell,m}}
\newcommand{\zlm}{Z_{\ell,m}}

\title[Mapping time-dependent magnetic topologies]{Mapping time-dependent magnetic topologies of active stars}

\author[B. Finociety and J.-F. Donati]{{B. Finociety$^{1}$ \thanks{E-mail: benjamin.finociety@irap.omp.eu},
J.-F. Donati$^{1}$}
\\
$^{1}$ Université de Toulouse, CNRS, IRAP, 14 av. Belin, 31400 Toulouse, France \\
}

\date{Accepted 2022 September 17. Received 2022 September 16; in original form 2022 May 24}

\pubyear{2022}

\begin{document}
\label{firstpage}
\pagerange{\pageref{firstpage}--\pageref{lastpage}}
\maketitle

\begin{abstract}
Throughout the last decades, Zeeman-Doppler Imaging (ZDI) has been intensively used to reconstruct large-scale magnetic topologies of active stars from time series of circularly polarized (Stokes~$V$) profiles. 
ZDI being based on the assumption that the topology to be reconstructed is constant with time (apart from being sheared by differential rotation), it fails at describing stellar magnetic fields that evolve on timescales similar to the observing period.
We present a new approach, called TIMeS (for Time-dependent Imaging of Magnetic Stars), to derive the time-dependent large-scale magnetic topologies of active stars, from time series of high-resolution Stokes~$V$ spectra. This new method uses the combined concepts of sparse approximation and Gaussian process regression to derive the simplest time-dependent magnetic topology consistent with the data. Assuming a linear relation between the Stokes~$V$ data and the reconstructed magnetic image, TIMeS is currently applicable to cases in which the magnetic field is not too strong (with an upper limit depending on $v\sin{i}$).
We applied TIMeS to several simulated data sets to investigate its ability to retrieve the poloidal and toroidal components of large-scale magnetic topologies. We find that the proposed method works best in conditions similar to those needed for ZDI, reconstructing reliable topologies with minor discrepancies at very low latitudes whose contribution to the data is small. We however note that TIMeS can fail at reconstructing the input topology when the field evolves on a timescale much shorter than the stellar rotation cycle.

\end{abstract}




\section{Introduction}

Indirect imaging of magnetic fields of low-mass stars ($M_*<1.5$~$M_\odot$) has been made possible thanks to the tomographic technique named Zeeman-Doppler Imaging (ZDI; \citealt{semel89,brown91,donatibrown97,donati06}). This method uses the principle of maximum-entropy image reconstruction \citep{skilling84} to invert time series of unpolarized (Stokes~$I$) and circularly polarized (Stokes~$V$) spectra into brightness and magnetic maps of active stars. ZDI can also be applied on linear polarization data (Stokes~$Q$ and $U$; \citealt{rosen15}) but it is less common given that the linearly polarized Zeeman signatures are generally an order of magnitude weaker than in their Stokes~$V$ equivalent. 

ZDI works best for rapid rotators \citep{donati89} that are neither viewed exactly pole-on (no modulation of the spectral lines from surface features) nor equator-on (north/south degeneracy), but can still be applied on slower rotators to reconstruct their global magnetic field (e.g. \citealt{petit08,klein21b}). In addition, using the Spherical Harmonics formalism proposed in \cite{donati06}, ZDI is able to reconstruct both simple and complex magnetic topologies and to characterize them in terms of their poloidal and toroidal magnetic components. 

For the last two decades, ZDI has been extensively used to study the large-scale magnetic topologies of low-mass stars ($<1.5\,M_\odot$, e.g. \citealt{morin08,morin10,donati13,hackman16,lavail18,donati19,yu19}). More recently, it has also been used to model the brightness maps of active pre-main sequence (PMS) stars and thereby the radial velocity signatures they generate, in order to filter them out and search for the potential presence of close-in planets through velocimetric techniques \citep{donati16,donati17,yu17}. 

Up to now, ZDI assumed that magnetic topologies are static and not subject to temporal evolution, apart from a potential shear caused by differential rotation (e.g. \citealt{donati2000}). However, magnetic topologies (and surface brightness maps) are known to be intrinsically variable, sometimes on timescales comparable to, or even shorter than the timespan on which data are collected, forcing observers to split their data into several shorter subsets on which ZDI can be applied (e.g. \citealt{donati17,yu19}).

\cite{yu19} implemented a simple (linear) time-dependence of the logarithmic relative brightness of each point at the surface of the star to fit their data and improve the filtering of the activity jitter. However, it yielded no more than marginal improvements with respect to the classical version of ZDI, illustrating the needs for a novel, more powerful approach to model intrinsic variability in a more efficient way.

In this paper, we outline a new method to recover time-dependent large-scale magnetic topologies from time series of high-resolution Stokes~$V$ profiles, including intrinsic variability. This new method, inspired from ZDI, combines sparse approximation and Gaussian process regression (GPR; \citealt{rasmussen06}), to retrieve the simplest large-scale magnetic field and the associated temporal evolution, consistent with the data. In Section~\ref{sec:method}, we present the process we implemented to fit the Stokes~$V$ profiles and reconstruct the associated magnetic topology. In Section~\ref{sec:simulations}, we illustrate the overall performances of the method. We then discuss the effects of most parameters in Section~\ref{sec:effects_parameters} before summarizing and discussing the results in Section~\ref{sec:summary}.

\section{Description of the method}
\label{sec:method}

We start with an overall description of the new method we propose.

\subsection{Description of the magnetic field}

The description we use for the magnetic field follows that of \cite{donati06}, albeit with a slight modification\footnote{These equations were actually used for virtually all ZDI studies based on the same code and published by the IRAP group since 2010. The radial and meridional fields are taken to be positive when pointing outwards and polewards, respectively. The azimuthal field is counted as positive when it is oriented in the same direction as the rotation velocity vector at the equator. The flattened polar view chosen to represent the ZDI magnetic reconstructions generally assumes that the star rotates anticlockwise about the star’s visible pole for display purposes.}:

\begin{equation}
    B_r(\theta,\phi) = \operatorname{Re}\left\{ \sum_{\ell=0}^{\ell_{\max}} \sum_{m=0}^{\ell}  \alm \ylm(\theta,\phi) \right\}
    \label{eq:radial}
\end{equation}

\begin{equation}
    B_\theta(\theta,\phi) = \operatorname{Re}\left\{ \sum_{\ell=0}^{\ell_{\max}} \sum_{m=0}^{\ell} \left[ (\alm + \blm) \zlm(\theta,\phi) + \glm \xlm(\theta,\phi) \right]  \right\}    \label{eq:meridional}                    
\end{equation}

\begin{equation}
\hspace*{-.2cm}
    B_\phi(\theta,\phi) =  \operatorname{Re}\left\{ \sum_{\ell=0}^{\ell_{\max}} \sum_{m=0}^{\ell} \left[ -(\alm + \blm) \xlm(\theta,\phi) + \glm \zlm(\theta,\phi) \right] \right\}
    \label{eq:azimuthal}
\end{equation}

where

\begin{equation}
    \ylm(\theta,\phi)= \sqrt{\frac{2\ell+1}{4\pi} \frac{(\ell-m)!}{(\ell+m)!}} \plm(\cos{\theta}) e^{im\phi}
\end{equation}

\begin{equation}
    \xlm(\theta,\phi)= \frac{1}{\ell+1} \frac{1}{\sin{\theta}}  \frac{\partial \ylm(\theta,\phi)}{\partial \phi}
\end{equation}

\begin{equation}
    \zlm(\theta,\phi)= \frac{1}{\ell+1}   \frac{\partial \ylm(\theta,\phi)}{\partial \theta}
\end{equation}

with $\theta$ and $\phi$ the colatitude and longitude at the surface of the star, $\ylm(\theta,\phi)$ and $\plm(\cos{\theta})$ the spherical harmonic mode and the associated Legendre polynomial of degree $\ell$ and order $m$. The complex coefficients\footnote{For $m=0$, the mode is axisymmetric and the coefficients are purely real.} $\alm$, $\blm$ and $\glm$ fully represent the field, with $\alm$ and $\blm$ describing the poloidal component of the field ($\blm$ charaterizing an additional contribution to of the meridional and azimuthal components with respect to the field description provided by $\alm$ alone) and $\glm$ describing the toroidal component. 

Even though the original equations are also valid, we chose this new formulation to ensure that the meridional and azimuthal components of the poloidal field are explicitly related to the radial one ($\blm$ in the previous set of equations being now replaced by $\alm+\blm$), allowing the code to reconstruct more consistent magnetic topologies, in the sense that the non-radial components of the poloidal field are automatically consistent with the radial one through the use of a single set of coefficients ($\alm$) whereas 2 sets of equal coefficients ($\alm=\blm$, with $\blm$ defined as in \citealt{donati06}) were required to achieve the same result with the previous formulation. With the new formulation, the $\blm$ coefficients are only used to describe potential departures of the poloidal component from a classical multipolar expansion. 
These equations are thus simpler, allowing one for instance to describe a dipole field with only one set of non-zero coefficients ($\alpha_{1,m}$) while two sets of equal non-zero coefficients were previously needed ($\alpha_{1,m}=\beta_{1,m}$). Changes in the sign of the 2 first field components (Eqs. \eqref{eq:radial} and \eqref{eq:meridional}) with respect to the formulation in \cite{donati06} have no impact on the reconstructed topology, only affecting the sign of the derived sets of coefficients. Such changes thus do not affect the results as long as the same relations between the field components and the reconstructed coefficients are used at all steps of the imaging process.

From a given temporal evolution of the coefficients, we can generate a magnetic field whose topology and strength vary with time, and compute the associated Stokes~$V$ profiles. In practice, we divide the surface of the star into a grid of 10 000 cells. We then derive the associated Stokes~$V$ profiles by integrating the local contribution of each cell, that we compute using Unno-Rachkovsky’s (UR) analytical solution to the polarized radiative transfer equations in a Milne–Eddington atmosphere (the $\beta$ parameter\footnote{This parameter refers to the slope of the Planck function with respect to the optical depth.} being set to 3; e.g. \citealt{landi04}). We also assume that the shape of the absorption profile is given by a Voigt function and slightly modified UR's equations to incorporate a tunable linear continuum limb-darkening.) 

In the following of this paper, what we call `a mode' corresponds in fact to either the real or imaginary part of one of the complex coefficients $\alm$, $\blm$ or $\glm$. Limiting ourselves up to a maximum degree $\ell_{\max}$, the maximum number of modes needed to describe a field topology at a given date is $p=3\ell_{\max}(\ell_{\max}+2)$.

\subsection{Overview}

As with ZDI, the idea of our method is to reconstruct an evolving magnetic topology using as few modes as possible. We therefore need to identify the modes that contribute most to the data (i.e. the Stokes~$V$ profiles) and estimate their temporal evolution over the time interval spanned by the observations.

The mode identification / selection process is divided into two main steps. We start with a first selection using sparse approximation to identify a small set of potential modes among the $p$ available modes (up to $\ell_{\max}$; see Sec.~\ref{sec:selection_process}). In a second iterative step based on a least-squares analysis, we remove the selected modes that do not significantly contribute to the data, thereby ensuring that the method yields the simplest magnetic topology consistent with the observed Stokes~$V$ profiles.

The time dependence of the selected modes is recovered using GPR thanks to which we can predict the strength of each mode at all time.

In a final step, we optimize our model by simply rescaling the time dependencies to achieve the best match to the data using least-squares minimization. We end up with a final set of $\alm$, $\blm$ and $\glm$ coefficients (with only few of them different from zero), that describe the magnetic topology and its evolution with time.

To summarize, our new approach consists in 5 steps outlined in detail further down:

\begin{enumerate}
    \item Use a sparse approximation to identify a small set of spherical harmonics modes 
    \item Use of an iterative process based on least-squares minimization to refine the selection and keep as few modes as possible
    \item Get a smooth temporal dependence of the modes using GPRs
    \item Scale the temporal dependencies to optimally match the observed Stokes~$V$ profiles
    \item Derive the time-dependent magnetic topology and the associated Stokes~$V$ profiles
\end{enumerate}

Sparsity is somewhat similar to the principle of maximum entropy used in ZDI (see e.g. \citealt{folsom18} for a description of the maximum entropy implementation in ZDI), as both methods aim at finding a solution that fits the data with a minimal amount of information in terms of modes (i.e. setting only few non-zero coefficients if they are truly needed to model the data). While maximum entropy tries to maximize a user-defined quantity, called entropy (e.g. quadratic sum of the coefficients to reconstruct), the sparse approximation used in this paper aims at minimizing the $L_1$-norm of the solution vector (i.e. sum of absolute values of the vector components) under a constraint on the $L_2$-norm (chi-squared fit of the model to the data), which is another way of achieving a similar goal. Even though MHD simulations suggest that magnetic topologies are characterized by a smooth power spectrum, our goal is not to reproduce all the details of the field topology but rather to find the modes contributing most to the overall field, and that are required to fit the data.

\subsection{Preliminary steps}
\label{sec:preliminary_steps}

Two preparatory steps are needed to speed up and simplify the reconstruction process, (i) the creation of a database of Stokes~$V$ profiles for a wide range of modes and phases on the one hand, and (ii) the estimation of the typical timescale on which the magnetic topology evolves on the other hand, using the simulated Stokes~$V$ observations on which TIMeS is to be applied.

\subsubsection{Creation of a database of Stokes~$V$ profiles}

We consider all $p$ modes up to a degree $\ell_{\rm max}$, compute the associated magnetic map using Eq.~\eqref{eq:radial} to \eqref{eq:azimuthal}, then derive the corresponding Stokes~$V$ profiles at 1000 different phases evenly spaced over a rotation cycle. These profiles make up our database of Stokes~$V$ profiles to be used in the main process.

This database must be customized to the star to be studied as the Stokes~$V$ profiles depend on (i) the inclination of the rotation axis to the line-of-sight $i$, (ii) the line-of-sight projected equatorial rotation velocity $v\sin{i}$, (iii) the spectral domain (e.g. limb-darkening coefficients), (iv) the assumed magnetic sensitivity (i.e. Landé factor) and (v) the instrument characteristics (e.g. spectral resolution).

\subsubsection{Estimation of the decay timescale of the longitudinal field}
\label{sec:longitudinal_field}

Assuming now that the observed Stokes~$V$ profiles are already collected, we compute the longitudinal field, $B_{\ell}$, associated with each observed profiles as the first moment of the Stokes~$V$ profile weighted by the surface brightness inhomogeneities, if any \citep{donati97}. We then use GPR with a quasi-periodic kernel (Eq.~\eqref{eq:quasi-periodic_kernel}) to model these data (e.g. \citealt{rajpaul15}). 

\begin{equation}
    k(t,t') = \theta_1^2 \exp{\left(-\frac{(t-t')^2}{2\,\theta_2^2} - \frac{\sin^2{\frac{\pi(t-t')}{\theta_3}}}{2\,\theta_4^2}\right)} 
    \label{eq:quasi-periodic_kernel}
\end{equation}

where $\theta_1$ is the amplitude of the GP, $\theta_2$ is the decay timescale (exponential timescale on which the longitudinal field model departs from a purely periodic signal), $\theta_3$ is the period of the GP (that should yield the stellar rotation period) and $\theta_4$ corresponds to the smoothing parameter describing the short-term variations induced by rotational modulation. To compensate for potential underestimates of the uncertainties, we add a term representing an excess of uncorrelated noise $s$. The log likelihood function to be maximized can then be written as:

\begin{equation}
    \log \mathcal{L} = -\frac{1}{2}\left(N_0 \log{2\pi} + \log{|\mathbf{K}+\mathbf{\Sigma}+\mathbf{S}| + \mathbf{y}^T(\mathbf{K}+\mathbf{\Sigma}+\mathbf{S})^{-1}\mathbf{y}}  \right)
    \label{eq:log_likelihood}
\end{equation}

where $\mathbf{y}$ corresponds to the measurements of the longitudinal field. $\mathbf{K}$, $\mathbf{\Sigma}$, $\mathbf{S}$ are the covariance matrix associated with the quasi-periodic kernel, the diagonal matrix containing the squared measurement uncertainties and the diagonal matrix $s^2\, \mathbf{I}$ where {\bf I} is the identity matrix.

In order to sample the posterior distribution of the 4 hyperparameters, we run a Markov Chain Monte Carlo (MCMC), using the \texttt{EMCEE PYTHON} module\footnote{The module can be found at \url{https://emcee.readthedocs.io/en/stable/}} \citep{emcee}. More specifically, we use 3500 iterations of 100 walkers and remove a burn-in period equal to five times the autocorrelation time. We consider the median of the posterior distributions as the optimal values and keep in particular the derived value of the decay timescale of the longitudinal field, hereafter noted $\theta_{B_\ell}$, as a proxy for the timescale on which the magnetic topology evolves. This value will be used in the main process (see Sec.~\ref{sec:gp}).

\subsection{Main process}
\label{sec:main_process}

We now describe in details the main process involving the database of Stokes~$V$ profiles and the decay timescale of the longitudinal field that were computed in the preliminary steps (see Sec.~\ref{sec:preliminary_steps}). 

\subsubsection{Selection of the modes}
\label{sec:selection_process}

The first step of the main process aims at selecting as few modes as possible to model the magnetic topology of the star, using sparse approximation. We begin by combining the sequence of observed Stokes~$V$ profiles in sliding subsets of $n$ consecutive profiles (with $n \ll q$, the total number of observed profiles) and we associate each group of profiles to the mean date over the corresponding subset. This date is then converted into a rotation cycle using the stellar rotation period, assumed to be known or derived from the GPR modeling of the longitudinal field (see Sec~\ref{sec:longitudinal_field}). We assume that the magnetic field we want to reconstruct is not too strong, to ensure that the relation between the Stokes~$V$ data and the magnetic map remains linear. We come back in Sec.~\ref{sec:summary} on this limitation. Our problem thus amounts to look for the simplest linear combination of database profiles that can reproduce the selected subset of observations to a given precision $\sigma$. In practice, this is achieved through sparse approximation (e.g.  \citealt{mallat93,tibshirani96,donoho03,donoho06}), this problem being known as Basis Pursuit denoising \citep{chen98} and formalized as follows:

\begin{equation}
    \rm minimize \,\,\, \lVert \mathbf{X}\rVert_1 \,\,\, \rm s.t. \,\,\, \lVert \mathbf{AX}-\mathbf{B}\rVert_2 \leq \sigma 
    \label{eq:BPDN}
\end{equation}

where $\mathbf{B}$ is a vector containing the $N$ data points in the group of $n$ profiles, $\mathbf{A}$ the dictionary, a $N\times p$ matrix for which each columns contains the spectral signatures of each mode (included in the database) taken at the observed phases, and $\sigma$ the level at which we wish the data to be fitted. The components $X_i$ of the vector $\mathbf{X}$ are directly associated with the modes to be reconstructed, i.e. proportional to the real or imaginary part of one of the $\alm$, $\blm$ and $\glm$ coefficients describing the field (up to a degree $\ell_{\max}$). Taking into account the noise in the data, the previous problem can be generalized by including a $N\times N$ diagonal matrix $\mathbf{W}$ containing the inverse of the errors on the spectral points. Problem \eqref{eq:BPDN} then becomes:

\begin{equation}
    \rm minimize \,\,\, \lVert \mathbf{X}\rVert_1 \,\,\, \rm s.t. \,\,\, \lVert \mathbf{W}(\mathbf{AX}-\mathbf{B})\rVert_2 \leq \tau 
    \label{eq:BPDN_W}
\end{equation}
 with $\tau^2$ being the chi-square ($\chi^2$) level at which we wish the data to be fitted.

To solve this problem, we use the SPGL1 solver\footnote{More specifically, we use the \texttt{spgl1 PYTHON} implementation that can be found at \url{https://spgl1.readthedocs.io/en/latest/index.html}} \citep{vandenberg08}. This solver requires the vector $\mathbf{WB}$ and the columns of $\mathbf{WA}$ to be normed. It also features some user-defined parameters such as weights to penalize the reconstruction of some coefficients in the vector $\mathbf{X}$, as well as a stopping criterion. In our case, we chose to penalize the reconstruction of the modes with a weight proportional to their degree $\ell$ and inversely proportional to the mean amplitude of their Stokes~$V$ profiles over a rotation cycle (computed from the profiles in the database). As the amplitudes of the Stokes~$V$ profiles associated with the coefficients $\blm$ and $\glm$ are lower than those associated with $\alm$, we ended up applying different penalizing weights to ensure that the field components associated with $\blm$ and $\glm$ are not too severely penalized with respect to those associated with $\alm$\footnote{In practice, noting $A_{\ell,m}$ the mean amplitude of the Stokes~$V$ profiles associated with a mode, the penalizing weights providing the best results are equal to $\ell/A_{\ell,m}$, $0.8\, \ell/A_{\ell,m}$ and $0.5\, \ell/A_{\ell,m}$ for the reconstruction of the modes associated with $\alm$, $\blm$ and $\glm$, respectively.}.
We also find that fixing the tolerance to $\tau^2=1.2\,N$ allows us to better retrieve the modes we are looking for. Relaxing the tolerance yields a solution that progressively diverges from the data (implying that some modes are missed), while tightening it further gives too much significance to spurious modes that are not really needed for modeling the data.

Running the SPGL1 solver for each group of profiles allows one to get the time dependence of vector $\mathbf{X}$ (and therefore of each mode). From that, we can compute the following relative powers for each coefficient $X_i$ as:

\begin{equation}
   \mathcal{P}(X_i) = \frac{\sum_t |X_i(t)| }{\max_{i}{\sum_t |X_i(t)|}}
\end{equation}

where $t$ refers to the mean rotation cycle associated with each group of profiles.
These powers reflect the contribution of each coefficient to the reconstructed field topology and range from 0 (no contribution) to 1 (maximum contribution). We only keep the modes whose relative power exceeds 5\% that of the dominant mode. We find that this threshold gives the best result and allows us to retrieve all significant modes (and no more than a few spurious ones).

As the value of some coefficients can be impacted by the presence of the other modes, we re-compute their temporal evolution, using only the selected modes in the process. We achieve this by simply solving the linear problem $\mathbf{AX}=\mathbf{B}$ for each group of profiles using an ordinary least-squares minimization, keeping only the columns of the dictionary associated with the selected modes. This least-squares approach is part of an iterative process in which we progressively remove the modes that contribute very little to our model. In practice, once a first solution is found, we compute the mean detection level of the coefficients $X_i$, noted $\mathcal{D}(X_i)$:

\begin{equation}
    \mathcal{D}(X_i) = \frac{1}{N_0} \sum_t\frac{|X_i(t)|}{\sigma_i(t)}
\end{equation}

where $N_0$ refers to the number of group of profiles (i.e. $N_0= q-n+1$) and $\sigma_i$ is the uncertainty on the value of coefficient $X_i$ estimated by least-squares minimization.

We then remove the mode with the lowest mean detection level and enter a new loop with the remaining modes, starting again with $\chi^2$ minimization then removing the least significant mode, until all components of $\mathbf{X}$ satisfy a mean detection level larger than 3 (i.e. for all $i$,  $\mathcal{D}(X_i)>3$). This criterion ensures that the values of the modes are significantly different from 0, i.e., that the selected modes contain relevant information for the model.

\subsubsection{Using GPR to model the time dependence}
\label{sec:gp}

In the previous step, we computed the time dependence of vector $\mathbf{X}$ whose coordinates are proportional to the real or imaginary part of one of the complex coefficient $\alm$, $\blm$ and $\glm$ describing the field. However, this time dependence is still somewhat noisy at this stage, e.g., as a result of crosstalk between modes. We thus need to minimize this noise component and make the time dependence as smooth as possible, yet without suppressing true signal.

To achieve this, we use GPR to model the time series of each coefficient $X_i$ taking into account their error bars as derived from least-squares minimization. We choose a squared exponential kernel for the GPR (e.g. \citealt{rajpaul15}), given by:

\begin{equation}
    k(t,t')=\theta_1^2\exp{\left(-\frac{(t-t')^2}{2\,\theta_2^2}\right)}
\end{equation}

where $\theta_1$ represents the amplitude of the GP and $\theta_2$ is the decay timescale, that we set to the value of $\theta_{B_\ell}$ found in Sec.~\ref{sec:longitudinal_field}. Note that imposing a lower (resp. larger) value for $\theta_2$ would increase (resp. reduce) the flexibility of the model, which would then make it prone to overfitting (resp. underfitting) the time series. As for the longitudinal field we introduce an excess of uncorrelated noise in our GP model, yielding the same log likelihood as Eq.~\eqref{eq:log_likelihood} with the covariance matrix corresponding to the squared exponential kernel. 

For each coefficient $X_i$, we sample the posterior distribution of the parameters $\theta_1$ and $s$, running a MCMC thanks to the \texttt{EMCEE PYTHON} module \citep{emcee} in the exact same way as in Sec.~\ref{sec:longitudinal_field}. We thus retrieve a smooth time dependence for each coefficient $X_i$, and more specifically their value at the rotation cycles associated with the individual observed Stokes~$V$ profiles. Note that the GPR prediction needs to be slightly extrapolated to infer the values for the first and last few profiles.

\subsubsection{Deriving synthetic profiles and associated maps}

In the final step of the process, we try to further optimize the fit by globally scaling up or down the time dependence of each mode through least-squares minimization. Once the scaling factors are obtained, we can compute the derived magnetic topology and its evolution with time, the corresponding set of synthetic Stokes~$V$ profiles and the associated reduced $\chi^2$ ($\chi^2_r$) with respect to the observations.

\begin{figure}
    \centering
    \includegraphics[scale=0.25,trim={10cm 7cm 10cm 7cm},clip]{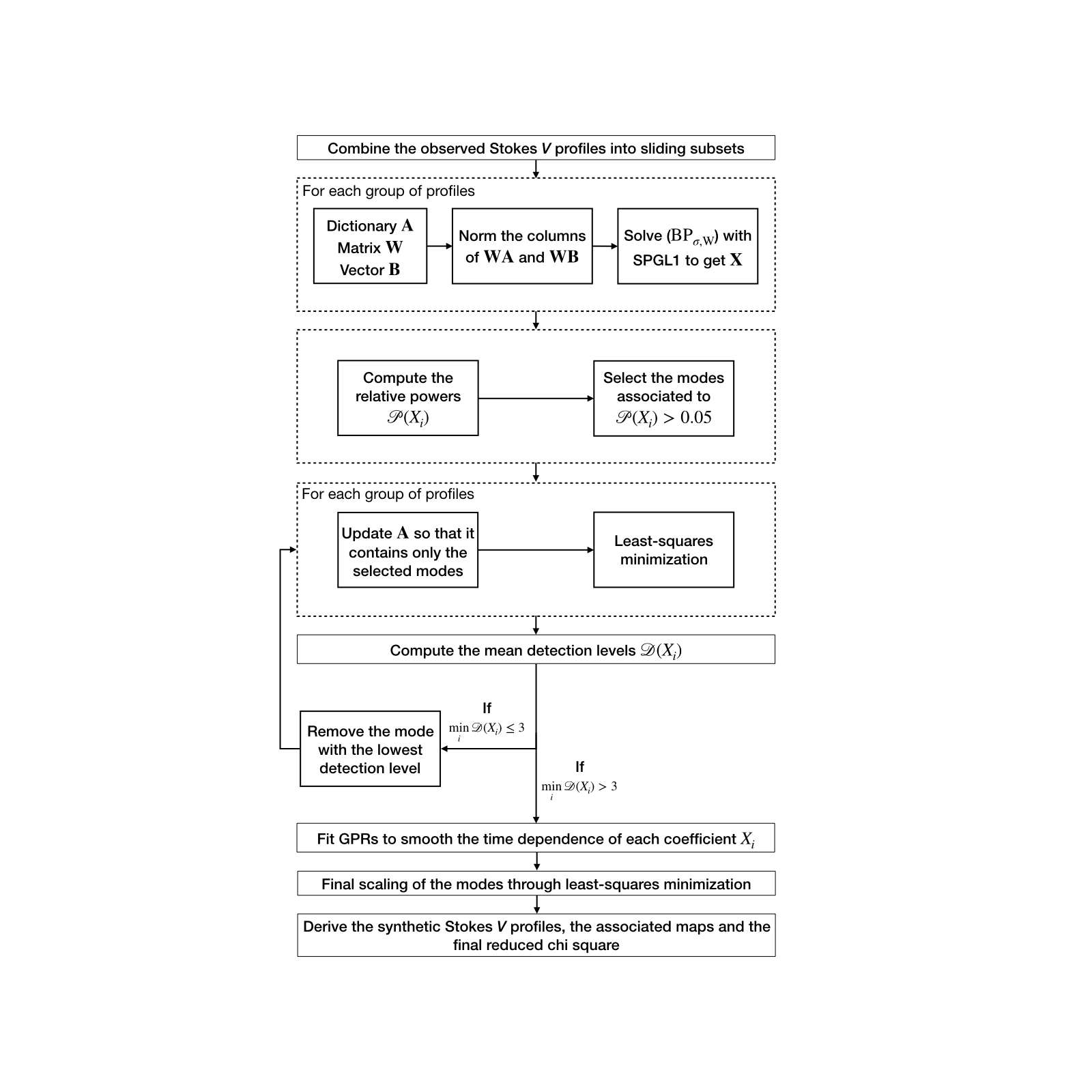}
    \caption{Schematic view of the main process. See Sec.~\ref{sec:main_process} for a detailed description.}
    \label{fig:process_total}
\end{figure}

A graphical layout of all steps of our imaging process, that we call TIMeS, for `Time-dependent Imaging of Magnetic Stars', is presented in Fig.~\ref{fig:process_total}.

\section{Simulations}
\label{sec:simulations}

We simulated several basic magnetic topologies described by a combination of modes up to $\ell=2$, with a strength typically reaching a maximum value of 500~G. We further assume that the filling factor $f$ is everywhere equal to 1.0, i.e., that each grid cell at the surface of the star is fully covered with the local magnetic field $B$. These topologies evolve over a time interval of 120~d during which 60 evenly-spread synthetic observations were generated. We further assume that the star is unspotted (i.e. no brightness features) and features a rotation period of 2.9~d (used to phase the profiles on the rotation cycle), an inclination of $i=60^\circ$ and $v\sin{i}=25$~\kms. The 60 observations correspond to Stokes~$V$ profiles collected in the near infrared (featuring a mean wavelength of 1700~nm, a limb-darkening coefficient of 0.3, a Doppler width of 1.8~\kms\ and a Landé factor of 1.2), with a spectral velocity bin of 2~\kms\ (similar to that of the near-infrared spectropolarimeter SPIRou; \citealt{donati20}) and a signal-to-noise ratio (SNR) of 5000. This very high SNR may not be representative of actual observations but was chosen to investigate the behavior of the method in optimal conditions.

In what follows, we apply our new method, TIMeS, on sliding subsets of size $n=6$ (covering about 3.5 rotation cycles, i.e. $\unsim10$~d), providing a sufficient sampling of the rotation period while minimizing the evolution of the field within each subset. Modes up to $\ell_{\max}=5$ are allowed in the reconstruction, implying that our code can choose between a total number of 105 coefficients at each time step to model the field topology and its temporal evolution.

\subsection{Poloidal field}
\label{sec:poloidal}

\begin{figure}
    \centering \hspace*{-0.2cm}
    \includegraphics[scale=0.2,trim={2.5cm 0.5cm 0cm 0cm},clip]{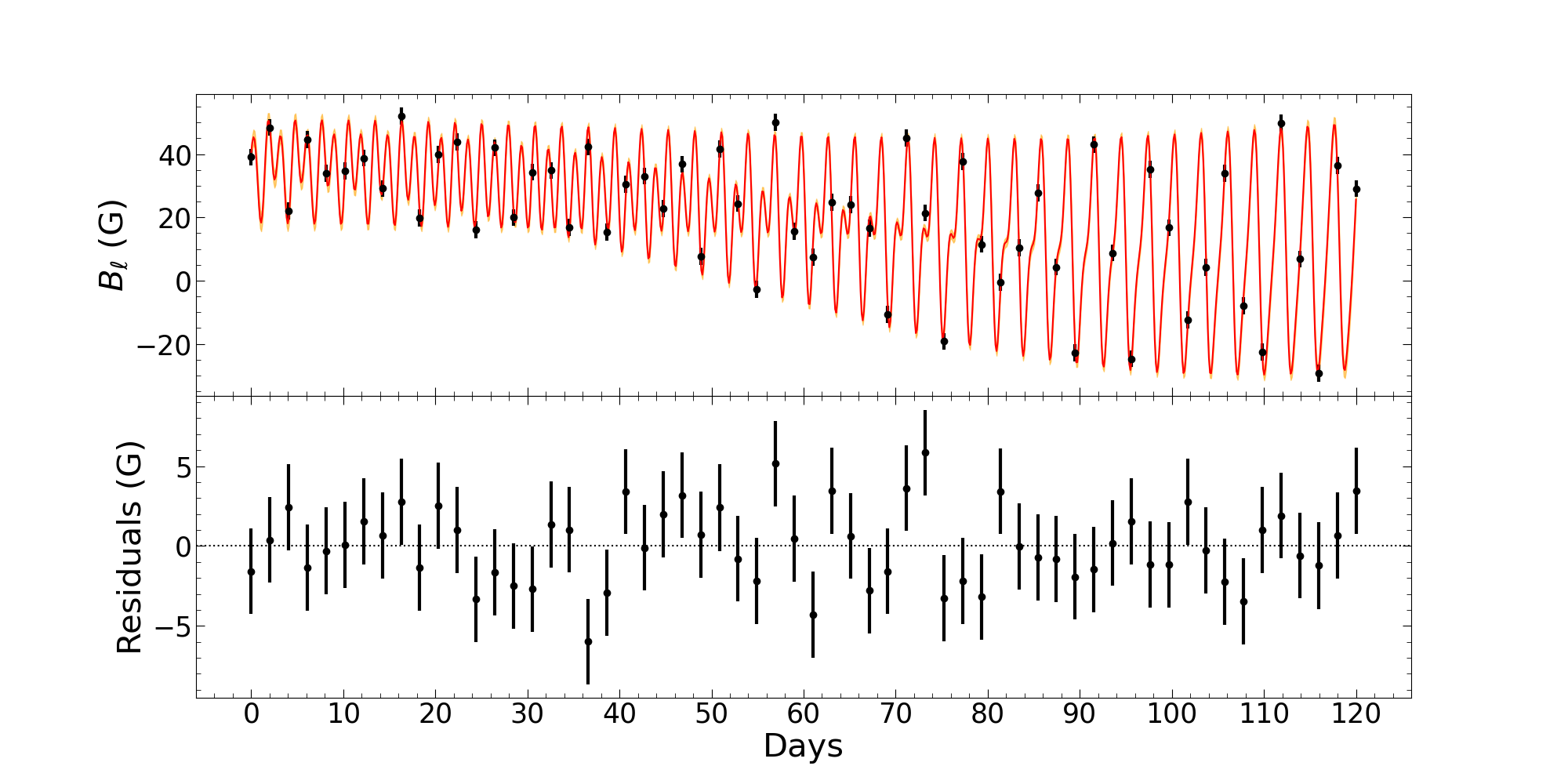}
    \hspace*{-0.2cm}
    \includegraphics[scale=0.2,trim={2.5cm 0.5cm 0cm 0cm},clip]{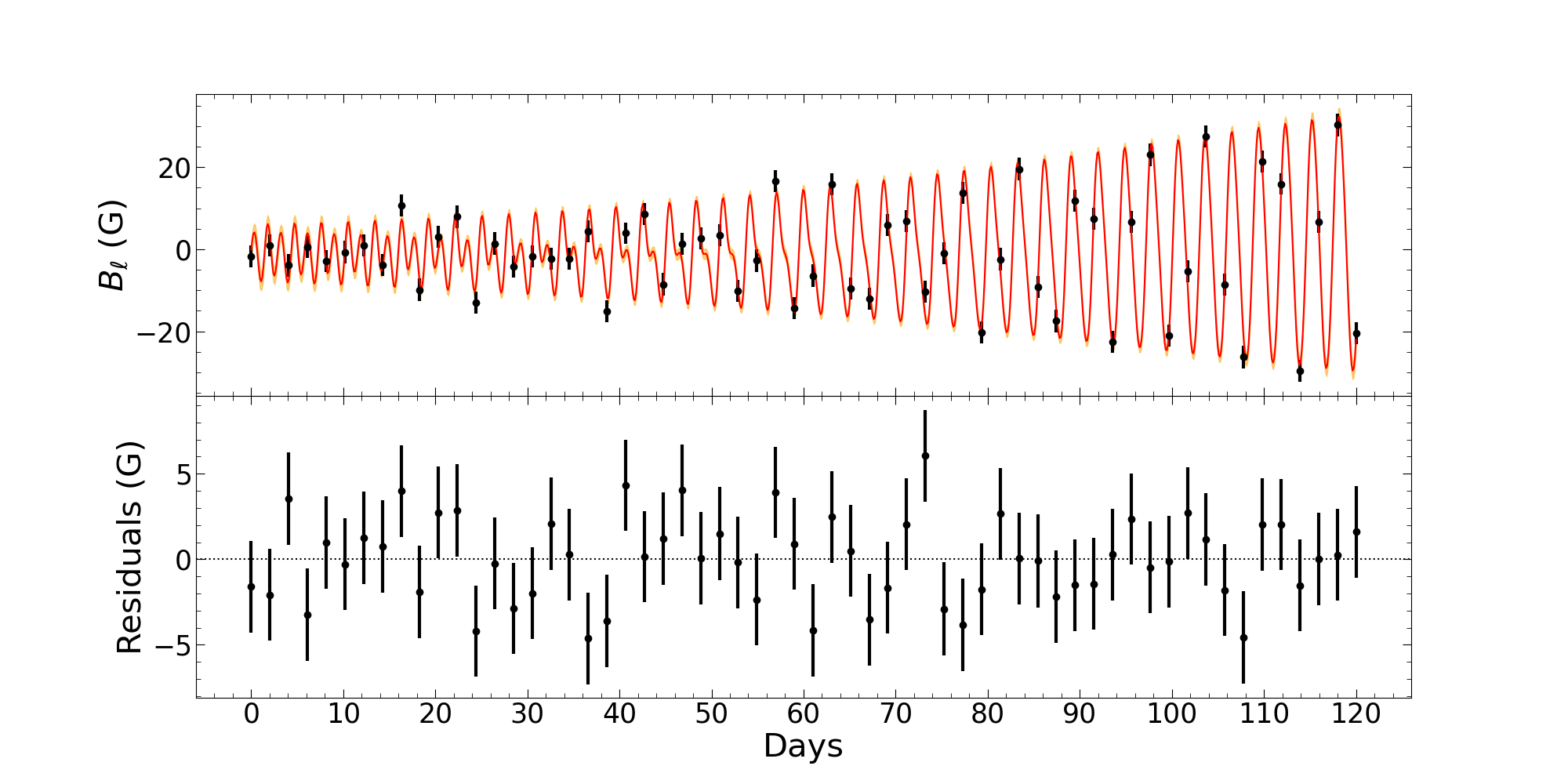}
    \hspace*{-0.2cm}
    \includegraphics[scale=0.2,trim={2.5cm 0.5cm 0cm 0cm},clip]{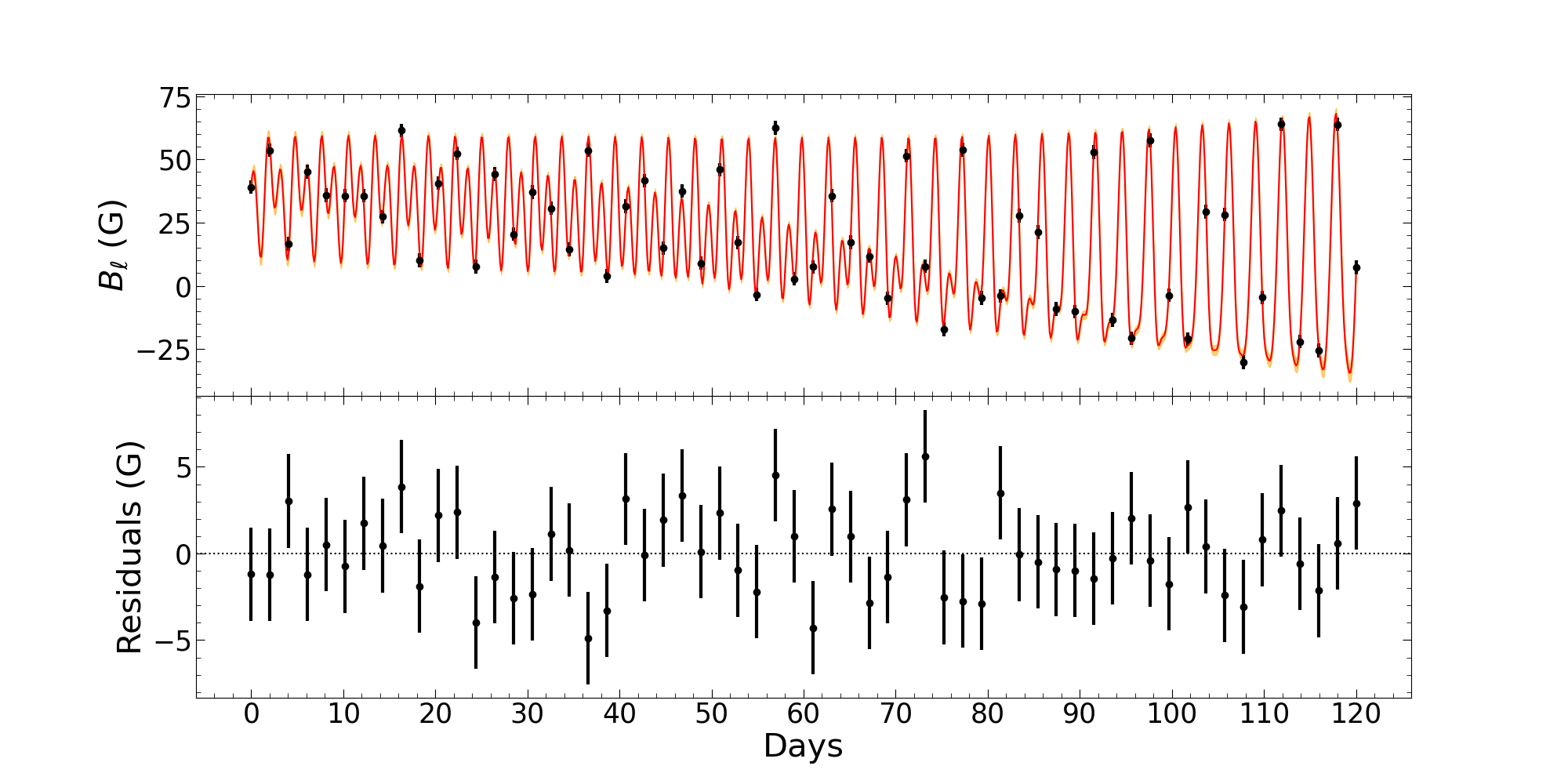}
    \caption{Longitudinal field associated with the simulated magnetic topologies presented in Sec.~\ref{sec:simulations}. For the three panels, the values associated with the 60 observations (black dots), and the GPR fit (red) along with its associated $1\sigma$ confidence interval in orange are shown at the top while the residuals between the fit and the measurements are shown at the bottom. \textit{First panel}: Purely poloidal field. The inferred decay timescale is found to be $74\pm11$~d and the residuals exhibit a RMS dispersion of 2.4~G. \textit{Second panel:} Purely toroidal field. The decay timescale is equal to $106 \pm 33$~d and the residuals show a RMS dispersion of 2.5~G. \textit{Third panel}: Field featuring both a poloidal and toroidal contribution. The decay timescale is found to be $69 \pm 10$~d and the residuals exhibit a RMS dispersion of 2.4~G} 
    \label{fig:longitudinal_field}
\end{figure}

\begin{figure*}
    \centering \hfill 
    \includegraphics[scale=0.078,trim={.5cm 1.cm 1cm 9cm},clip]{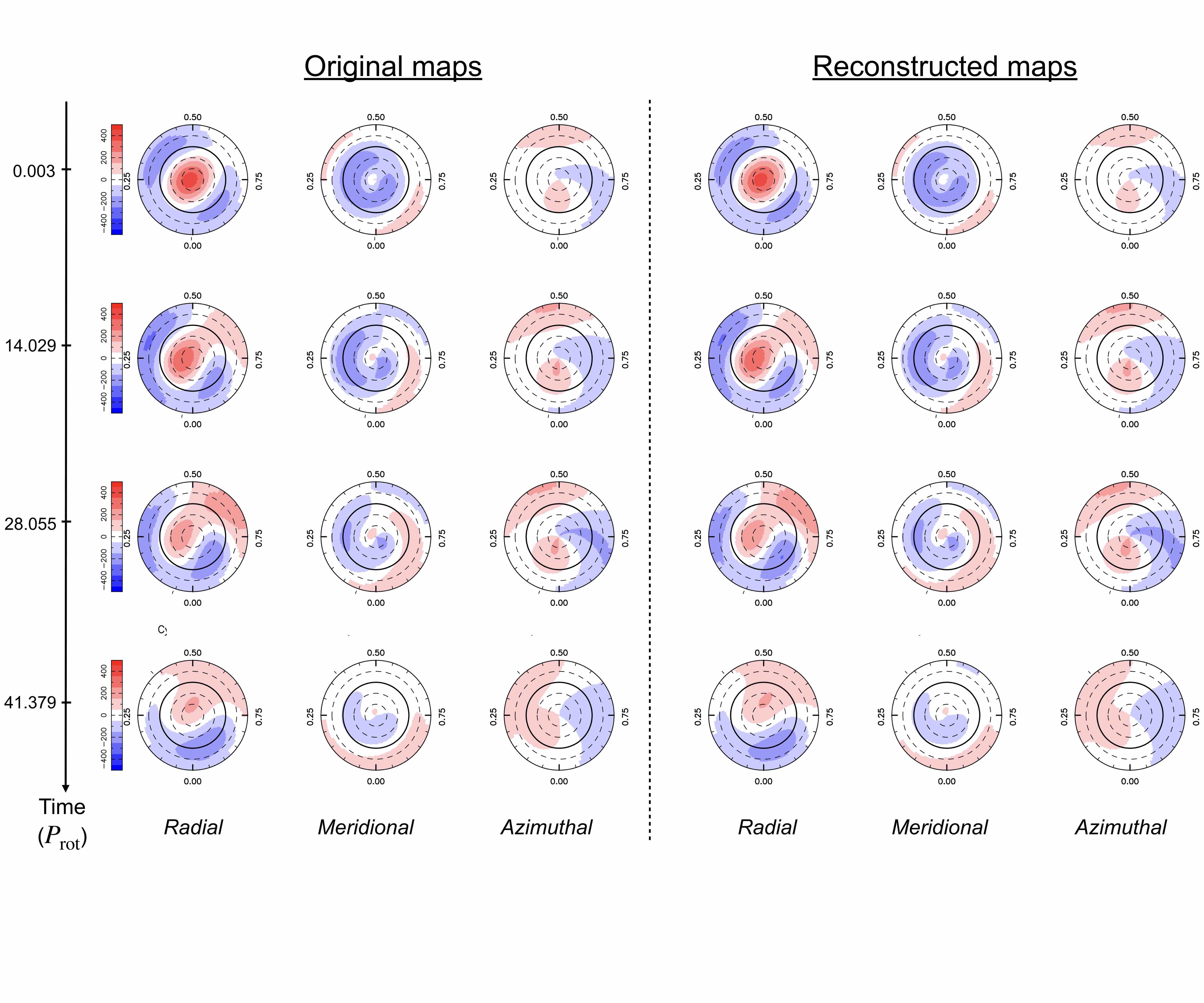}
    \hfill \vspace*{-2cm}
    \includegraphics[scale=0.14,trim={3cm 5.5cm 5cm 5cm},clip]{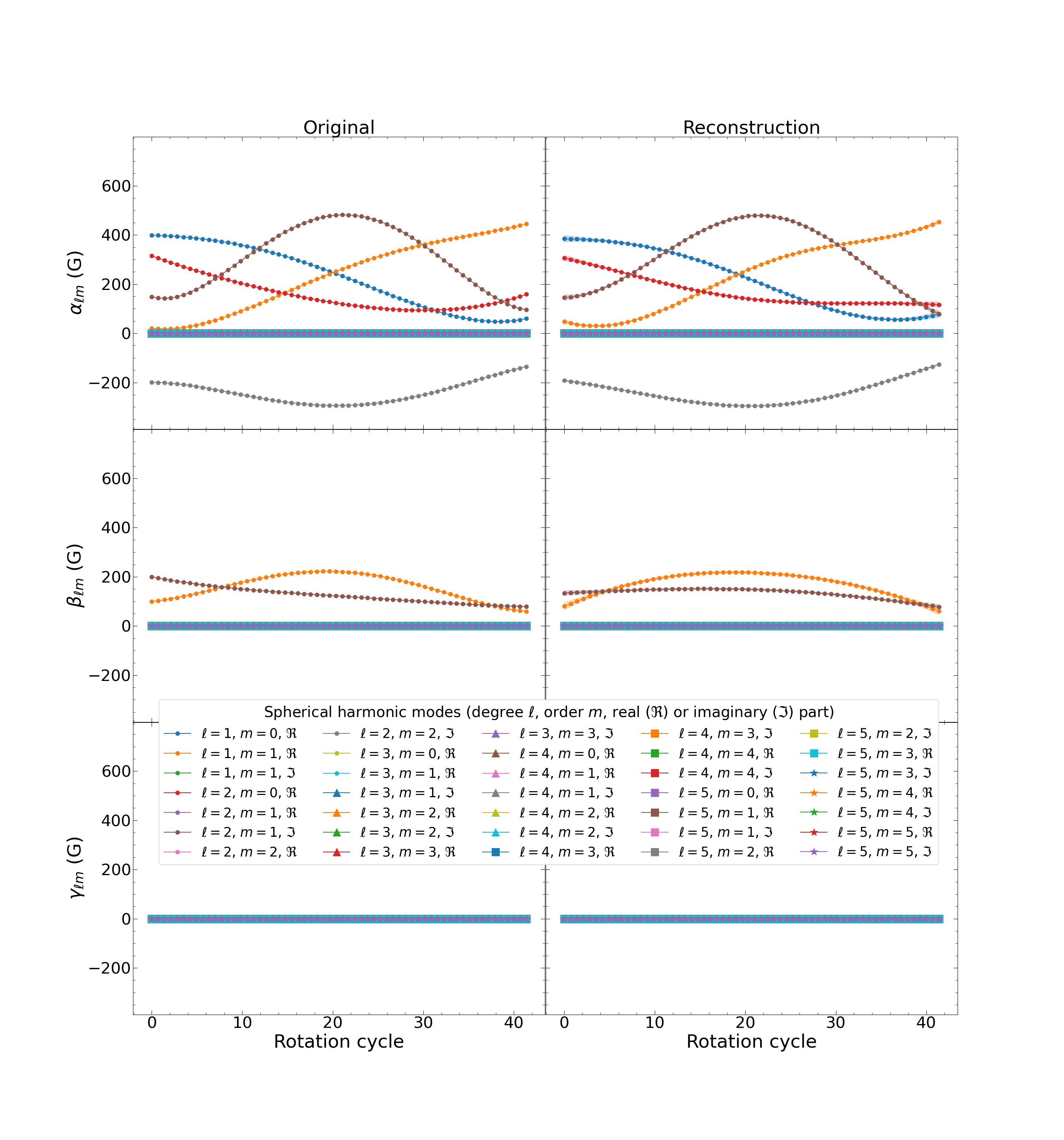}
    \includegraphics[scale=0.16,trim={0cm 0.5cm 0cm 0cm},clip]{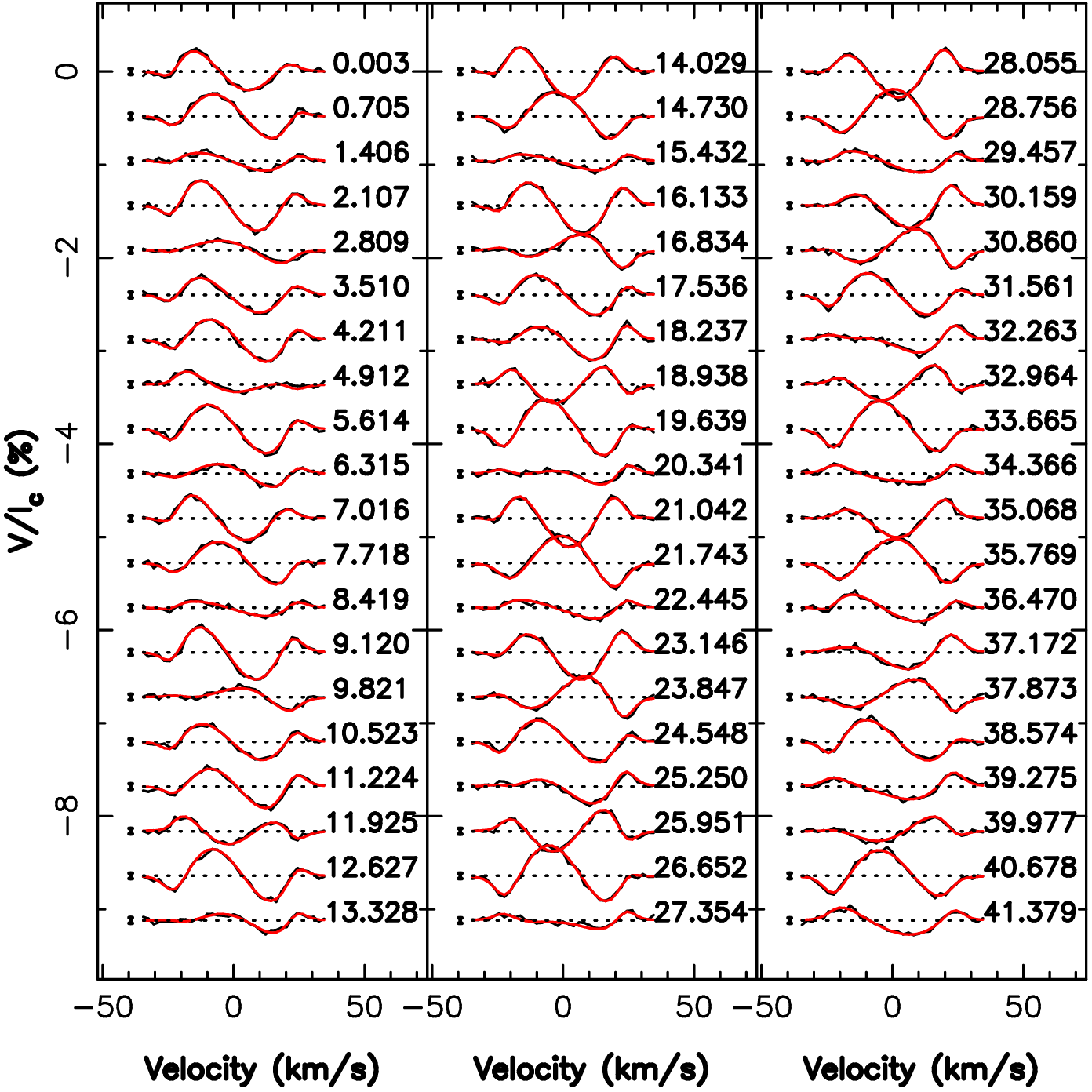}
    \caption{Reconstruction of a purely poloidal field. \textit{Top:} Maps of the the simulated (first to third columns) and reconstructed (fourth to sixth columns) radial, meridional and azimuthal magnetic field components for different cycles over the observation period (timeline going from the first to the last row) in the case of a purely poloidal magnetic field. The maps are shown in a flattened polar projection with the pole at the center, the equator represented as a bold black line and the $60^\circ$ and $30^\circ$ latitude parallels shown as dashed lines. The ticks around the reconstructed star correspond to the phase of the simulated spectropolarimetric observations. In these plots, red indicates positive radial, meridional and azimuthal fields (in G) that point outwards, polewards and counterclockwise respectively. The maps are shown down to latitudes of $-60^\circ$, beyond which nothing is visible to a putative observer given the assumed orientation of the rotation axis with respect to the line of sight ($i=60^\circ$). \textit{Bottom left}: Evolution with time of the spherical harmonics coefficients, for the original image (first column) and the reconstructed one (second column). Each curve (color/symbol) corresponds to the real ($\Re$) or imaginary ($\Im$) part of a spherical harmonic mode of degree $\ell$ and order $m$. The filled colored area around each curve of the right panels represents the $1\sigma$ confidence level derived from GPR. \textit{Bottom right}: Stokes~$V$ profiles for a purely poloidal field. The simulated data are shown in black while the reconstructed profiles are plotted in red. The rotation cycle is mentioned on the right of each profile and the $3\sigma$ error bars are displayed on their left.}
    \label{fig:maps_poloidal}
\end{figure*}

We first simulated a purely poloidal field described by $\alm$ and $\blm$ coefficients, evolving independently from each other. We however make no prior assumption on the field in the reconstruction process, letting all coefficients (including $\glm$) to be reconstructed if needed.  

We started by computing the longitudinal field associated with the 60 observations, then fitted these measurements and determined the decay timescale $\theta_{B_\ell}$ as described in Sec.~\ref{sec:longitudinal_field} (see the evolution of $B_\ell$ and the associated fit in the first panel of Fig.~\ref{fig:longitudinal_field}). We find $\theta_{B_{\ell}}=74\pm11$~d.

With our new method, the Stokes~$V$ profiles (shown in Fig.~\ref{fig:maps_poloidal}) are fitted down to  $\chi^2_r =1.02$. We are able to retrieve the input field with the modes that were injected (no spurious toroidal field is recovered). Our method yields time dependencies of the $\alm$ and $\blm$ coefficients that are similar to the injected ones (bottom left of Fig.~\ref{fig:maps_poloidal}). The synthetic maps corresponding to the retrieved coefficients are shown at the top of Fig.~\ref{fig:maps_poloidal}, demonstrating that the evolution of the reconstructed topology is fully consistent with the original one.

\subsection{Toroidal field}
\label{sec:toroidal}

We then simulated a purely toroidal field described by a few $\glm$ coefficients only, with other modes than in the purely poloidal field case.

We now find a higher (though still consistent) value for the decay timescale of the longitudinal field ($106\pm33$~d). Second panel of Fig.~\ref{fig:longitudinal_field} shows the measurements of the longitudinal field along with the associated GPR fit.

\begin{figure*}
    \centering \hspace*{-0.5cm}
    \includegraphics[scale=0.09,trim={.5cm 20.cm 0cm 9cm},clip]{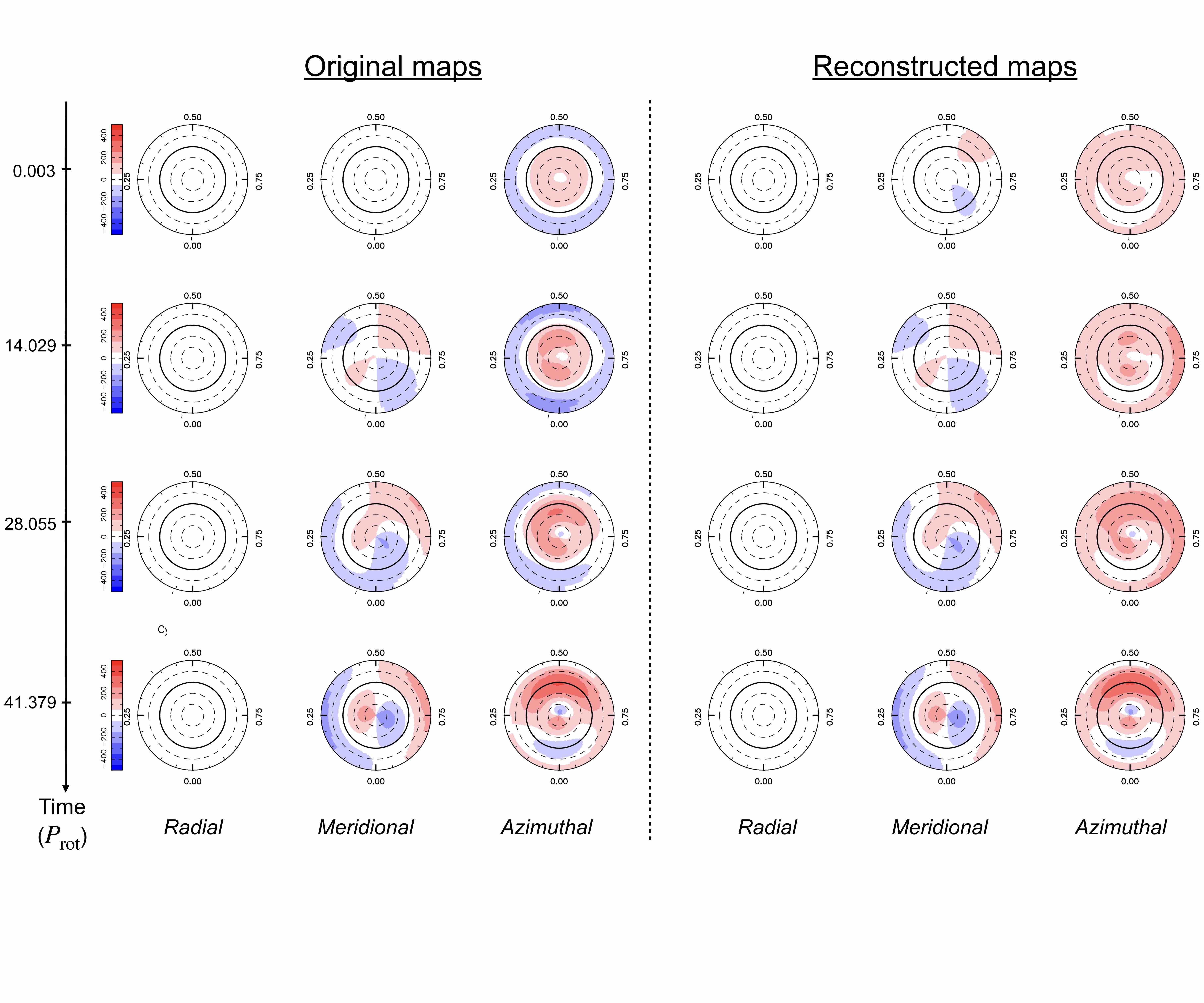} \hspace*{-0.8cm}
    \includegraphics[scale=0.16,trim={3cm 5.5cm 5cm 6cm},clip]{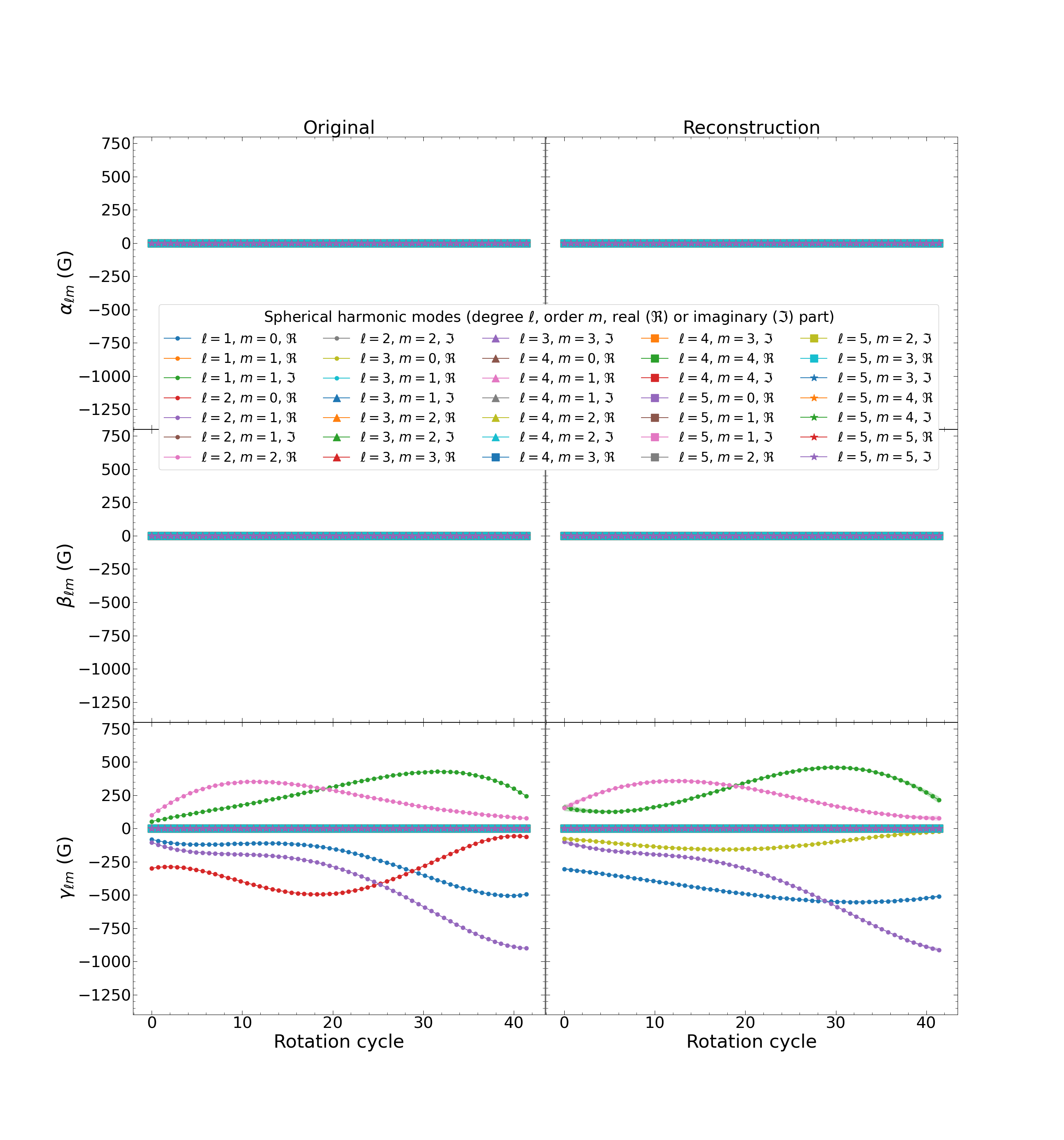} 
    \includegraphics[scale=0.17,trim={0cm 0.5cm 0cm 0cm},clip]{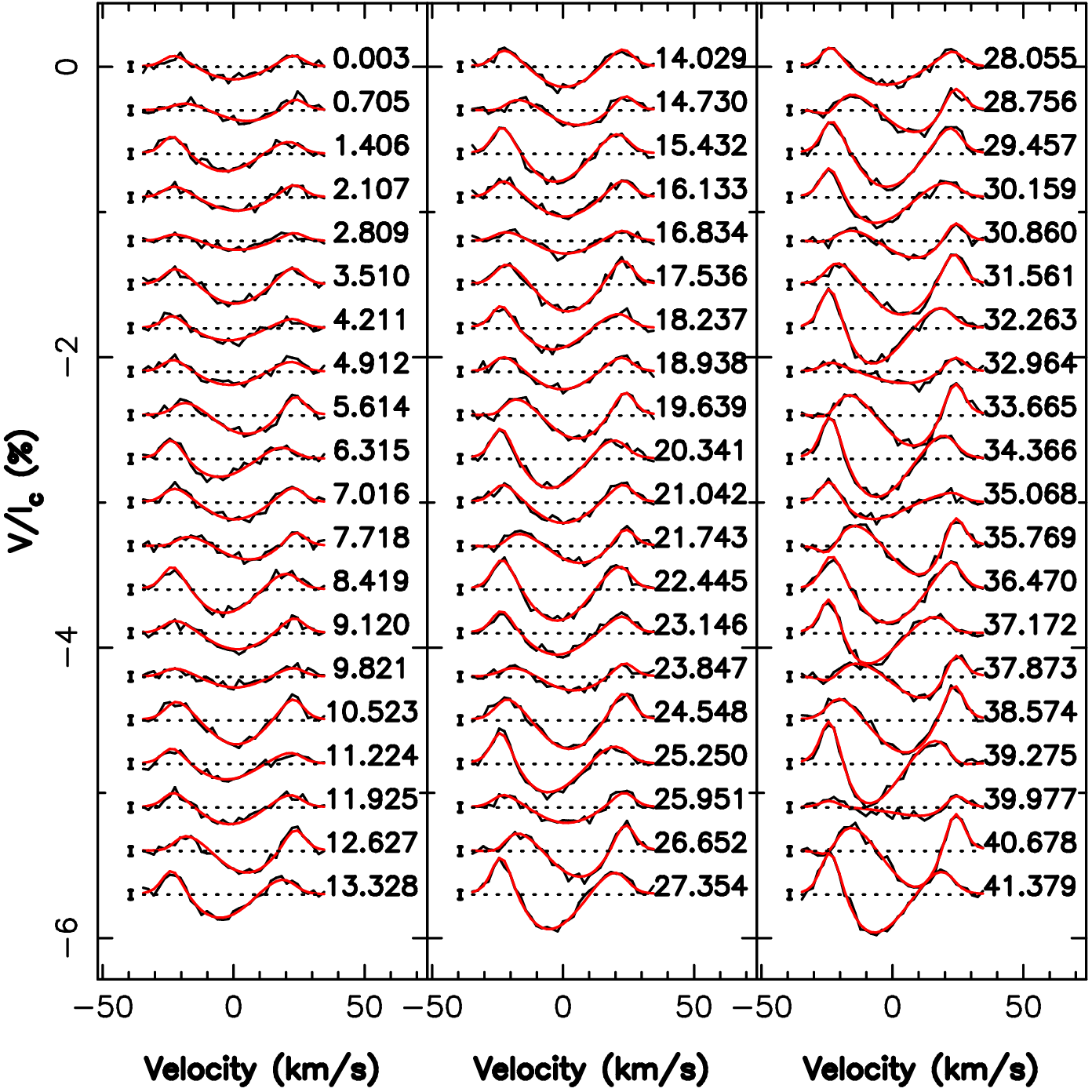}
    \caption{Same as Fig.~\ref{fig:maps_poloidal} for a purely toroidal field.}
    \label{fig:maps_toroidal}
\end{figure*}

Unlike in the poloidal case, the method does not exactly retrieve the original modes and we observe cross-talk between $\gamma_{1,0}$, $\gamma_{2,0}$ and $\gamma_{3,0}$. We nevertheless see that the strength and time dependencies of the reconstructed modes are consistent with the simulated ones and no spurious poloidal field is recovered (bottom left of Fig.~\ref{fig:maps_toroidal}). The Stokes~$V$ profiles are fitted down to $\chi^2_r=1.03$, and the modeled topology is very similar to the one associated with the simulated data, although some differences can be seen mainly at the southernmost latitudes, whose actual visibility to Earth-based observers is quite limited (see Fig.~\ref{fig:maps_toroidal}). It demonstrates that the reported cross-talk between modes reflects a genuine ambiguity between some of the axisymmetric modes describing the toroidal field, whose Stokes~$V$ signatures are quite similar.

\subsection{Poloidal + Toroidal field}
\label{sec:poloidal_toroidal}

We finally simulated an input magnetic topology featuring both the poloidal and toroidal components described in the two previous examples. The decay timescale of this field is found to be equal to $69\pm10$~d (third panel of Fig.~\ref{fig:longitudinal_field}).

\begin{figure*}
    \centering \hspace*{-0.5cm}
    \includegraphics[scale=0.09,trim={.5cm 20.cm 0cm 9cm},clip]{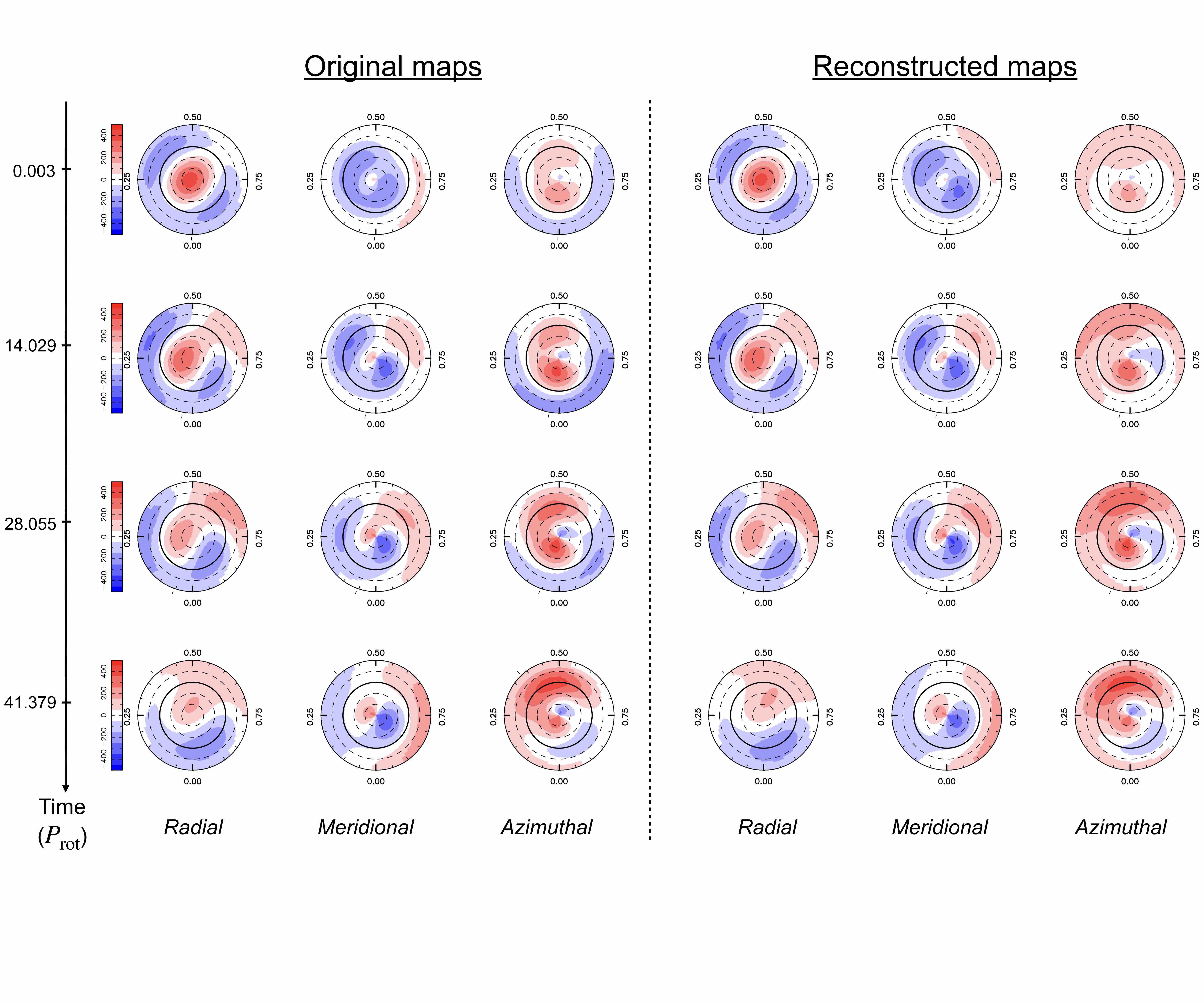} \hspace*{-0.8cm}
    \includegraphics[scale=0.16,trim={3cm 5.5cm 5cm 5cm},clip]{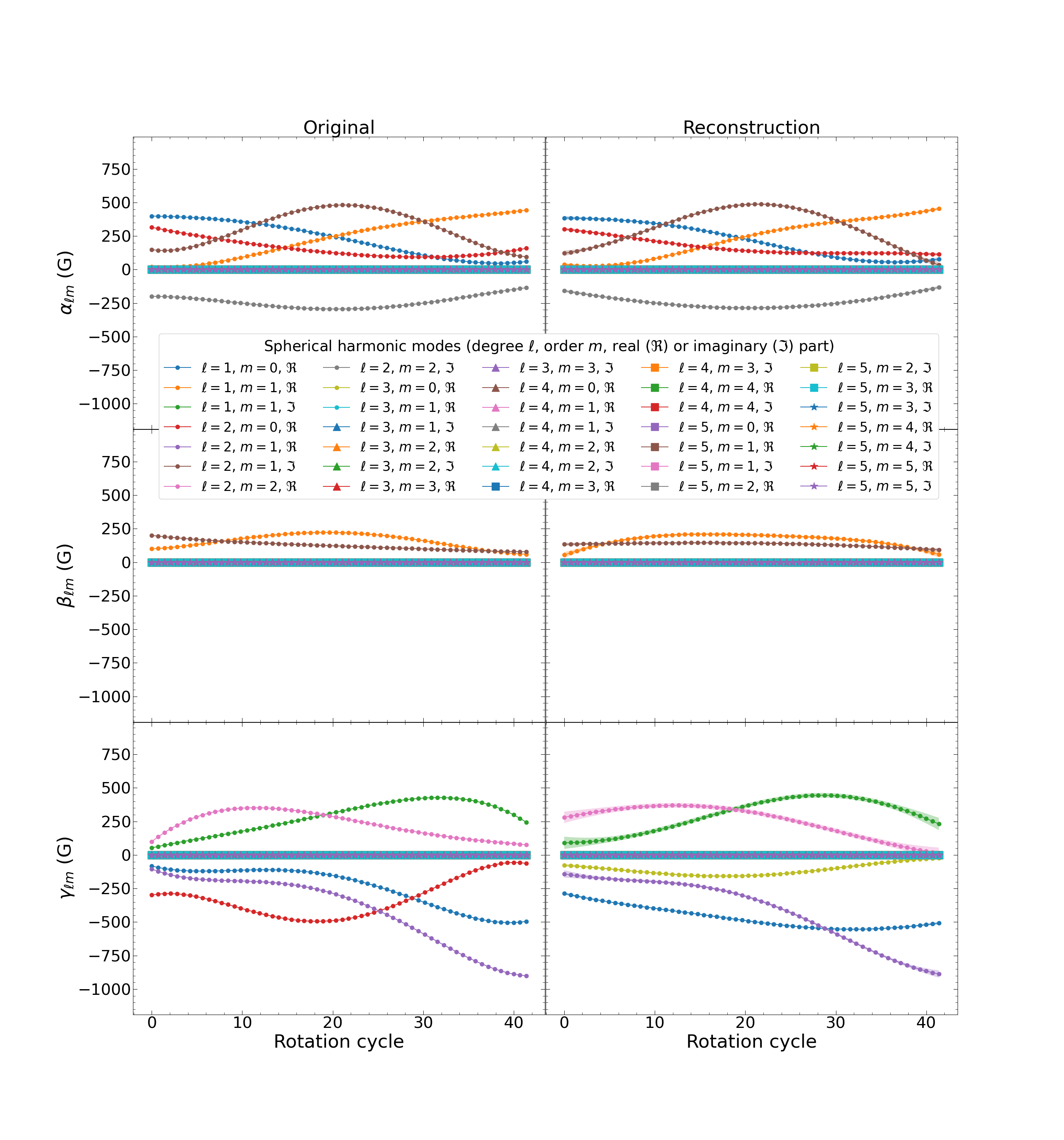} 
    \includegraphics[scale=0.17,trim={0cm 0.5cm 0cm 0cm},clip]{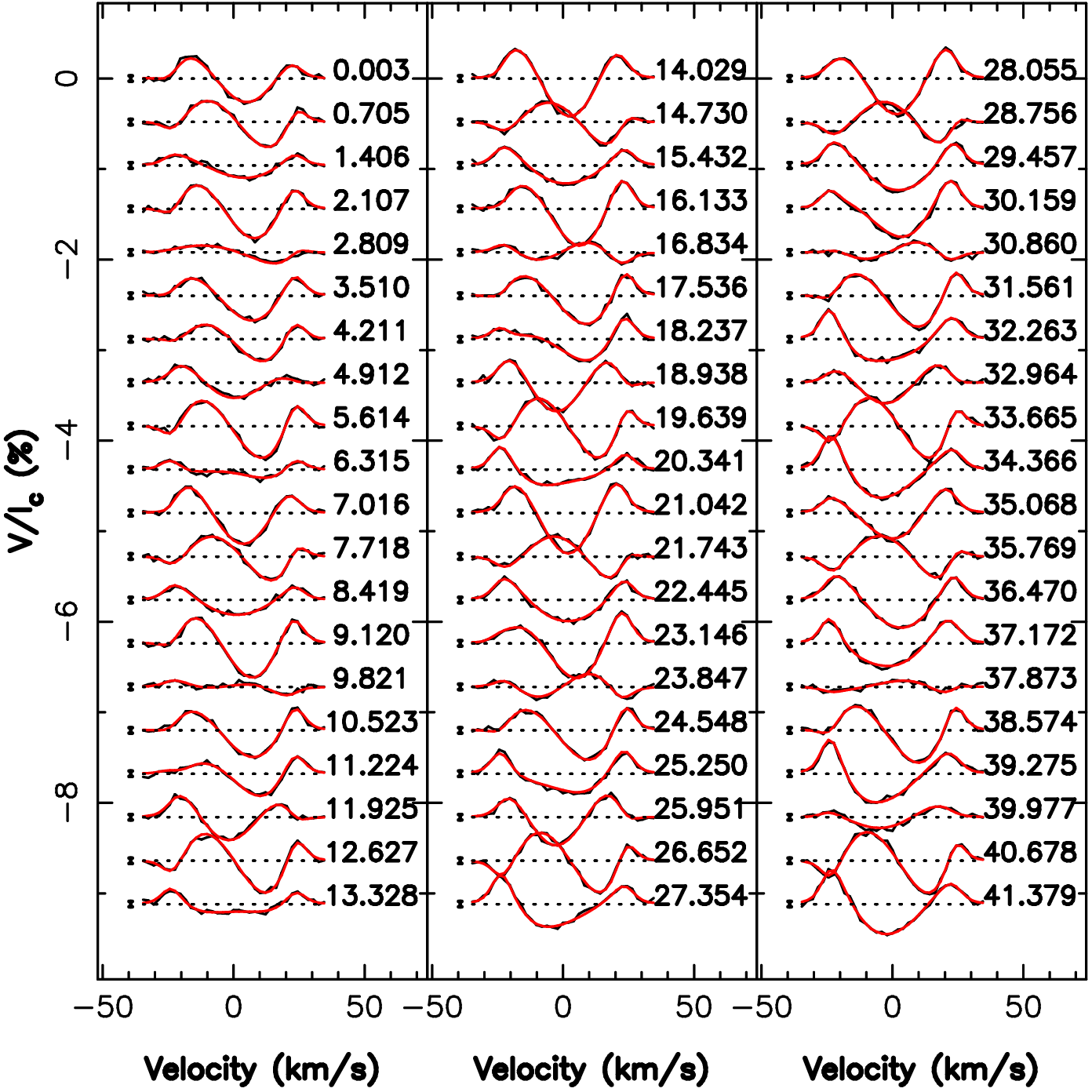}
    \caption{Same as Fig.~\ref{fig:maps_poloidal} for the magnetic topology described by a poloidal and a toroidal field.}
    \label{fig:maps_poloidal_toroidal}
\end{figure*}

The evolution of the coefficients describing the field, as well as the associated Stokes~$V$ profiles, are shown at the bottom of Fig.~\ref{fig:maps_poloidal_toroidal}. We obtain a \chisqr\ of 1.02, thus similar to those obtained in the two previous examples, while ZDI fits the data down to $\chi^2_r = 8.5$ which clearly illustrates the benefits of TIMeS over ZDI. Once again, the reconstructed maps are very similar to the original ones (top of Fig.~\ref{fig:maps_poloidal_toroidal}), and our method reconstructs well the input poloidal field whereas the inferred toroidal field suffers from from the same cross-talk between modes as those mentioned above.

Applying TIMeS on the same topology but assuming now a filling factor of $f=0.3$ instead of $f=1.0$ (implying that 30\% of each surface cell contains a magnetic field of local strength $B/f$ while the rest of the cell is non magnetic), we find virtually identical results, demonstrating that the assumed linearity is not an issue for this field flux and $v\sin{i}$, even with low values of the filling factor that imply stronger values of the local field strength.

\section{Influence of the parameters}
\label{sec:effects_parameters}
In this Section, we discuss the effects of the various parameters involved in the process. The results mentioned in this Section were obtained from simulations assuming a stellar magnetic topology similar to that described in Sec.~\ref{sec:poloidal_toroidal}.

\subsection{Effects of the stellar and observation parameters}

We first review the behaviour of TIMeS for parameters that are not directly related to the method itself. This includes the stellar parameters (inclination, $v\sin{i}$, rotation period) but also the observational ones, like the quality of the data (i.e. SNR) and the temporal sampling.

\subsubsection{Stellar inclination}

We now consider a star with $i=30^\circ$ and $v\sin{i}=25$~\kms. In this configuration, the northern pole of the star is better seen which modifies the way the radial, meridional and azimuthal components contributes to the Stokes~$V$ profiles and strongly restricts the contribution of features located in the southern hemisphere.
In addition, the amplitude of the Stokes~$V$ profiles in the database changes, which directly affects the weights used in the SPGL1 solver. 
For $i$ $\unsim30^\circ$, we only see minor changes in the reconstruction affecting mainly low latitude features that actually do not contribute much to the observed data (Fig.~\ref{fig:inc30}).
Unlike in the reference case, TIMeS only retrieves the injected modes with no cross-talk. The Stokes~$V$ profiles are still fitted down to a similar \chisqr.

For nearly equator-on configurations (e.g. $i$ $\unsim80^\circ$), the contributions to the Stokes~$V$ profiles from both hemispheres start to partly cancel out for some modes. As a result, these modes were not retrieved and the fit of the Stokes~$V$ profiles is slightly degraded ($\chi^2_r=1.17$) although the overall topology remains consistent with the simulated one (Fig.~\ref{fig:inc80}), showing that the method still behaves reasonably well.

\subsubsection{$v\sin{i}$}
\label{sec:vsini}
For these tests, we adjusted the SNR of each simulation to ensure that the Stokes~$V$ signatures are detected with the same precision, taking into account that the depth, width and number of velocity bins of the synthesized Stokes~$V$ profiles all vary when $v\sin{i}$ is modified. Typically, the ratio between the noise and the amplitude is in the range 3--6\% in the different cases. As the number of points in the profiles increases with $v\sin{i}$, it compensates more or less exactly for the difference in amplitude.

Let us consider a star with a $v\sin{i}$ of 15~\kms\ and an inclination of $60^\circ$. In this case, the Stokes~$V$ profiles are deeper and narrower, thus reducing the amount of details that can be reconstructed at the surface of the star. The fit to the data (now featuring a SNR of 3000) is only slightly degraded ($\chi^2_r=1.11$) with respect to the reference case ($v\sin{i}=25$~\kms). We however note that this time the reconstruction of both the poloidal and toroidal components suffer from cross-talks (e.g. $\beta_{2,1}$ coefficient replaced by a combination of $\alpha_{3,1}$ and $\beta_{1,1}$; Fig.~\ref{fig:coeff_poloidal_toroidal_vsini15}), with minor impact on the reconstructed topology. 

For larger $v\sin{i}$ (e.g. 50~\kms), the Stokes~$V$ profiles are shallower and broader. This time, the data (featuring a SNR of 10,000) are fitted down to $\chi^2_r=1.02$. We obtain results very similar to those of the reference case of Sec.~\ref{sec:poloidal_toroidal} with only minor changes in the retrieved time dependencies (Fig.~B2, available as supplementary material), as expected from the better spatial resolution provided by the higher $v\sin{i}$ (allowing a better identification of modes of higher degrees). 

We also tested the method at $v\sin{i}=5$~\kms, with a SNR decreased to 2000. In this case, the reconstructed maps are less detailed as expected from the loss of spatial resolution at the surface of the star that results from the lower $v\sin{i}$. Despite some modes are missed by TIMeS (Fig.~\ref{fig:coeff_poloidal_toroidal_vsini5}), we are nevertheless able to fit the Stokes~$V$ profiles down to $\chi^2_r=1.18$ showing that the method is still able to reconstruct the overall topology in the case of slower rotators, despite some loss of information. 

Applying TIMeS on the same cases, but assuming now a filling factor of $f=0.3$, yields similar results, except for the $v\sin{i} = 5$~\kms\ case, for which the assumed linearity between the reconstructed image and the Stokes~$V$ profiles starts to break downn preventing the code to properly identify the correct modes. We come back on the limits of this assumption in Sec.~\ref{sec:summary}.

\subsubsection{Complex magnetic field}
\label{sec:complex_field}
\begin{figure*}
    \centering \hspace*{-0.5cm}
    \includegraphics[scale=0.09,trim={.5cm 27.cm 0cm 9cm},clip]{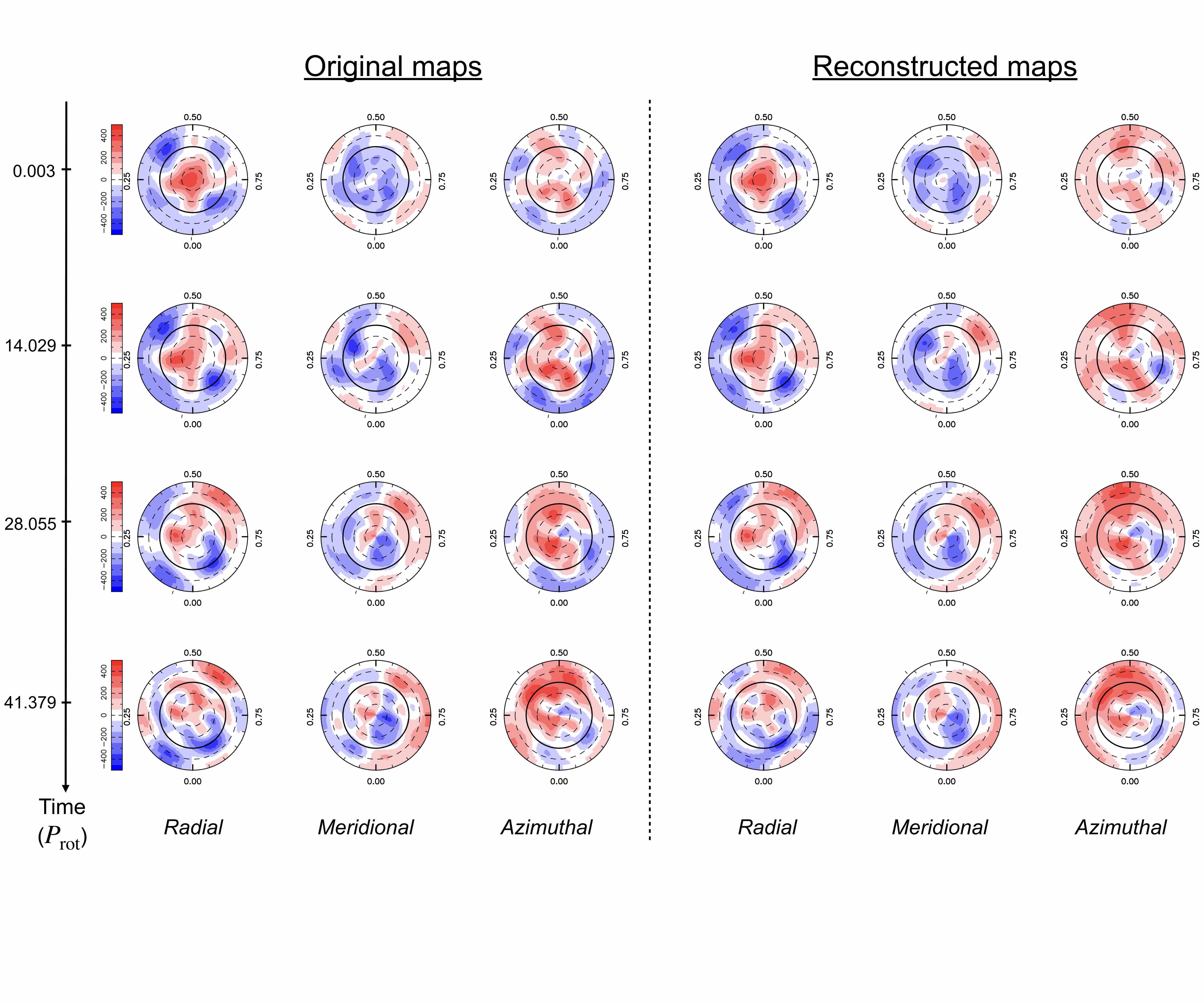} \hspace*{-0.8cm}
    \includegraphics[scale=0.12,trim={3cm 8cm 3cm 6cm},clip]{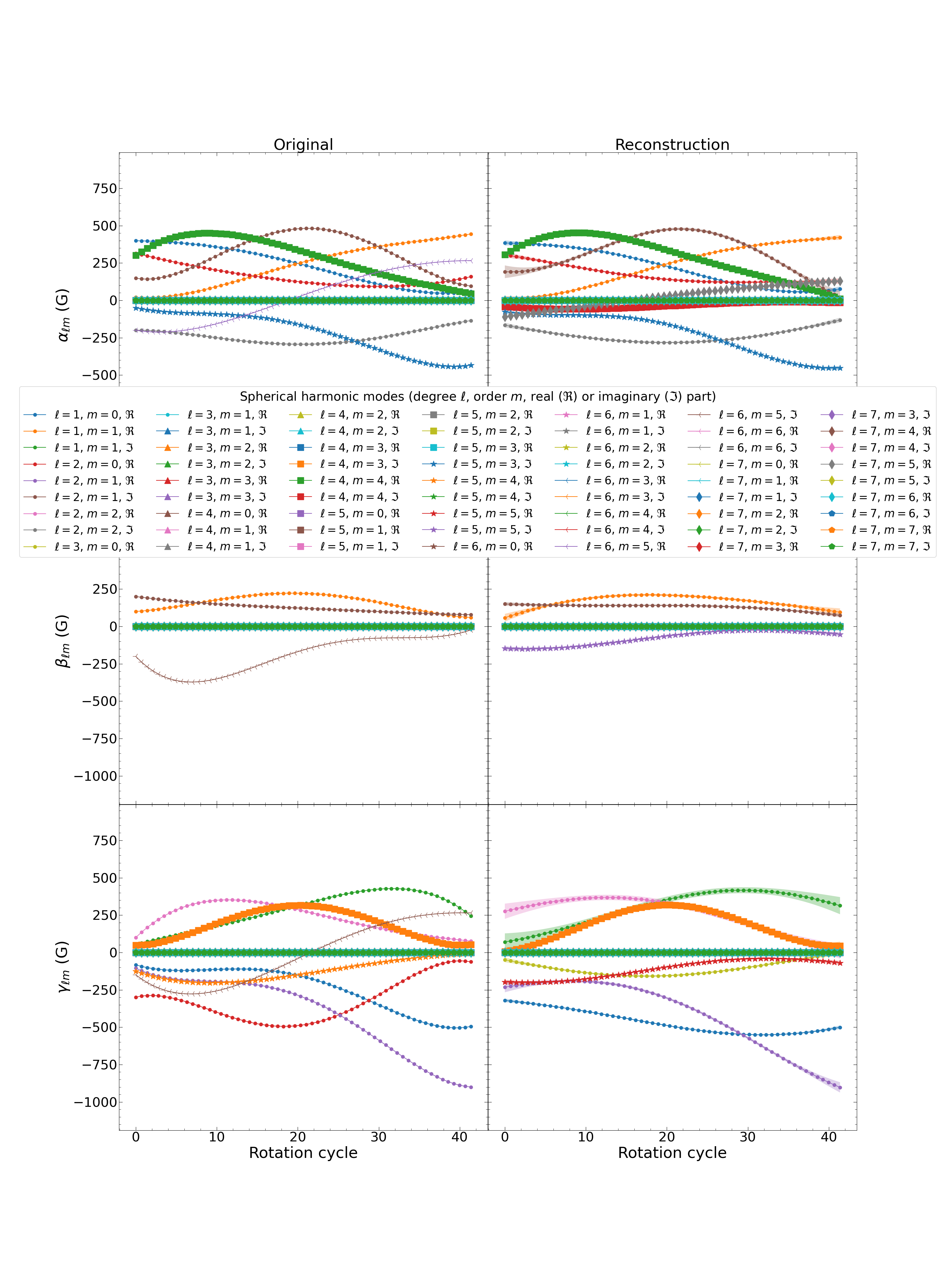}
    \includegraphics[scale=0.17,trim={0cm 0.5cm 0cm 0cm},clip]{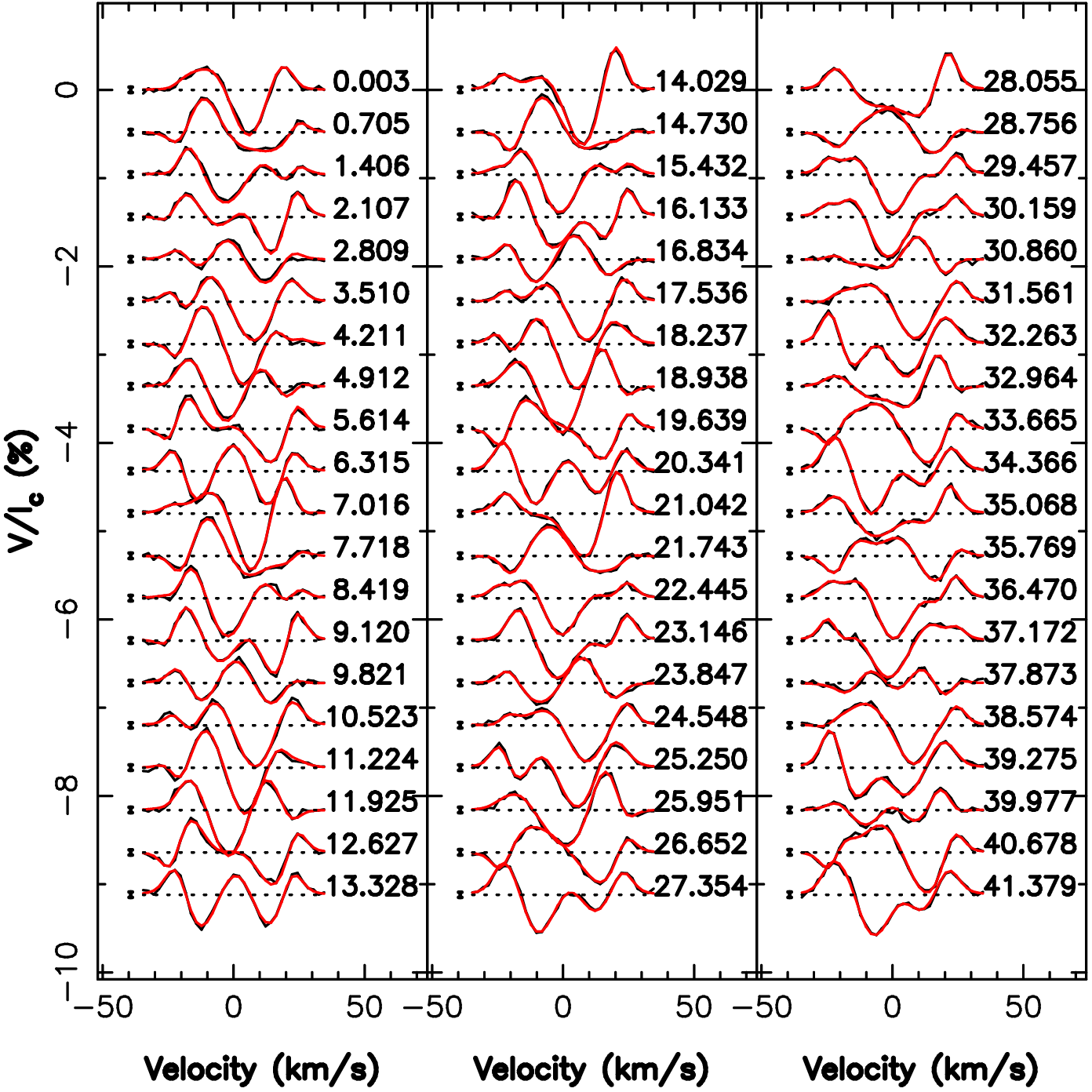}
    \caption{Same as Fig.~\ref{fig:maps_poloidal} for a complex topology described by spherical harmonics modes up to $\ell=6$ for a star featuring $i=60^\circ$ and $v\sin{i}=25$~\kms. The Stokes~$V$ profiles are fitted down to $\chi^2_r=1.19$.}
    \label{fig:vsini25}
\end{figure*}

We also simulated the case of a more complex magnetic topology described by spherical harmonic modes up to $\ell=6$. We therefore allowed modes up to a higher $\ell_{\max}$ (set to 10) than in the reference case (set to 5) in TIMeS, which increased the potential cross-talks between the modes in the mode identification / selection process. The fit of the Stokes~$V$ profiles is improved when considering a slightly larger $n$ (i.e. number of consecutive profiles in the sliding subsets used in the selection process), set to 10 for this Section, thereby increasing the sampling of the stellar rotation. The $v\sin{i}$ also plays a role in the reconstruction of complex magnetic fields as a higher $v\sin{i}$ increases the spatial resolution at the surface of the stars and thus allows the reconstruction of smaller features (i.e. modes of higher degree). We show the reconstructed topologies when assuming $i=60^\circ$ and $v\sin{i}$ equal to 25, 15 and 50~\kms\ in Figs.~\ref{fig:vsini25}, C1 and C2 (available as supplementary material), respectively, with differences mainly appearing for features having the weakest contribution to the Stokes~$V$ profiles (i.e. the smallest features and those located at low latitudes). 

\subsubsection{Stellar rotation period}
\label{sec:rotation_period}

We also applied our method to the same star as in Sec~\ref{sec:poloidal_toroidal} but with longer rotation periods up to 29~d. Fitting the Stokes~$V$ profiles with a consistent modeled topology requires an increase in the number of consecutive profiles considered in the selection process (i.e. $n\,\unsim10$). These $n$ profiles should efficiently sample the stellar rotation cycle so that our tomographic method can identify and select the modes consistent with the data. For the longest period, the field evolves significantly over one rotation cycle, meaning that the static field assumption is no longer verified even for the subsets of $n$ profiles used in the SPGL1 solver, therefore slightly degrading the final \chisqr. In this particular case, ensuring that the $n$ profiles sample about one rotation cycle avoids spurious modes to be selected by the reconstruction process.

\subsubsection{Quality of the data: SNR}
\label{sec:snr}
In this Section, we consider that all parameters are set to the value of the reference case (Sec.~\ref{sec:poloidal_toroidal}) except for the SNR, now decreased to 1000. For such a value, the smallest contributions to the Stokes~$V$ profiles (e.g. from the smallest features and those located at low latitudes) are hidden in the noise, which results in a less precise fit of the Stokes~$V$ profiles ($\chi^2_r=1.16$) and less detailed reconstructed maps. We nevertheless observe that the overall magnetic topology is well reconstructed (Fig.~\ref{fig:snr1000}).

We note that 2 $\alm$ and 3 $\glm$ coefficients are missed by our method and that all $\blm$ coefficients are not reconstructed, most likely due to their weak contribution to the Stokes~$V$ profiles and to the penalization weights used in the SPGL1 solver. We also see that the amplitude of the modeled time dependencies starts to depart from the injected ones as shown in Fig.~\ref{fig:snr1000}. These differences arise from the worse quality of the data, preventing the method to identify all the modes and their actual strength.

\subsubsection{Temporal sampling}
\label{sec:observational_strategy}

\begin{figure*}
    \centering \hspace*{-0.5cm}
    \includegraphics[scale=0.09,trim={.5cm 25.cm 0cm 9cm},clip]{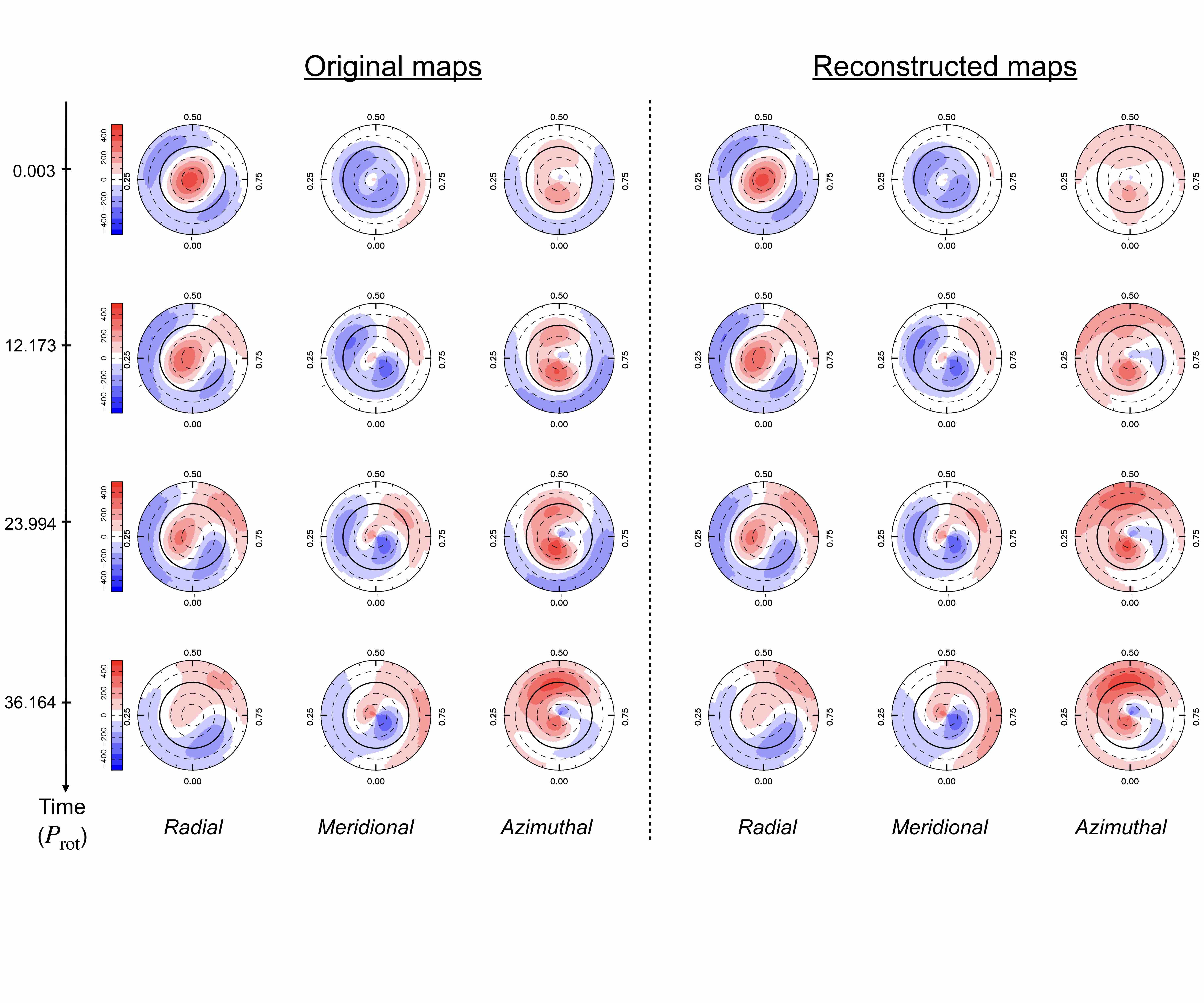} \hspace*{-0.8cm}
    \includegraphics[scale=0.16,trim={3cm 5.5cm 5cm 6cm},clip]{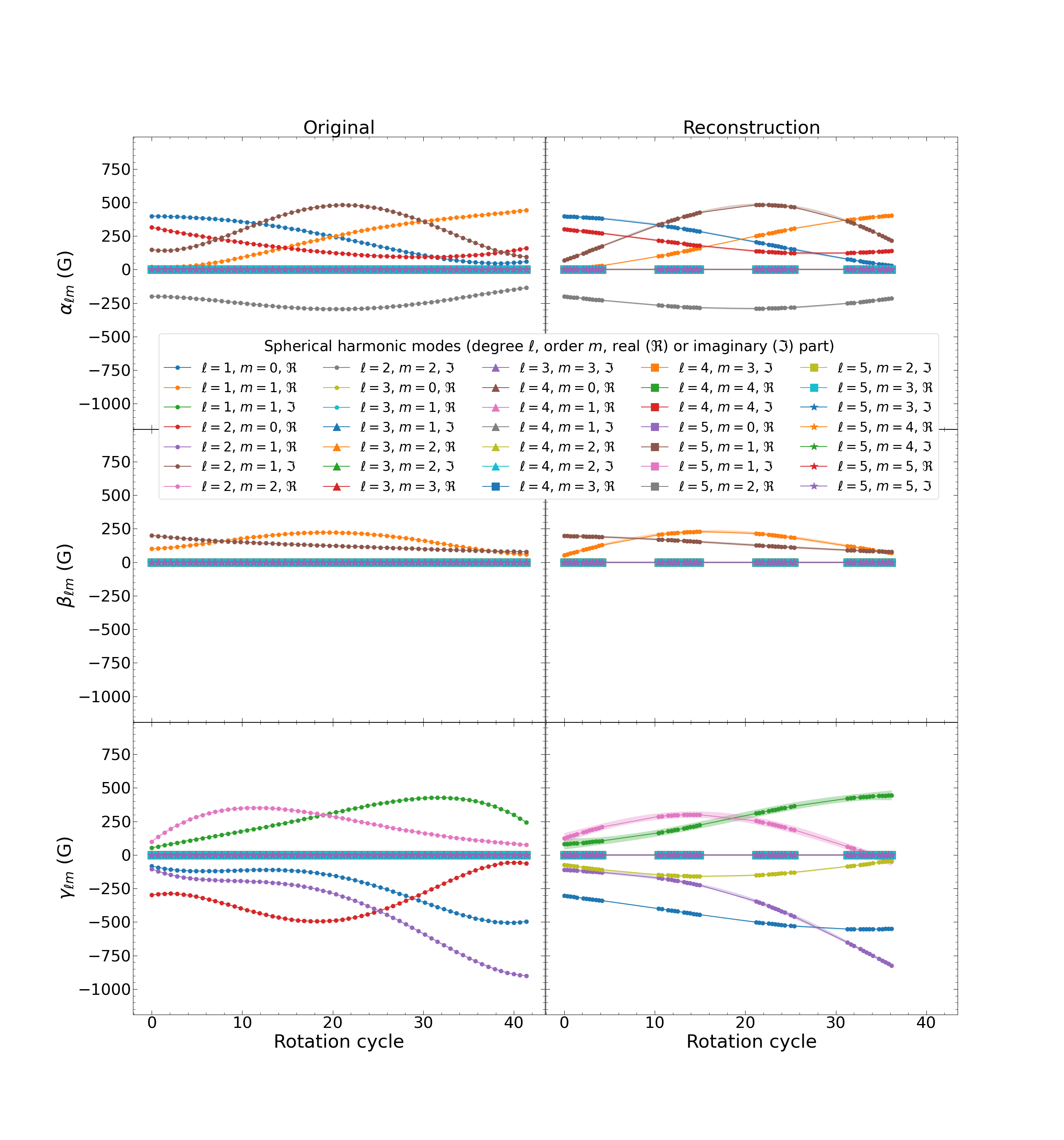}
     \includegraphics[scale=0.17,trim={0cm 0.5cm 0cm 0cm},clip]{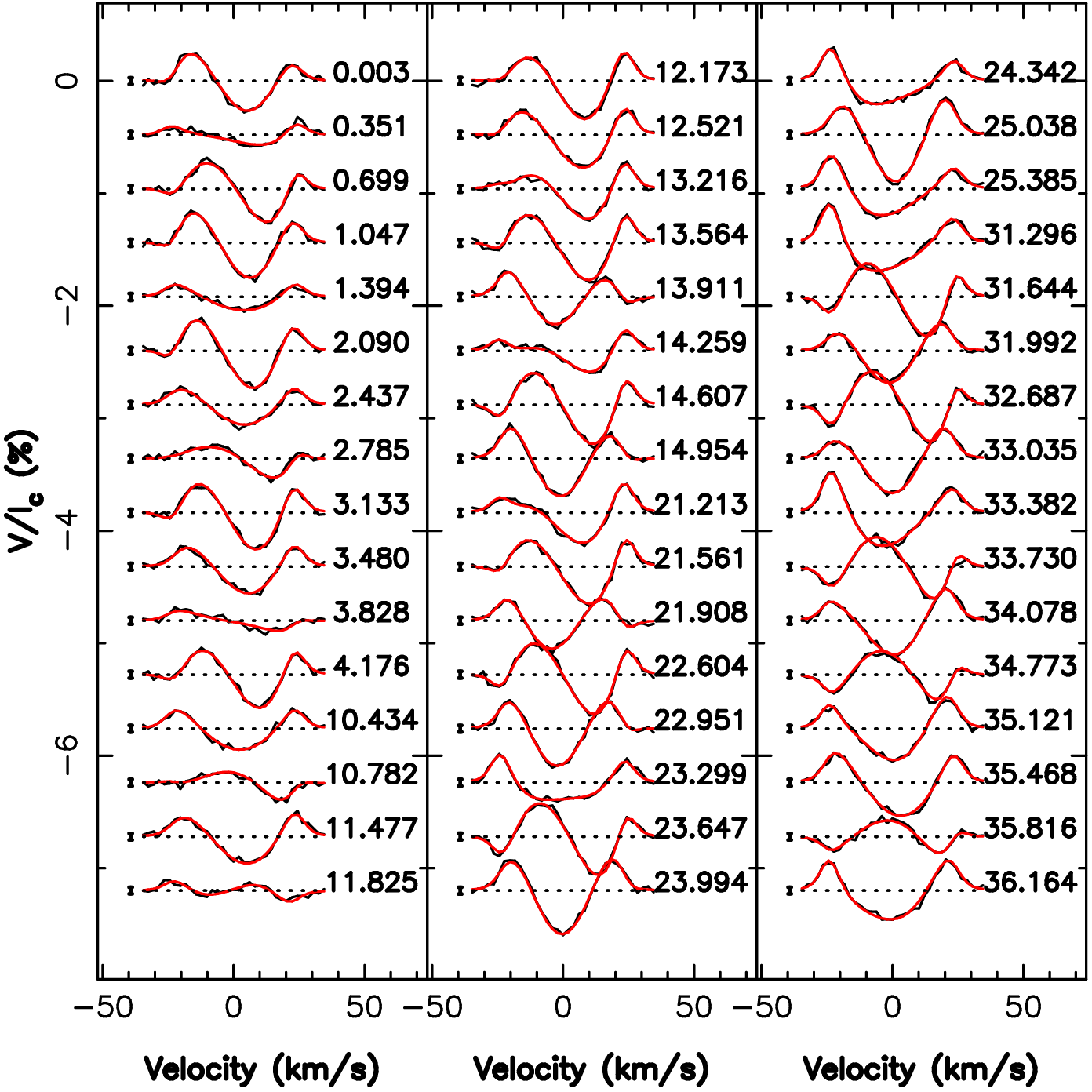}
    \caption{Same as Fig.~\ref{fig:maps_poloidal} for the star and magnetic field described in Sec.~\ref{sec:poloidal_toroidal} but with observations collected during 4 periods of 15~d (as in a typical SPIRou monitoring campaign).} 
    \label{fig:spirou}
\end{figure*}

We simulated a more realistic observational campaign of the star described in Sec.~\ref{sec:poloidal_toroidal}, mimicking a typical monitoring with SPIRou over a period of 4 months. This monitoring consists in 60 observations collected during 4 periods of 15~d (corresponding to bright time) separated by intervals of 15~d with no observation (corresponding to dark time), yielding an uneven sampling with large gaps. We also incorporated weather hazards by assuming a probability of 20\% that an observation is not validated or not carried out, yielding a full set of 48 Stokes~$V$ profiles with a SNR of 5000.

Despite these limitations, we are still able to estimate the decay timescale of the longitudinal field, found to be equal to $76\pm16$~d. 
For such a temporal sampling, we also ensure that the $n$ profiles of each subsets are not spread over 2 different observing periods since the magnetic field is expected to significantly evolve during the gaps. 
Covering $\unsim3.5$ rotation periods per subset as in the reference simulations of Sec.~\ref{sec:simulations}, requires to rise the number of profiles per subset to $n=10$. We are able to fit the Stokes~$V$ profiles down to $\chi^2_r=1.05$ with a magnetic topology similar to the simulated one, demonstrating that a realistic temporal sampling does not hinder our ability to reconstruct the evolving magnetic topology (Fig.~\ref{fig:spirou}).

\subsection{Effects of the method parameters}

We now discuss the impact of 2 main method parameters involved in the selection process: the penalization weights and the number of profiles $n$ in each subsets, both used in the SPGL1 solver.

\subsubsection{Choice of the penalization weights}

In Sec.~\ref{sec:selection_process}, we used different penalization weights for the reconstruction of the modes associated with the $\alm$, $\blm$ and $\glm$ coefficients. As Stokes~$V$ profiles mostly reflect the contribution of the radial field, described by the $\alm$ coefficients only, the amplitudes of the spectral signatures corresponding to the $\blm$ and $\glm$ coefficients in the database are typically twice smaller than those associated with $\alm$ coefficients; we therefore need to impose different proportionality factors to avoid the reconstruction of these modes to be too severely penalized (as mentioned in Sec.~\ref{sec:selection_process}). Our choice favors the simplest poloidal and toroidal components (i.e., with the lowest $\ell$ values) consistent with the data, adding non-zero $\blm$ coefficients only if these coefficients are truly necessary. 

Setting all the proportionality factors to 1 (instead of 0.8 and 0.5 for $\blm$ and $\glm$, respectively; see Sec.~\ref{sec:selection_process}) in the case of a purely toroidal field (Sec.~\ref{sec:toroidal}) yields a reconstructed topology that is not fully consistent with the input one (see Fig.~E1, available as supplementary material), as a weak poloidal field is also reconstructed, illustrating the need to adapt the weights for the modes associated with a $\blm$ or $\glm$ coefficient.

In addition, the weights we imposed in the selection process are proportional to the degree $\ell$ of the modes as this choice provides the best results. Using a higher power of $\ell$ can prevent the selection of significant modes of high degree while using a lower power can allow several spurious modes of high degree to be selected, making the modeled topology more complex.

\subsubsection{Number of profiles in each subsets}
\label{sec:number_profiles_subsets}

Let us now focus on another key parameter, the number $n$ of consecutive profiles in the subsets (see Sec.~\ref{sec:selection_process}). Our tomographic method requires that $n$ is large enough to sample the rotation cycle, but small enough to ensure that the field variation is minimal over the time interval spanned by the $n$ profiles.

For instance, when considering the reference case (Sec.~\ref{sec:poloidal_toroidal}), we found that $n=6$~--~10 provides reconstructed magnetic maps similar to the injected ones (with similar values of $\chi^2_r$). Increasing $n$ results in a better sampling of the rotation cycle but with all profiles providing information that is no longer self consistent due to the evolution of the field, yielding models that start to diverge from the input topology.
We show the reconstructed maps obtained when setting $n$ to 3 (i.e. 1.4 rotation cycles or 4.1~d) and 12 (i.e. 7.7 rotation cycles or 22.3~d) in Figs.~F1 and F2 (available as supplementary material), illustrating that our method fails when (i) the sampling of the stellar rotation period is not sufficient or (ii)~the field strongly changes during the time interval spanned by the $n$ profiles. 

More generally, fixing $n$ so that the $n$ profiles cover 10--20\% of the decay timescale $\theta_{B_\ell}$ while still sampling at least one rotation cycle, is found to yield the best identification of the modes describing the field. 

\section{Summary and discussion}
\label{sec:summary}

We presented a new method, called TIMeS, for reliably reconstructing stellar large-scale magnetic topologies, that evolve with time over the observing period, from time series of Stokes~$V$ spectra, assuming that the field to be reconstructed is not too strong to ensure a linear relation between the Stokes $V$ data and the evolving magnetic map. This new method uses (i) sparse approximation to select as few modes as possible to describe the simplest magnetic topology consistent with the data and (ii) GPR to model the evolution of these modes with time.

We intentionally showed the results for a simple topology ($\ell_{\max}=2$) to start with (Sec.~\ref{sec:simulations}). Such topologies are actually not unusual, even for rapid rotators for which the reconstructed magnetic energy is frequently concentrated in low order modes (e.g. V374~Peg, \citealt{donati06b}; LkCa~4,  \citealt{donati14}). We also applied our new method to a more complex topology ($\ell_{\max}=6$; Sec.~\ref{sec:complex_field}) so that all the results presented in this paper provide a fair description of what TIMeS is capable of.

Our simulations show that our method is able to reliably reconstruct magnetic topologies evolving with time, the magnetic strength being recovered with a precision down to a fraction of percent in optimal conditions (up to a few percents for low $v\sin{i}$), and succeeds at disentangling their poloidal and toroidal components. Applying the classical version of ZDI to the reference case (Sec.~\ref{sec:poloidal_toroidal}) yields inconsistent model with a \chisqr\ about 8 times larger than the one we obtain with our new method, clearly demonstrating the improvement brought by the proposed imaging scheme.

The magnetic reconstructed topology can suffer from cross-talk between modes, especially when the input magnetic field is complex, i.e. characterized by modes of high degree. Such cross-talks generally occur between spherical harmonics modes of order $m$ ($\geq0$, i.e. not only axisymmetric modes) and degree $\ell-1$, $\ell$ and $\ell+1$ (or even $\ell+2$). However, these cross-talks only have a minor impact on the reconstructed magnetic topology, affecting mainly low-latitude features whose contribution to the data is very small (as a result of both limb darkening and limited visibility).

Applying TIMeS to data sets for various stellar parameters, temporal samplings or method parameters shows that it works best in conditions similar to those needed for ZDI (i.e. rapid rotators, no extreme inclinations). However, the input Stokes~$V$ are never fitted down to exactly a unit \chisqr, indicating that small discrepancies remain between our reconstructed and input Stokes~$V$ profiles. These discrepancies mainly reflect errors in the process that result from the evolution of the field within the sliding subsets of profiles used to reconstruct the parent time-dependent magnetic topology.

When the field becomes too strong, the assumption of a linear relation between the Stokes~$V$ profiles and the reconstructed image starts to break down preventing one to apply TIMeS in its current implementation. We estimate conservative upper limits for the magnetic flux $B$ of $\unsim1.3$, $\unsim2.0$, $\unsim2.8$ and $\unsim4.3$~kG, for a filling factor $f=1$ and $v\sin{i}=5$, 15, 25 and 50~\kms, respectively, corresponding to thresholds at which the maximal amplitude of the Stokes~$V$ profiles associated with the reconstructed image is 25 times larger than the rms of the difference between the true Stokes~$V$ profiles and those approximated by TIMeS. These upper limits are reduced to 0.5, 0.7, 0.9 and 1.3~kG when assuming a filling factor $f=0.3$ and remain above the maximum value of $B$ in our simulations but for one case, confirming that the linear approximation underlying TIMeS is not an issue for the topologies considered in this paper, except when $f=0.3$ and $v\sin{i}=5$~\kms\ (Secs.~\ref{sec:poloidal_toroidal} and \ref{sec:vsini}). The results obtained in this extreme case further indicate that the derived upper limits are reliable. TIMeS can therefore be applied to most data sets of low-mass stars studied so far with ZDI, even those featuring strong magnetic fields like AU~Mic \citep{klein21}. However, for magnetic fluxes stronger than the limits mentioned above, the hypothesis of linearity is no longer valid; for instance, for a 10~kG field and a filling factor $f=1.0$, we find that the maximum amplitudes of the Stokes~$V$ profiles are only 7 and 0.7 times larger than the rms of the difference between the true and approximated Stokes~$V$ profiles, for $v\sin{i}=50$ and 5~\kms, respectively, thus too small for TIMeS to behave reliably.

A method whose mode selection process relied on the principle of maximum entropy was also implemented and tested on the same reference cases. In practice, we identified and added the modes to the model through an iterative process until finding the best fit to the observed Stokes~$V$ profiles, unlike TIMeS that directly selects the smallest number of modes consistent with the data using sparse approximation. This alternative approach however suffers from a longer computation time and usually yields more complex models that do not significantly improve the results.

To pursue the work initiated in this paper, TIMeS will be applied on actual spectropolarimetric data to model and characterize the magnetic field, and its evolution on short and intermediate timescales, of active PMS stars such as V1298~Tau and AU~Mic. In addition, a complementary approach based on Principal Component Analysis (PCA) will now be envisaged to improve the sampling of the stellar rotation period and therefore overcome the major limitation of our method (Sec.~\ref{sec:number_profiles_subsets}). Finally, we will adapt TIMeS to the reconstruction of the brightness distribution at the surface of active stars. This should allow us in particular to improve the modeling and filtering of the activity jitter in the radial velocity curves of such stars, thereby enhancing our ability at detecting potential close-in planets whose velocimetric signatures are still hidden in the activity jitter of their host stars.

\section*{Acknowledgements}
We acknowledge funding from the European Research Council (ERC) under the H2020 research \& innovation programme (grant agreement \#740651 NewWorlds).
We thank Nathan Hara for insightful comments about sparse approximation that helped improve the robustness of our method.
We thank the referee for valuable
comments and suggestions that improved the manuscript
\section*{Data Availability}

The simulated data underlying this article will be shared on reasonable request to the corresponding author.
The code developed in this study is still in development and not yet publicly available.

\bibliographystyle{mnras}
\bibliography{main} 

\begin{thebibliography}{}
\makeatletter
\relax
\def\mn@urlcharsother{\let\do\@makeother \do\$\do\&\do\#\do\^\do\_\do\%\do\~}
\def\mn@doi{\begingroup\mn@urlcharsother \@ifnextchar [ {\mn@doi@}
  {\mn@doi@[]}}
\def\mn@doi@[#1]#2{\def\@tempa{#1}\ifx\@tempa\@empty \href
  {http://dx.doi.org/#2} {doi:#2}\else \href {http://dx.doi.org/#2} {#1}\fi
  \endgroup}
\def\mn@eprint#1#2{\mn@eprint@#1:#2::\@nil}
\def\mn@eprint@arXiv#1{\href {http://arxiv.org/abs/#1} {{\tt arXiv:#1}}}
\def\mn@eprint@dblp#1{\href {http://dblp.uni-trier.de/rec/bibtex/#1.xml}
  {dblp:#1}}
\def\mn@eprint@#1:#2:#3:#4\@nil{\def\@tempa {#1}\def\@tempb {#2}\def\@tempc
  {#3}\ifx \@tempc \@empty \let \@tempc \@tempb \let \@tempb \@tempa \fi \ifx
  \@tempb \@empty \def\@tempb {arXiv}\fi \@ifundefined
  {mn@eprint@\@tempb}{\@tempb:\@tempc}{\expandafter \expandafter \csname
  mn@eprint@\@tempb\endcsname \expandafter{\@tempc}}}

\bibitem[\protect\citeauthoryear{{Brown}, {Donati}, {Rees}  \& {Semel}}{{Brown}
  et~al.}{1991}]{brown91}
{Brown} S.~F.,  {Donati} J.~F.,  {Rees} D.~E.,   {Semel} M.,  1991, A\&A, \href
  {https://ui.adsabs.harvard.edu/abs/1991A&A...250..463B} {250, 463}

\bibitem[\protect\citeauthoryear{Chen, Donoho  \& Saunders}{Chen
  et~al.}{1998}]{chen98}
Chen S.~S.,  Donoho D.~L.,   Saunders M.~A.,  1998, \mn@doi [SIAM Journal on
  Scientific Computing] {10.1137/S1064827596304010}, 20, 33

\bibitem[\protect\citeauthoryear{{Donati} \& {Brown}}{{Donati} \&
  {Brown}}{1997}]{donatibrown97}
{Donati} J.~F.,  {Brown} S.~F.,  1997, A\&A, \href
  {https://ui.adsabs.harvard.edu/abs/1997A&A...326.1135D} {326, 1135}

\bibitem[\protect\citeauthoryear{{Donati}, {Semel}  \& {Praderie}}{{Donati}
  et~al.}{1989}]{donati89}
{Donati} J.~F.,  {Semel} M.,   {Praderie} F.,  1989, A\&A, \href
  {https://ui.adsabs.harvard.edu/abs/1989A&A...225..467D} {225, 467}

\bibitem[\protect\citeauthoryear{{Donati}, {Semel}, {Carter}, {Rees}  \&
  {Collier Cameron}}{{Donati} et~al.}{1997}]{donati97}
{Donati} J.-F.,  {Semel} M.,  {Carter} B.~D.,  {Rees} D.~E.,   {Collier
  Cameron} A.,  1997, \mn@doi [MNRAS] {10.1093/mnras/291.4.658}, \href
  {https://ui.adsabs.harvard.edu/abs/1997MNRAS.291..658D} {291, 658}

\bibitem[\protect\citeauthoryear{{Donati}, {Mengel}, {Carter}, {Marsden},
  {Collier Cameron}  \& {Wichmann}}{{Donati} et~al.}{2000}]{donati2000}
{Donati} J.~F.,  {Mengel} M.,  {Carter} B.~D.,  {Marsden} S.,  {Collier
  Cameron} A.,   {Wichmann} R.,  2000, \mn@doi [\mnras]
  {10.1046/j.1365-8711.2000.03570.x}, \href
  {https://ui.adsabs.harvard.edu/abs/2000MNRAS.316..699D} {316, 699}

\bibitem[\protect\citeauthoryear{{Donati}, {Forveille}, {Collier Cameron},
  {Barnes}, {Delfosse}, {Jardine}  \& {Valenti}}{{Donati}
  et~al.}{2006a}]{donati06b}
{Donati} J.-F.,  {Forveille} T.,  {Collier Cameron} A.,  {Barnes} J.~R.,
  {Delfosse} X.,  {Jardine} M.~M.,   {Valenti} J.~A.,  2006a, \mn@doi [Science]
  {10.1126/science.1121102}, \href
  {https://ui.adsabs.harvard.edu/abs/2006Sci...311..633D} {311, 633}

\bibitem[\protect\citeauthoryear{{Donati} et~al.,}{{Donati}
  et~al.}{2006b}]{donati06}
{Donati} J.~F.,  et~al., 2006b, \mn@doi [MNRAS]
  {10.1111/j.1365-2966.2006.10558.x}, \href
  {https://ui.adsabs.harvard.edu/abs/2006MNRAS.370..629D} {370, 629}

\bibitem[\protect\citeauthoryear{{Donati} et~al.,}{{Donati}
  et~al.}{2013}]{donati13}
{Donati} J.~F.,  et~al., 2013, \mn@doi [\mnras] {10.1093/mnras/stt1622}, \href
  {https://ui.adsabs.harvard.edu/abs/2013MNRAS.436..881D} {436, 881}

\bibitem[\protect\citeauthoryear{{Donati} et~al.,}{{Donati}
  et~al.}{2014}]{donati14}
{Donati} J.~F.,  et~al., 2014, \mn@doi [MNRAS] {10.1093/mnras/stu1679}, \href
  {https://ui.adsabs.harvard.edu/abs/2014MNRAS.444.3220D} {444, 3220}

\bibitem[\protect\citeauthoryear{{Donati} et~al.,}{{Donati}
  et~al.}{2016}]{donati16}
{Donati} J.~F.,  et~al., 2016, \mn@doi [Nature] {10.1038/nature18305}, \href
  {https://ui.adsabs.harvard.edu/abs/2016Natur.534..662D} {534, 662}

\bibitem[\protect\citeauthoryear{{Donati} et~al.,}{{Donati}
  et~al.}{2017}]{donati17}
{Donati} J.~F.,  et~al., 2017, \mn@doi [MNRAS] {10.1093/mnras/stw2904}, \href
  {https://ui.adsabs.harvard.edu/abs/2017MNRAS.465.3343D} {465, 3343}

\bibitem[\protect\citeauthoryear{{Donati} et~al.,}{{Donati}
  et~al.}{2019}]{donati19}
{Donati} J.~F.,  et~al., 2019, \mn@doi [\mnras] {10.1093/mnrasl/sly207}, \href
  {https://ui.adsabs.harvard.edu/abs/2019MNRAS.483L...1D} {483, L1}

\bibitem[\protect\citeauthoryear{{Donati} et~al.,}{{Donati}
  et~al.}{2020}]{donati20}
{Donati} J.~F.,  et~al., 2020, \mn@doi [MNRAS] {10.1093/mnras/stz3368}, \href
  {https://ui.adsabs.harvard.edu/abs/2020MNRAS.491.5660D} {491, 5660}

\bibitem[\protect\citeauthoryear{Donoho \& Elad}{Donoho \&
  Elad}{2003}]{donoho03}
Donoho D.~L.,  Elad M.,  2003, \mn@doi [Proceedings of the National Academy of
  Sciences] {10.1073/pnas.0437847100}, 100, 2197

\bibitem[\protect\citeauthoryear{Donoho, Elad  \& Temlyakov}{Donoho
  et~al.}{2006}]{donoho06}
Donoho D.,  Elad M.,   Temlyakov V.,  2006, \mn@doi [IEEE Transactions on
  Information Theory] {10.1109/TIT.2005.860430}, 52, 6

\bibitem[\protect\citeauthoryear{{Folsom} et~al.,}{{Folsom}
  et~al.}{2018}]{folsom18}
{Folsom} C.~P.,  et~al., 2018, \mn@doi [\mnras] {10.1093/mnras/stx3021}, \href
  {https://ui.adsabs.harvard.edu/abs/2018MNRAS.474.4956F} {474, 4956}

\bibitem[\protect\citeauthoryear{{Foreman-Mackey}, {Hogg}, {Lang}  \&
  {Goodman}}{{Foreman-Mackey} et~al.}{2013}]{emcee}
{Foreman-Mackey} D.,  {Hogg} D.~W.,  {Lang} D.,   {Goodman} J.,  2013, \mn@doi
  [PASP] {10.1086/670067}, \href
  {https://ui.adsabs.harvard.edu/abs/2013PASP..125..306F} {125, 306}

\bibitem[\protect\citeauthoryear{{Hackman}, {Lehtinen}, {Ros{\'e}n},
  {Kochukhov}  \& {K{\"a}pyl{\"a}}}{{Hackman} et~al.}{2016}]{hackman16}
{Hackman} T.,  {Lehtinen} J.,  {Ros{\'e}n} L.,  {Kochukhov} O.,
  {K{\"a}pyl{\"a}} M.~J.,  2016, \mn@doi [\aap] {10.1051/0004-6361/201527320},
  \href {https://ui.adsabs.harvard.edu/abs/2016A&A...587A..28H} {587, A28}

\bibitem[\protect\citeauthoryear{{Klein}, {Donati}, {H{\'e}brard}, {Zaire},
  {Folsom}, {Morin}, {Delfosse}  \& {Bonfils}}{{Klein}
  et~al.}{2021a}]{klein21b}
{Klein} B.,  {Donati} J.-F.,  {H{\'e}brard} {\'E}.~M.,  {Zaire} B.,  {Folsom}
  C.~P.,  {Morin} J.,  {Delfosse} X.,   {Bonfils} X.,  2021a, \mn@doi [\mnras]
  {10.1093/mnras/staa3396}, \href
  {https://ui.adsabs.harvard.edu/abs/2021MNRAS.500.1844K} {500, 1844}

\bibitem[\protect\citeauthoryear{{Klein} et~al.,}{{Klein}
  et~al.}{2021b}]{klein21}
{Klein} B.,  et~al., 2021b, \mn@doi [\mnras] {10.1093/mnras/staa3702}, \href
  {https://ui.adsabs.harvard.edu/abs/2021MNRAS.502..188K} {502, 188}

\bibitem[\protect\citeauthoryear{{Landi Degl'Innocenti} \& {Landolfi}}{{Landi
  Degl'Innocenti} \& {Landolfi}}{2004}]{landi04}
{Landi Degl'Innocenti} E.,  {Landolfi} M.,  2004, Polarisation in Spectral
  Lines.
Kluwer Academic Publishers, Dordrecht

\bibitem[\protect\citeauthoryear{{Lavail}, {Kochukhov}  \& {Wade}}{{Lavail}
  et~al.}{2018}]{lavail18}
{Lavail} A.,  {Kochukhov} O.,   {Wade} G.~A.,  2018, \mn@doi [\mnras]
  {10.1093/mnras/sty1825}, \href
  {https://ui.adsabs.harvard.edu/abs/2018MNRAS.479.4836L} {479, 4836}

\bibitem[\protect\citeauthoryear{Mallat \& Zhang}{Mallat \&
  Zhang}{1993}]{mallat93}
Mallat S.,  Zhang Z.,  1993, \mn@doi [IEEE Transactions on Signal Processing]
  {10.1109/78.258082}, 41, 3397

\bibitem[\protect\citeauthoryear{{Morin} et~al.,}{{Morin}
  et~al.}{2008}]{morin08}
{Morin} J.,  et~al., 2008, \mn@doi [\mnras] {10.1111/j.1365-2966.2008.13809.x},
  \href {https://ui.adsabs.harvard.edu/abs/2008MNRAS.390..567M} {390, 567}

\bibitem[\protect\citeauthoryear{Morin, Donati, Petit, Delfosse, Forveille  \&
  Jardine}{Morin et~al.}{2010}]{morin10}
Morin J.,  Donati J.-F.,  Petit P.,  Delfosse X.,  Forveille T.,   Jardine
  M.~M.,  2010, \mn@doi [MNRAS] {10.1111/j.1365-2966.2010.17101.x}, 407, 2269

\bibitem[\protect\citeauthoryear{{Petit} et~al.,}{{Petit}
  et~al.}{2008}]{petit08}
{Petit} P.,  et~al., 2008, \mn@doi [\mnras] {10.1111/j.1365-2966.2008.13411.x},
  \href {https://ui.adsabs.harvard.edu/abs/2008MNRAS.388...80P} {388, 80}

\bibitem[\protect\citeauthoryear{{Rajpaul}, {Aigrain}, {Osborne}, {Reece}  \&
  {Roberts}}{{Rajpaul} et~al.}{2015}]{rajpaul15}
{Rajpaul} V.,  {Aigrain} S.,  {Osborne} M.~A.,  {Reece} S.,   {Roberts} S.,
  2015, \mn@doi [MNRAS] {10.1093/mnras/stv1428}, \href
  {https://ui.adsabs.harvard.edu/abs/2015MNRAS.452.2269R} {452, 2269}

\bibitem[\protect\citeauthoryear{Rasmussen \& Williams}{Rasmussen \&
  Williams}{2006}]{rasmussen06}
Rasmussen C.,  Williams C.,  2006, Gaussian Processes for Machine Learning.
Adaptive Computation and Machine Learning, MIT Press, Cambridge, MA, USA

\bibitem[\protect\citeauthoryear{{Ros{\'e}n}, {Kochukhov}  \&
  {Wade}}{{Ros{\'e}n} et~al.}{2015}]{rosen15}
{Ros{\'e}n} L.,  {Kochukhov} O.,   {Wade} G.~A.,  2015, \mn@doi [\apj]
  {10.1088/0004-637X/805/2/169}, \href
  {https://ui.adsabs.harvard.edu/abs/2015ApJ...805..169R} {805, 169}

\bibitem[\protect\citeauthoryear{{Semel}}{{Semel}}{1989}]{semel89}
{Semel} M.,  1989, A\&A, \href
  {https://ui.adsabs.harvard.edu/abs/1989A&A...225..456S} {225, 456}

\bibitem[\protect\citeauthoryear{{Skilling} \& {Bryan}}{{Skilling} \&
  {Bryan}}{1984}]{skilling84}
{Skilling} J.,  {Bryan} R.~K.,  1984, \mn@doi [MNRAS]
  {10.1093/mnras/211.1.111}, \href
  {https://ui.adsabs.harvard.edu/abs/1984MNRAS.211..111S} {211, 111}

\bibitem[\protect\citeauthoryear{Tibshirani}{Tibshirani}{1996}]{tibshirani96}
Tibshirani R.,  1996, \mn@doi [Journal of the Royal Statistical Society: Series
  B (Methodological)] {https://doi.org/10.1111/j.2517-6161.1996.tb02080.x}, 58,
  267

\bibitem[\protect\citeauthoryear{{Yu} et~al.,}{{Yu} et~al.}{2017}]{yu17}
{Yu} L.,  et~al., 2017, \mn@doi [MNRAS] {10.1093/mnras/stx009}, \href
  {https://ui.adsabs.harvard.edu/abs/2017MNRAS.467.1342Y} {467, 1342}

\bibitem[\protect\citeauthoryear{{Yu} et~al.,}{{Yu} et~al.}{2019}]{yu19}
{Yu} L.,  et~al., 2019, \mn@doi [MNRAS] {10.1093/mnras/stz2481}, \href
  {https://ui.adsabs.harvard.edu/abs/2019MNRAS.489.5556Y} {489, 5556}

\bibitem[\protect\citeauthoryear{van~den Berg \& Friedlander}{van~den Berg \&
  Friedlander}{2009}]{vandenberg08}
van~den Berg E.,  Friedlander M.~P.,  2009, \mn@doi [SIAM Journal on Scientific
  Computing] {10.1137/080714488}, 31, 890

\makeatother
\end{thebibliography}

\appendix

\section{Impact of stellar inclination}

\begin{figure*}
    \centering \hspace*{-0.5cm}
    \includegraphics[scale=0.09,trim={.5cm 27.cm 0cm 9cm},clip]{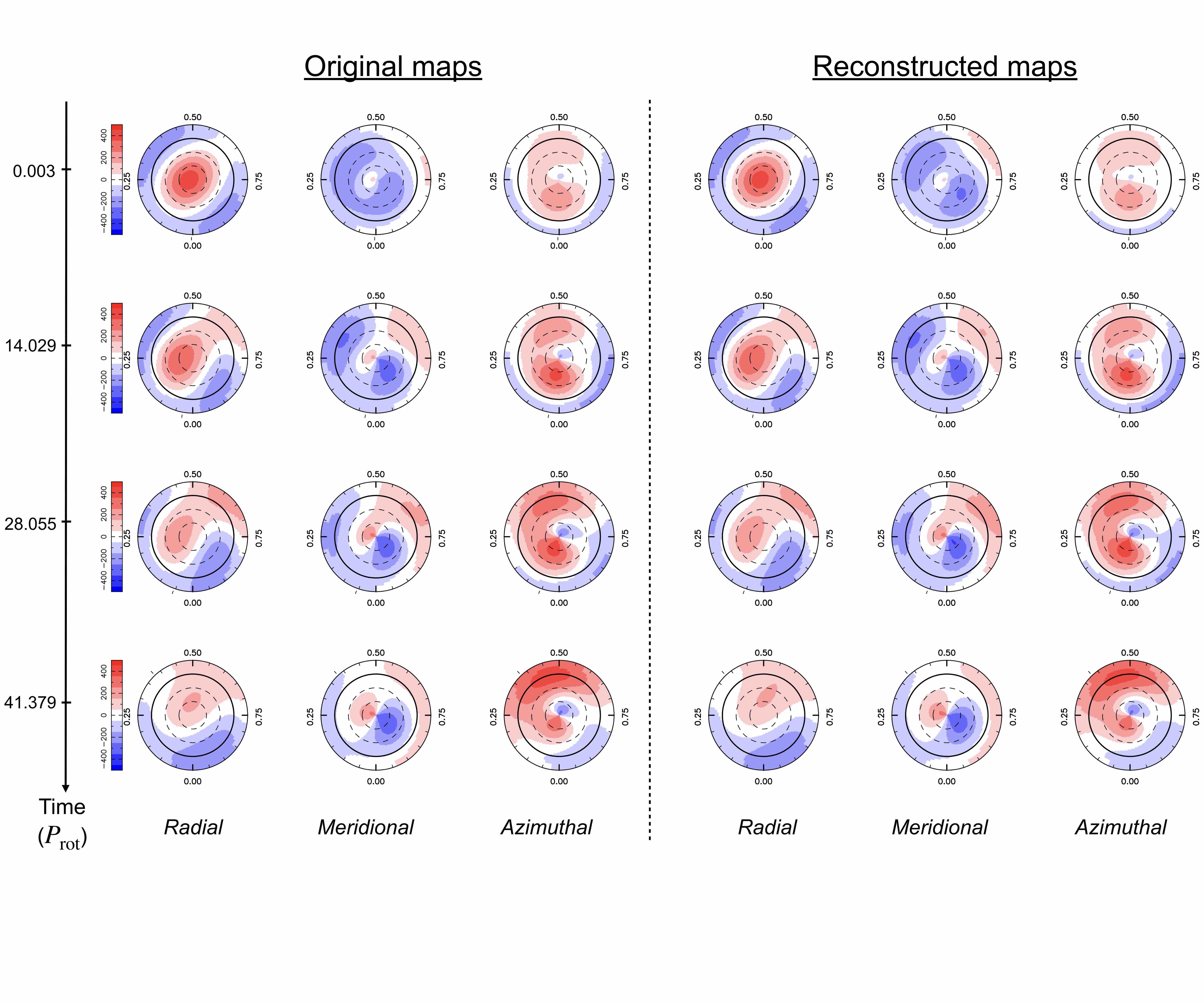} \hspace*{-0.8cm}
    \includegraphics[scale=0.16,trim={3cm 5.5cm 5cm 6cm},clip]{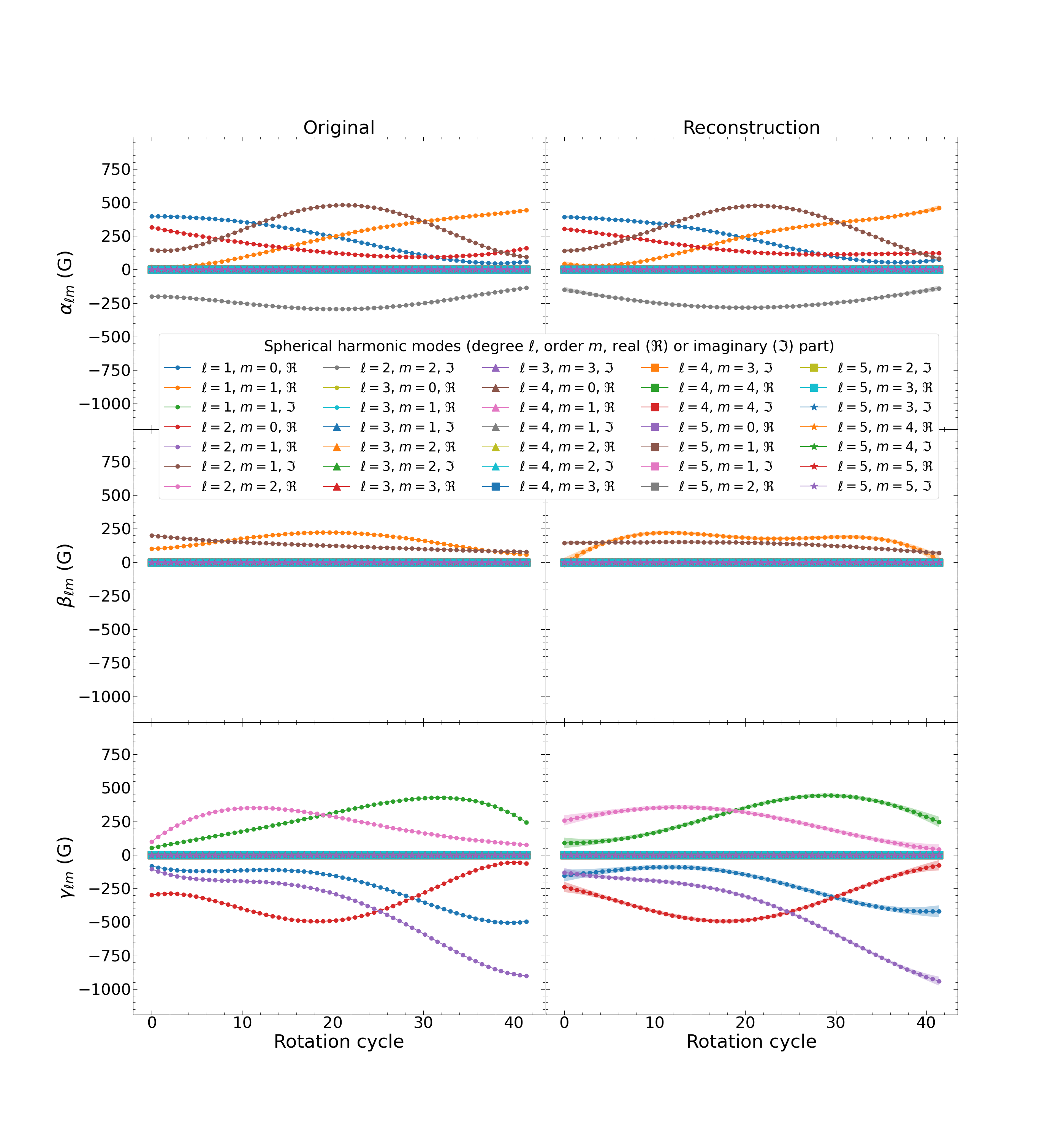}
    \includegraphics[scale=0.17,trim={0cm 0.5cm 0cm 0cm},clip]{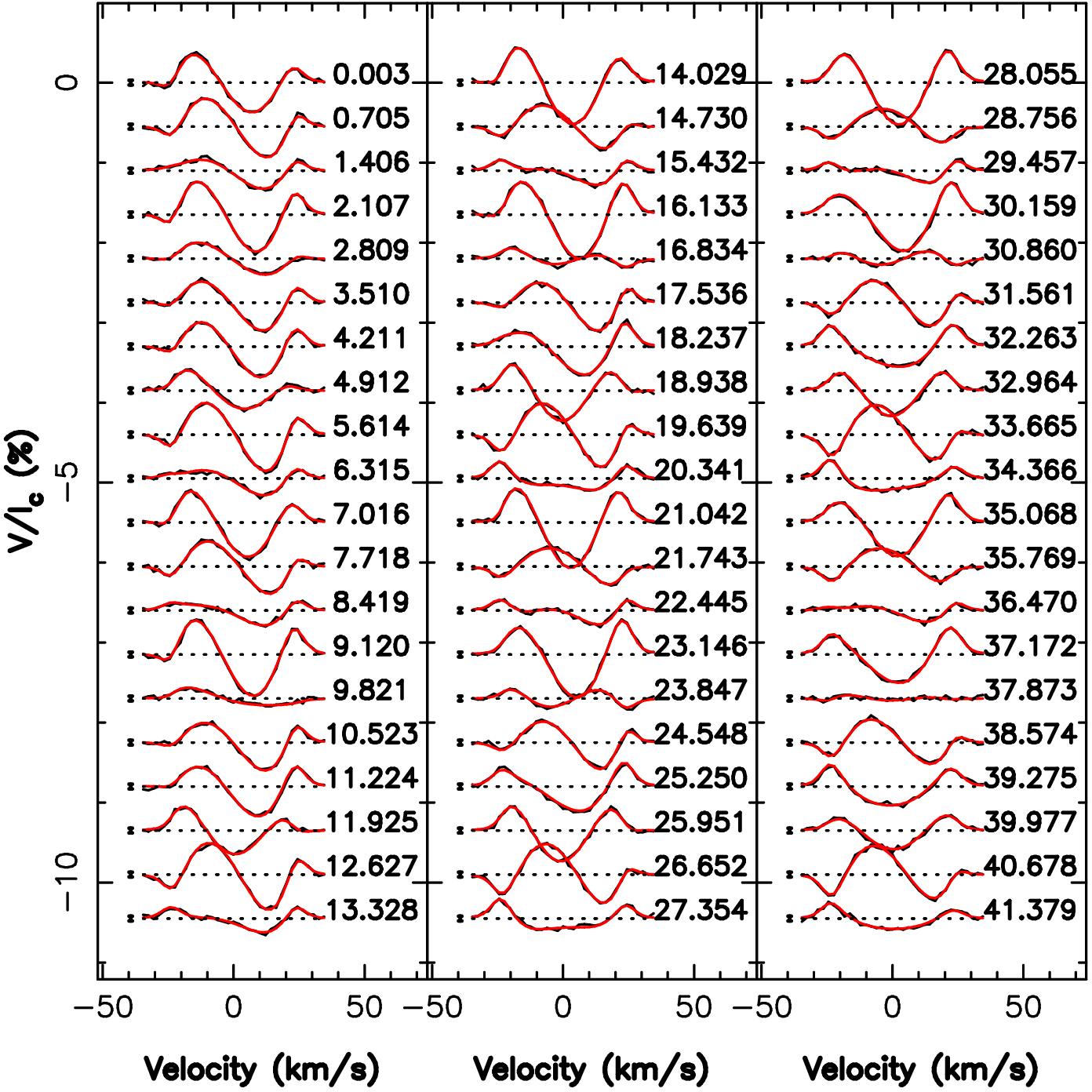}
    \caption{Reconstruction of the same topology as in the reference case (Sec.~\ref{sec:poloidal_toroidal}) for a star with $i=30^\circ$. \textit{Top}: Reconstructed maps with the star shown in a flattened polar view. \textit{Bottom left}: Comparison of the time dependencies of the input and reconstructed coefficients describing the field. \textit{Bottom right}: Observed (black) and reconstructed (red) Stokes~$V$ profiles, along with the $3\sigma$ error bars on the left of each profile. } 
    \label{fig:inc30}
\end{figure*}

\begin{figure*}
    \centering \hspace*{-0.5cm}
    \includegraphics[scale=0.09,trim={.5cm 20.cm 0cm 9cm},clip]{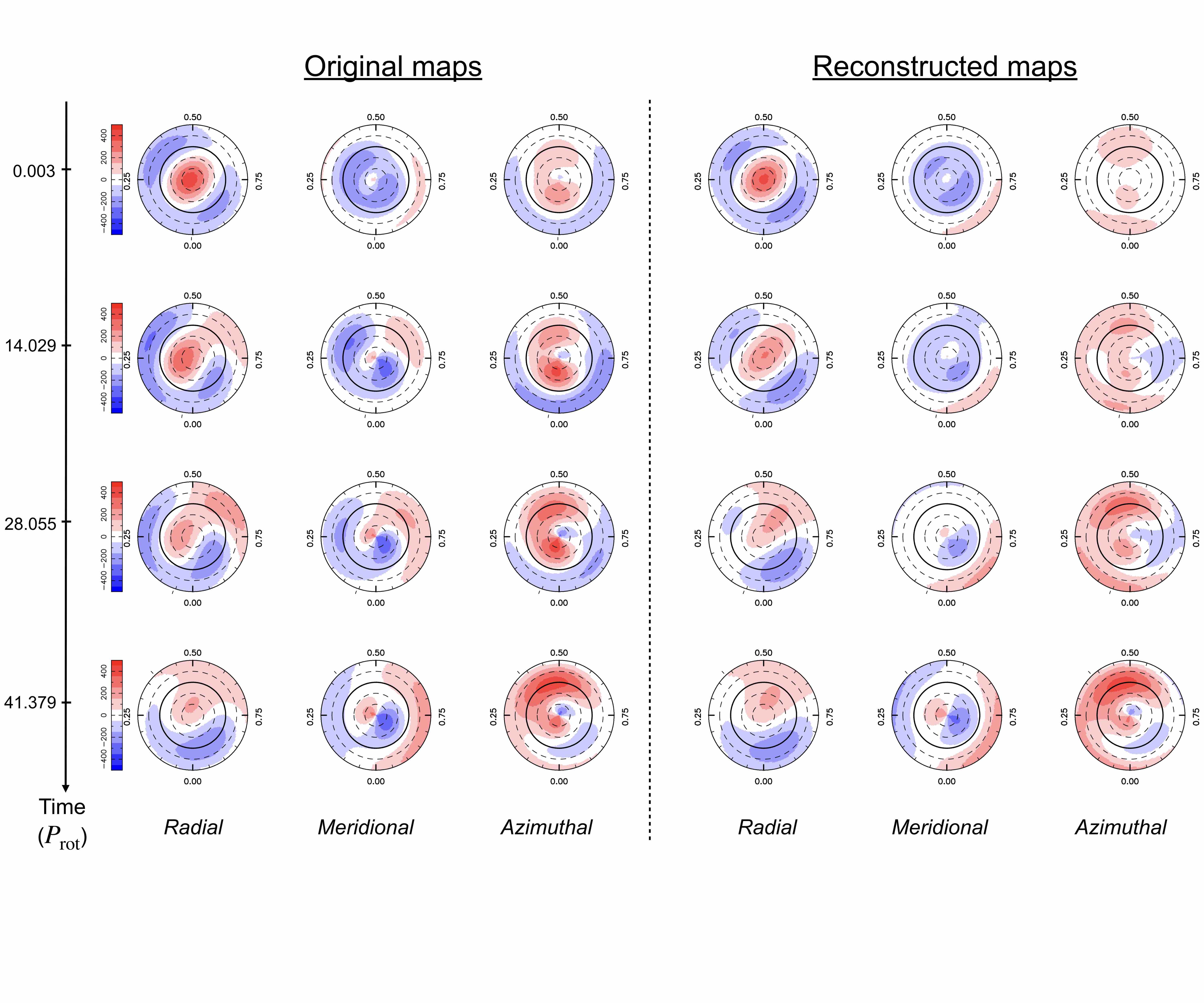} \hspace*{-0.8cm}
    \includegraphics[scale=0.16,trim={3cm 5.5cm 5cm 6cm},clip]{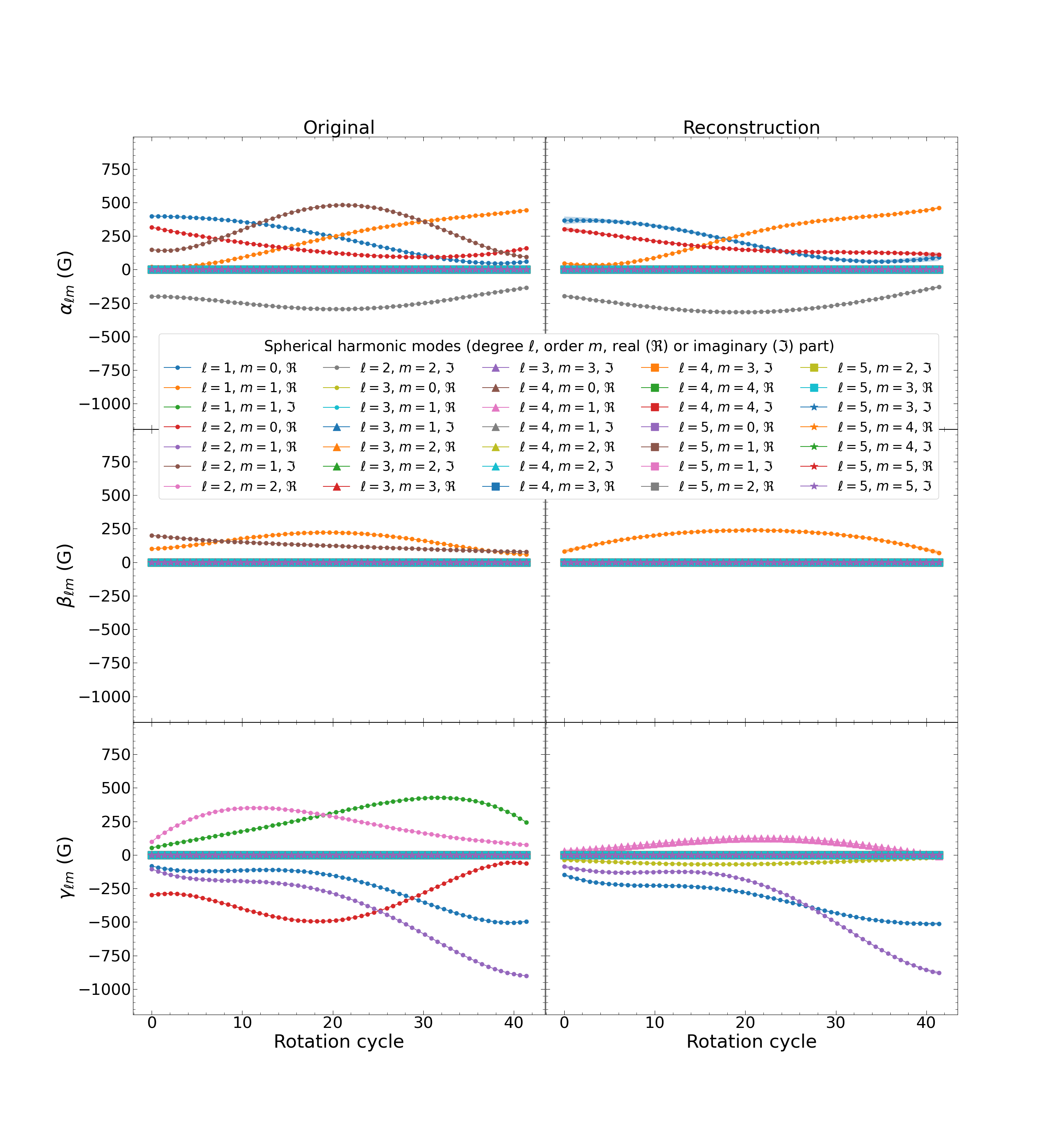}
    \includegraphics[scale=0.17,trim={0cm 0.5cm 0cm 0cm},clip]{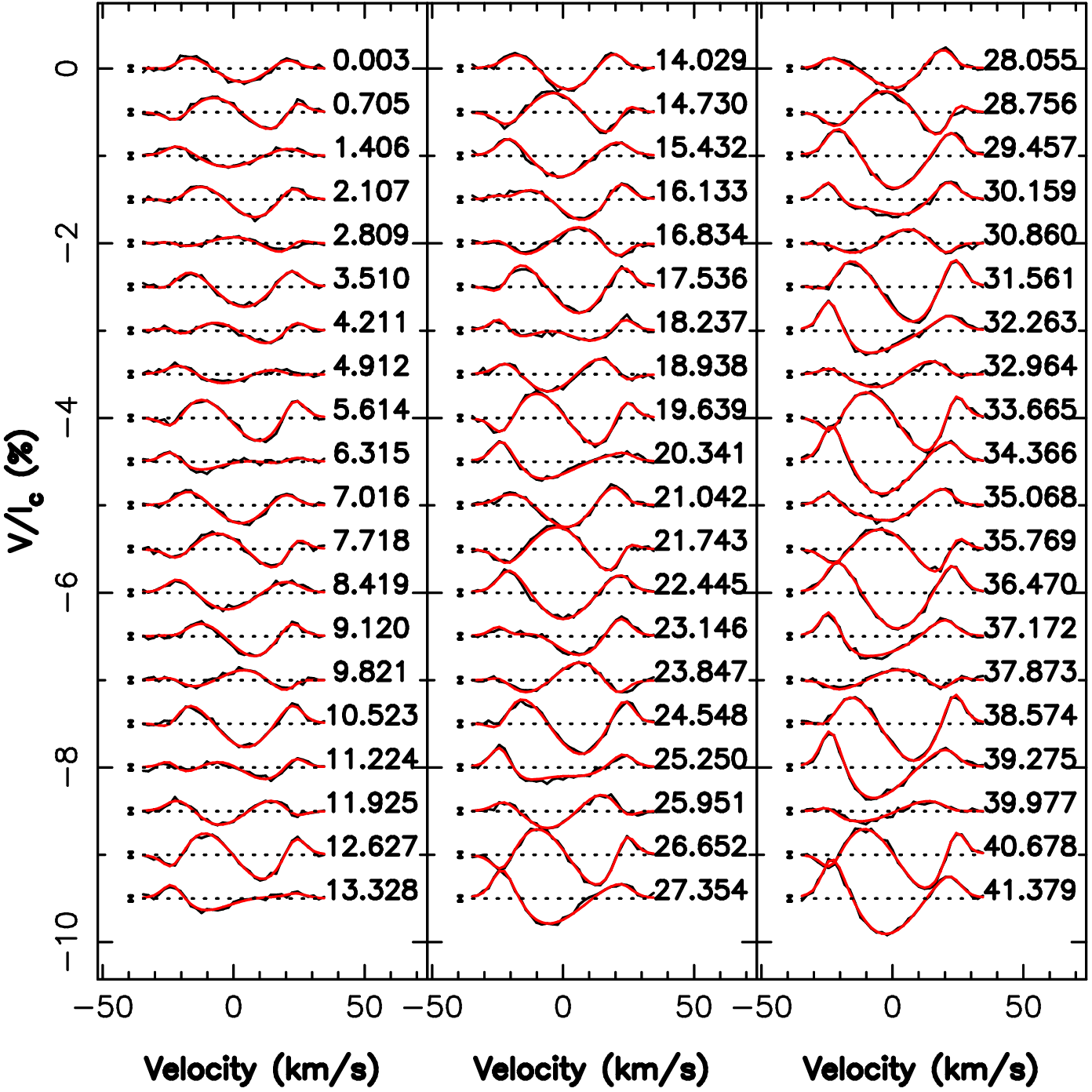}
    \caption{Same as Fig.~\ref{fig:inc30} for a stellar inclination of $80^\circ$.}
    
    \label{fig:inc80}
\end{figure*}

In this Section, we present the magnetic topology reconstructed for different stellar inclinations (setting the SNR of the Stokes~$V$ to 5000). The simulated data corresponds to the topology described in Sec.~\ref{sec:poloidal_toroidal} for a star featuring a $v\sin{i}=25$~\kms and an inclination of $30^\circ$ (Fig.~\ref{fig:inc30}) and $80^\circ$ (Fig.~\ref{fig:inc80}).

\section{Impact of $\lowercase{v}\sin{\lowercase{i}}$}

\begin{figure*}
    \centering \hspace*{-0.5cm}
    \includegraphics[scale=0.09,trim={.5cm 20.cm 0cm 9cm},clip]{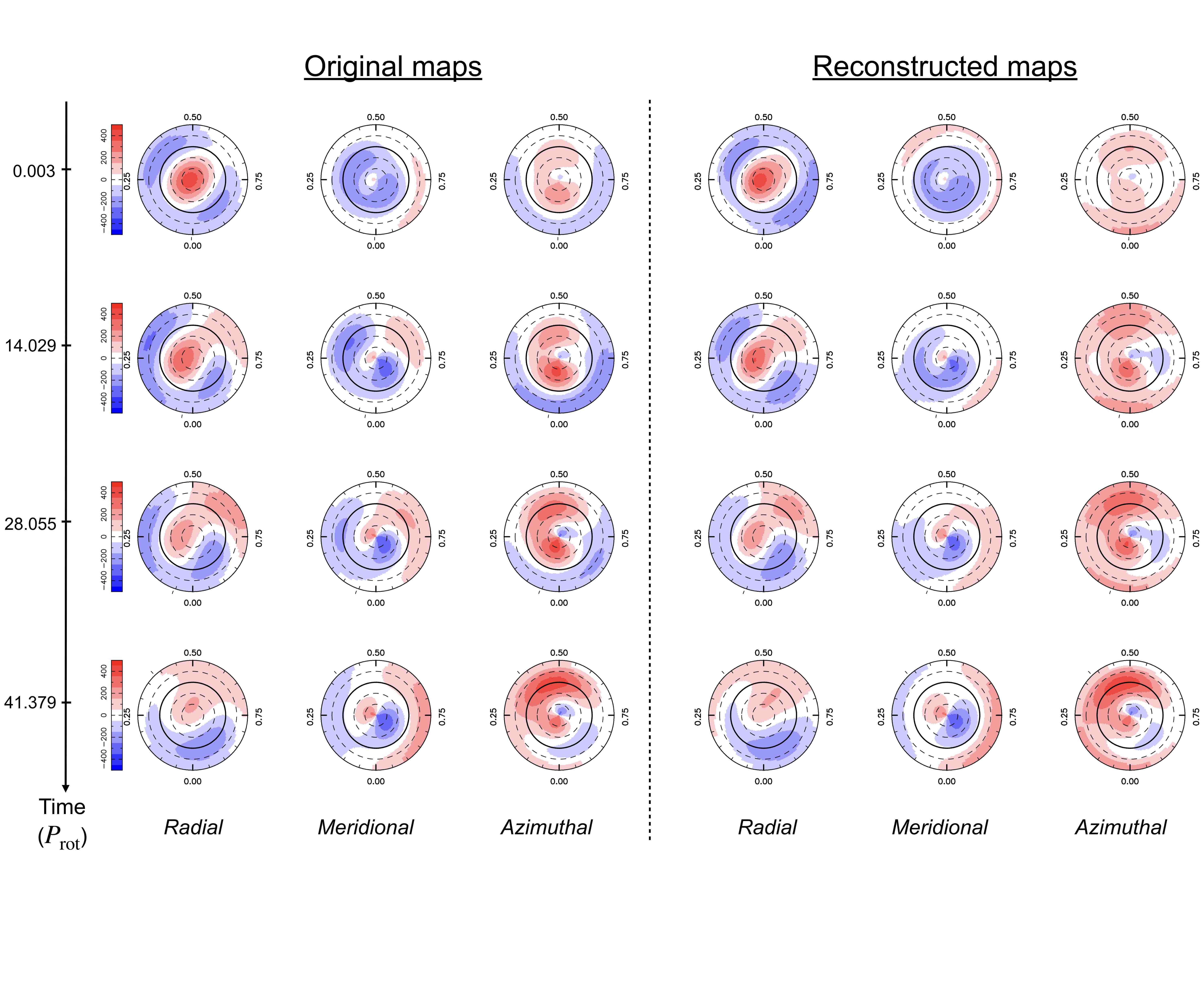} \hspace*{-0.8cm}
    \includegraphics[scale=0.16,trim={3cm 5.5cm 5cm 6cm},clip]{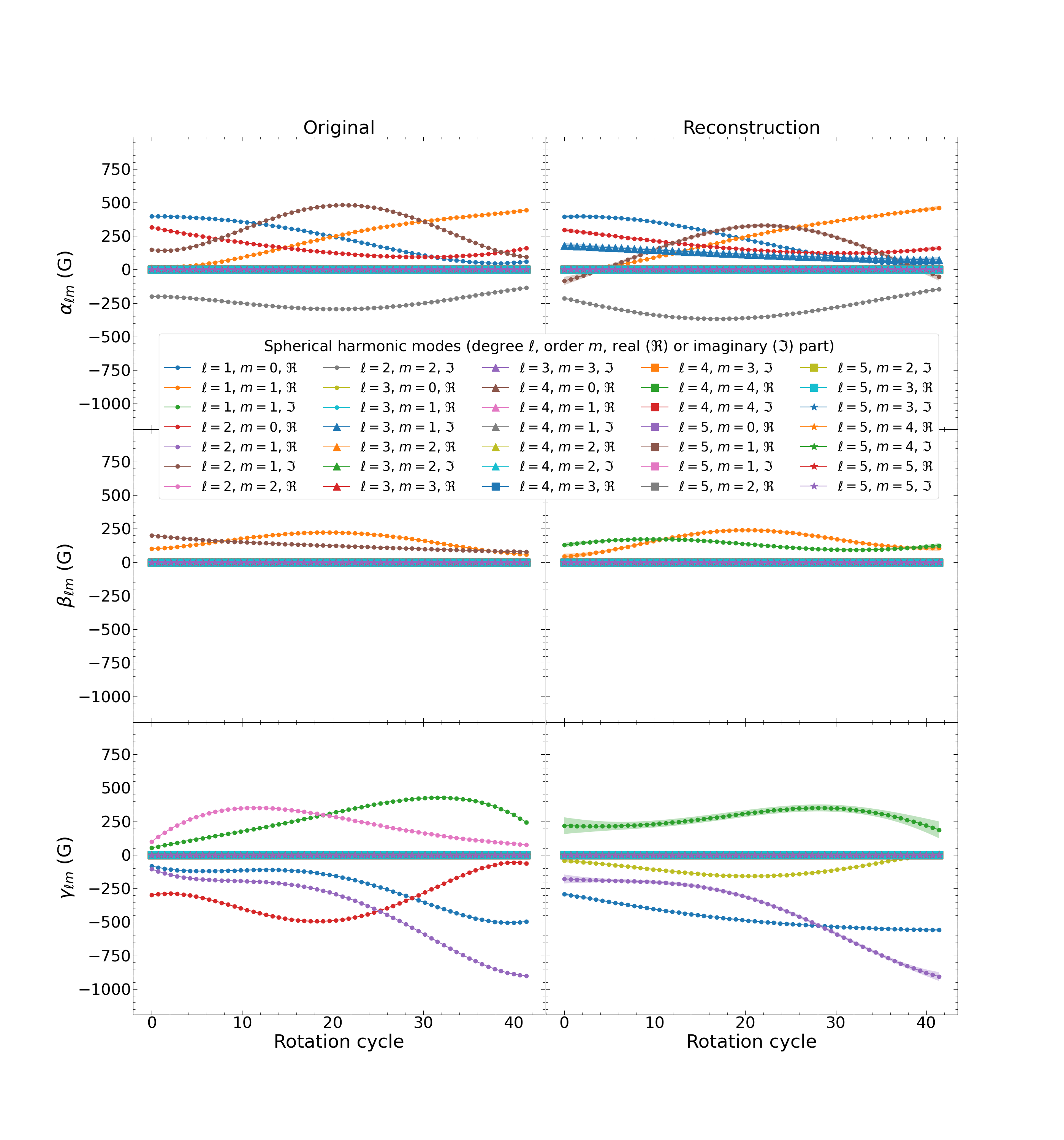} 
    \includegraphics[scale=0.17,trim={0cm 0.5cm 0cm 0cm},clip]{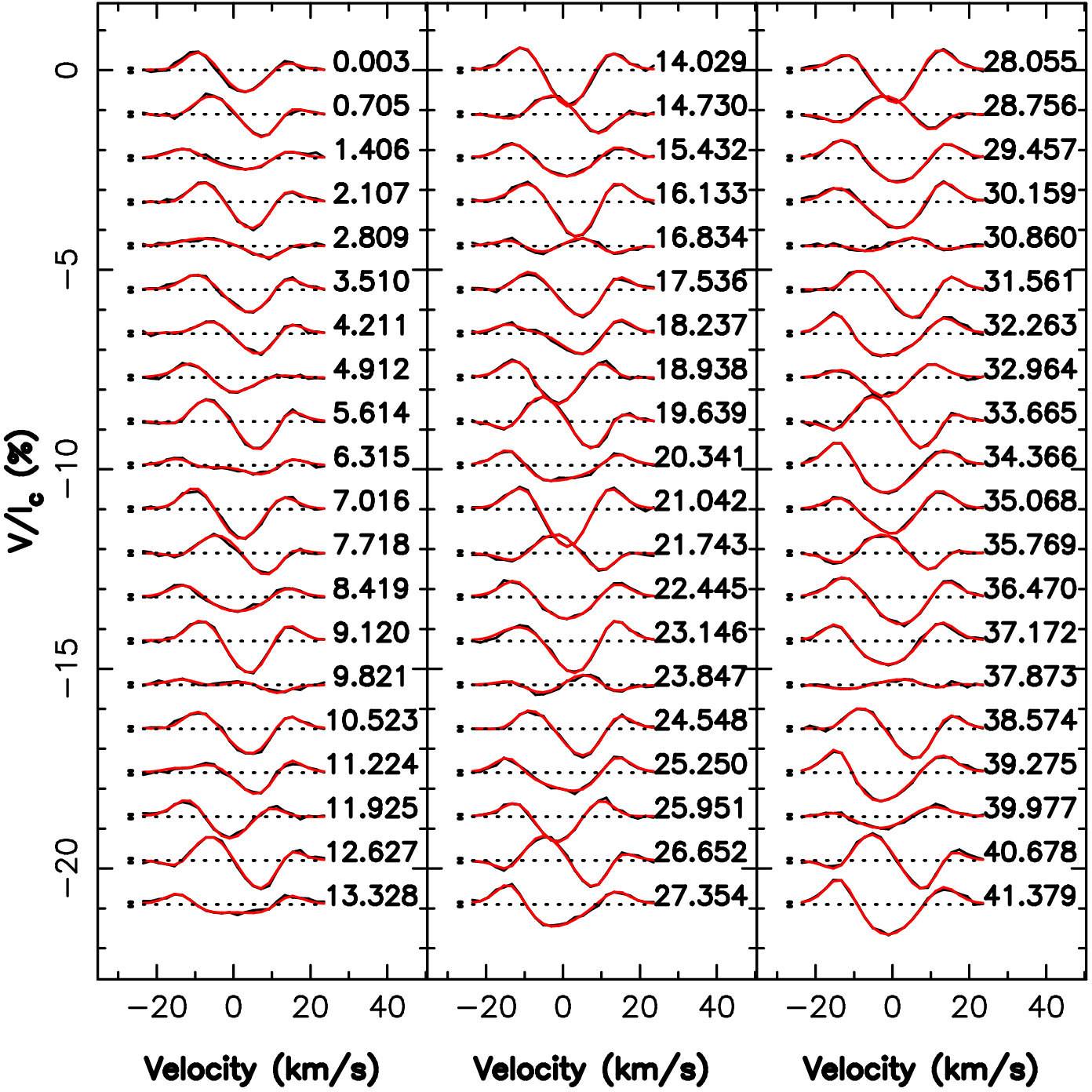}
   \caption{Same as Fig.~\ref{fig:inc30} for the same star as in the reference case (Sec.~\ref{sec:poloidal_toroidal}) except for the $v\sin{i}=15$~\kms.}
    \label{fig:coeff_poloidal_toroidal_vsini15}
\end{figure*}

We present the reconstructed topologies and coefficients when considering the same topology as in the reference case (Sec.~\ref{sec:poloidal_toroidal}) for a star featuring a $v\sin{i}=15$ (Fig.~\ref{fig:coeff_poloidal_toroidal_vsini15}), 50 (Fig.~B2, available as supplementary material) and 5~\kms (Fig.~\ref{fig:coeff_poloidal_toroidal_vsini5}). The SNR of the Stokes~$V$ profiles were set to 3000, 10000 and 2000, respectively, to take into account the differences in the amplitude of the Stokes~$V$ profiles as well as in the number of spectral points per profile. 

\renewcommand{\thefigure}{B3}
\begin{figure*}
    \centering \hspace*{-0.5cm}
    \includegraphics[scale=0.09,trim={.5cm 20.cm 0cm 9cm},clip]{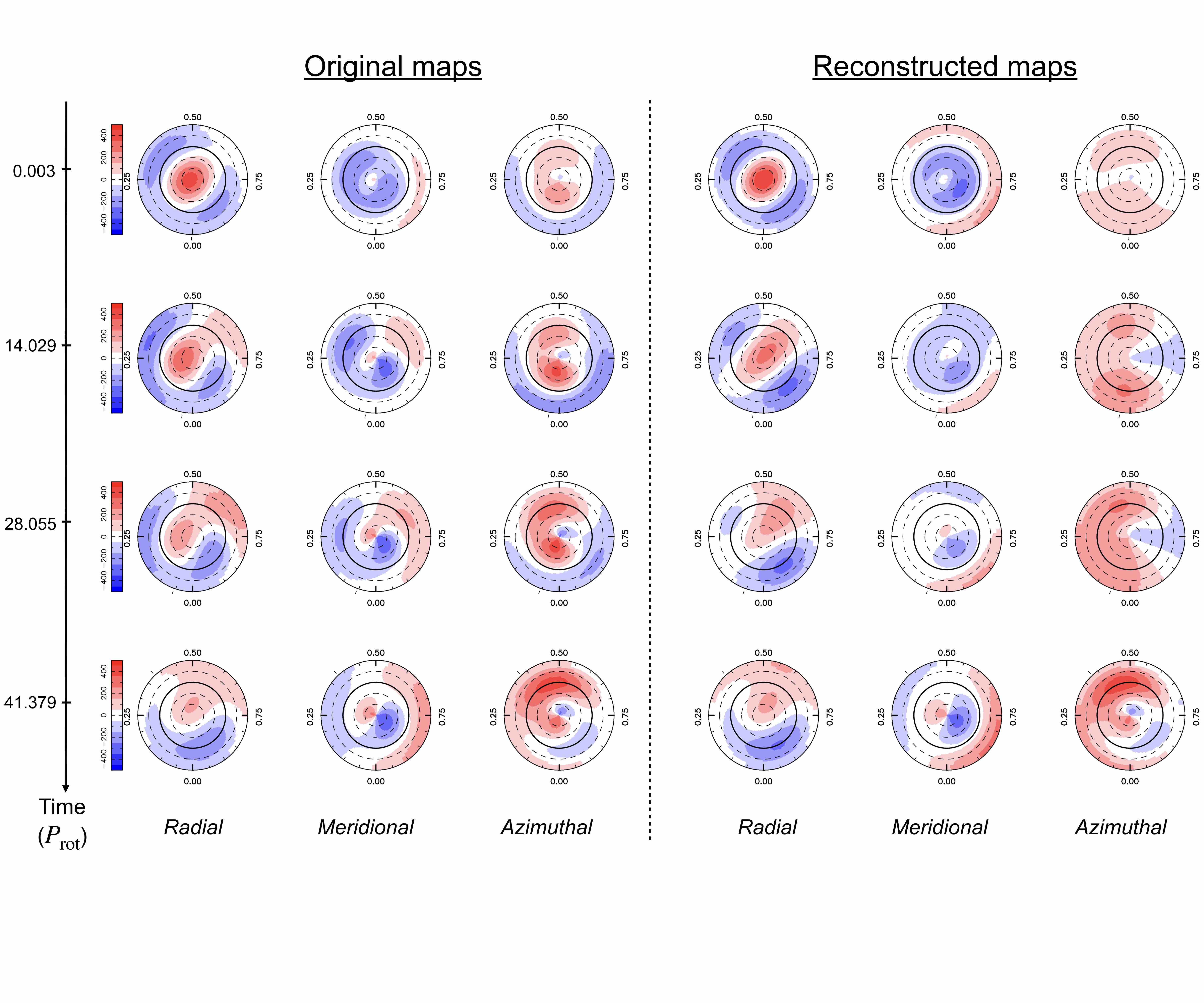} \hspace*{-0.8cm}
    \includegraphics[scale=0.16,trim={3cm 5.5cm 5cm 6cm},clip]{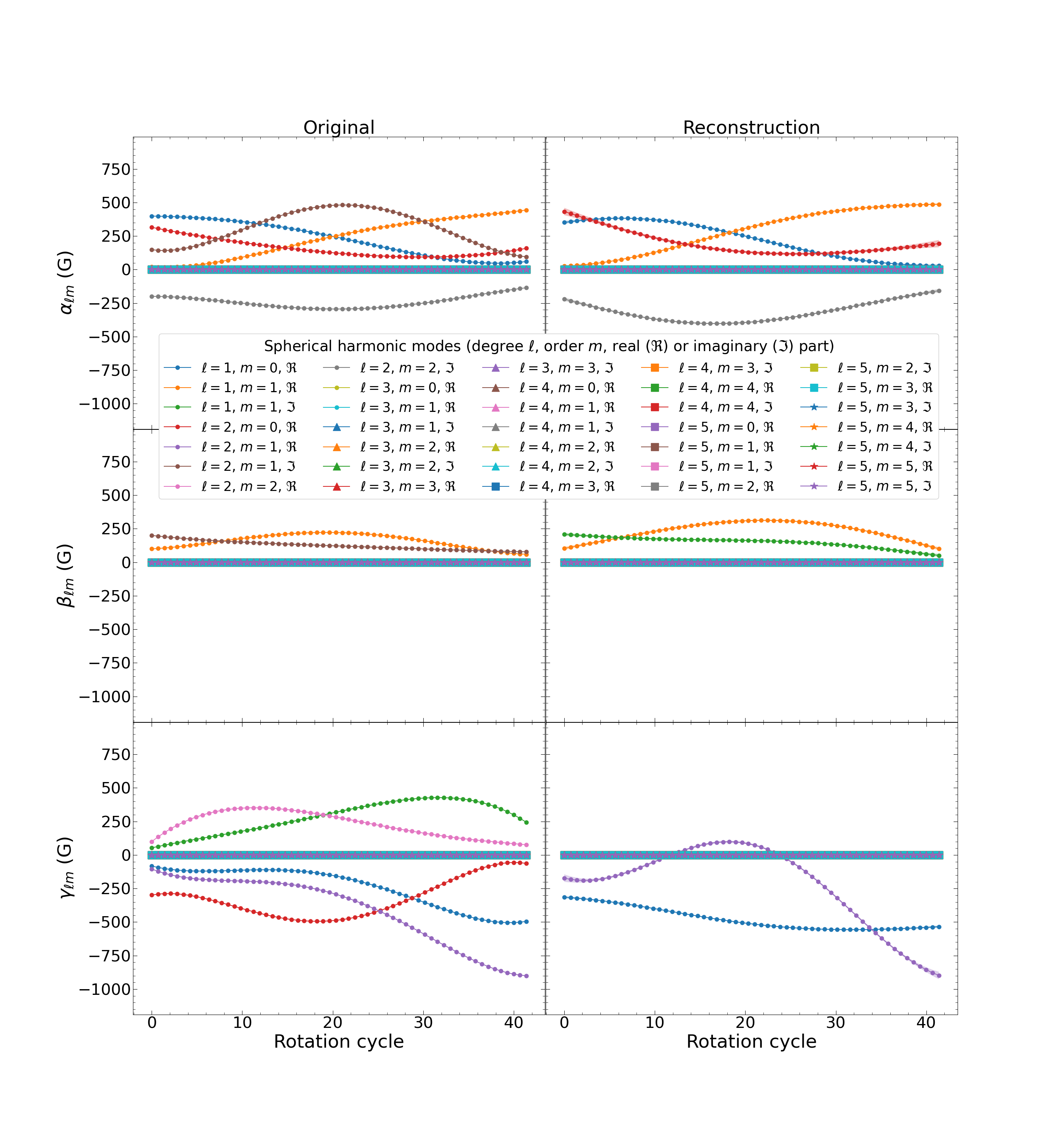}
    \includegraphics[scale=0.17,trim={0cm 0.5cm 0cm 0cm},clip]{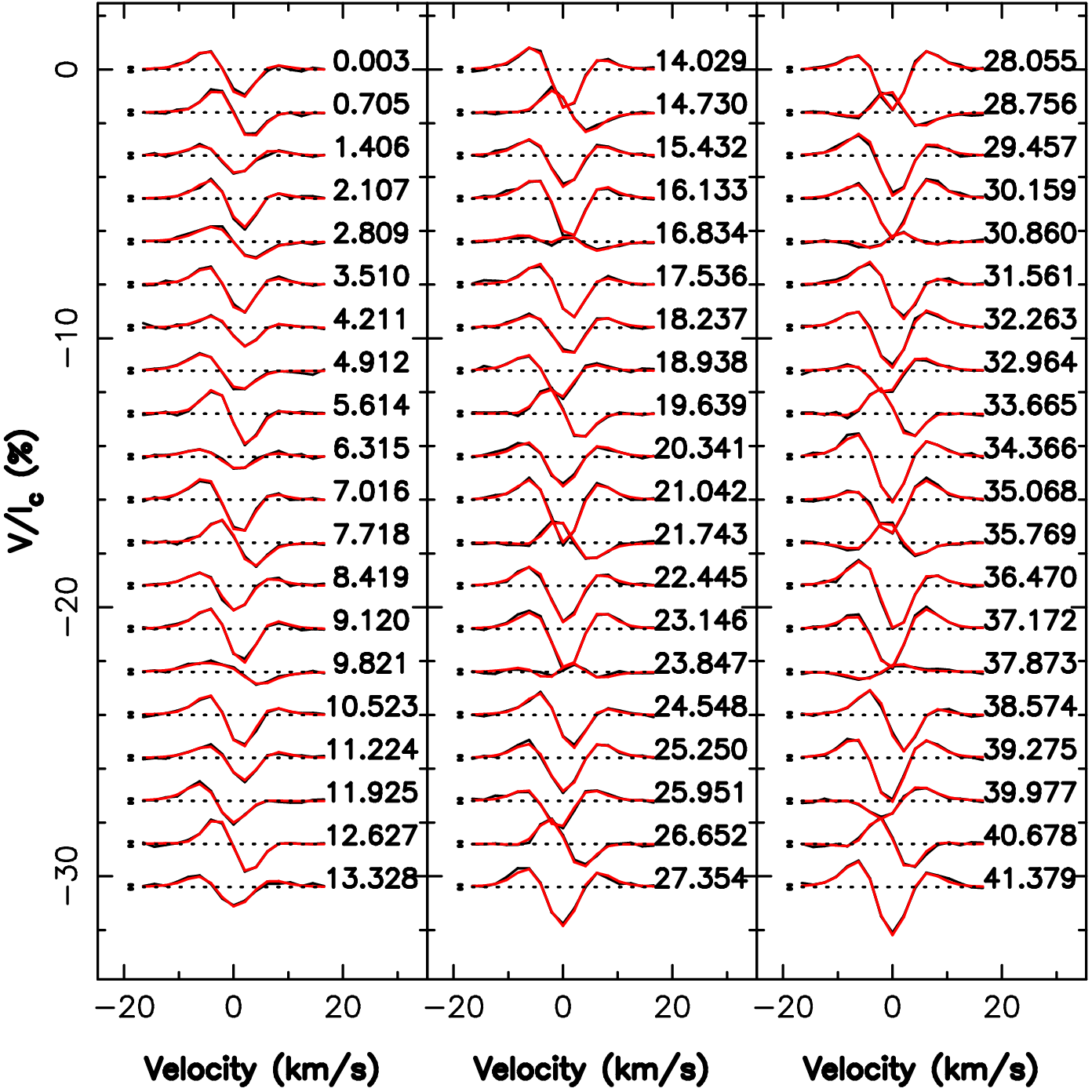}
   \caption{Same as Fig.~\ref{fig:inc30} for the same star as in the reference case (Sec.~\ref{sec:poloidal_toroidal}) except for the $v\sin{i}=5$~\kms.}
    \label{fig:coeff_poloidal_toroidal_vsini5}
\end{figure*}

\section{Reconstruction of a complex field}
\label{appendix:vsini}

In order to clearly see the effect of the $v\sin{i}$ on the reconstructed topology, we simulated a complex magnetic topology described by spherical harmonics modes up to a degree $\ell=6$. We show the reconstructed maps for $v\sin{i}= 15$ (Fig.~C1, available as supplementary material) and 50~\kms (Fig.~C2, available as supplementary material) using $n=10$ in TIMeS. The SNRs were fixed to 3000 and 10000 to ensure that the Stokes~$V$ are still detected with the same precision when varying the $v\sin{i}$.

\section{Impact of a lower SNR}

Fig.~\ref{fig:snr1000} illustrates the reconstruction of the reference case (Sec.~\ref{sec:poloidal_toroidal}) when the SNR of the Stokes~$V$ profiles is decreased down to 1000. In this case, the data are fitted down to $\chi^2_r=1.16$. Some modes are missed by TIMeS due to the highest noise level implying a loss of information.

\renewcommand{\thefigure}{D1}
\begin{figure*}
    \centering \hspace*{-0.5cm}
    \includegraphics[scale=0.09,trim={.5cm 20.cm 0cm 9cm},clip]{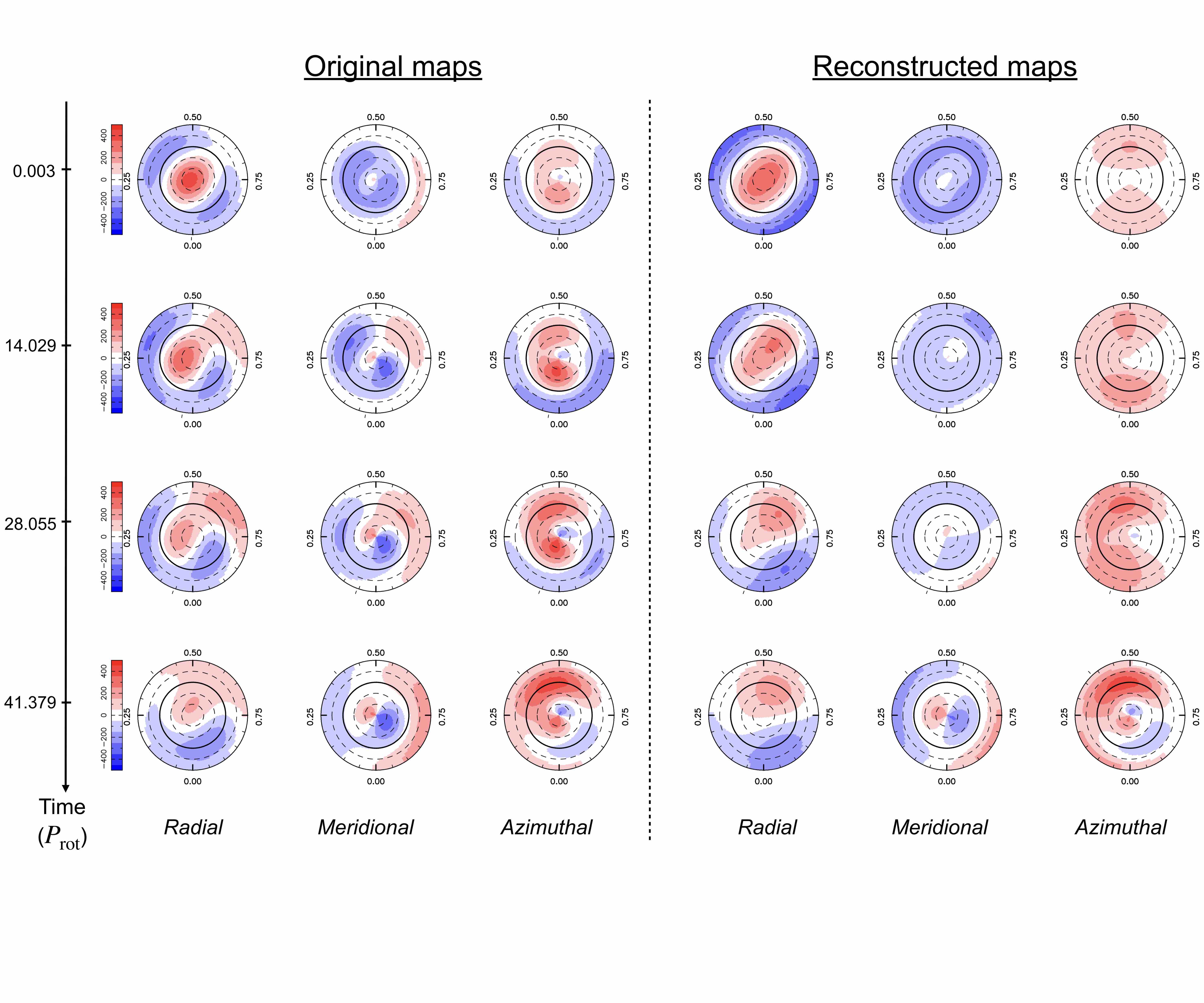} \hspace*{-0.8cm}
    \includegraphics[scale=0.16,trim={3cm 5.5cm 5cm 6cm},clip]{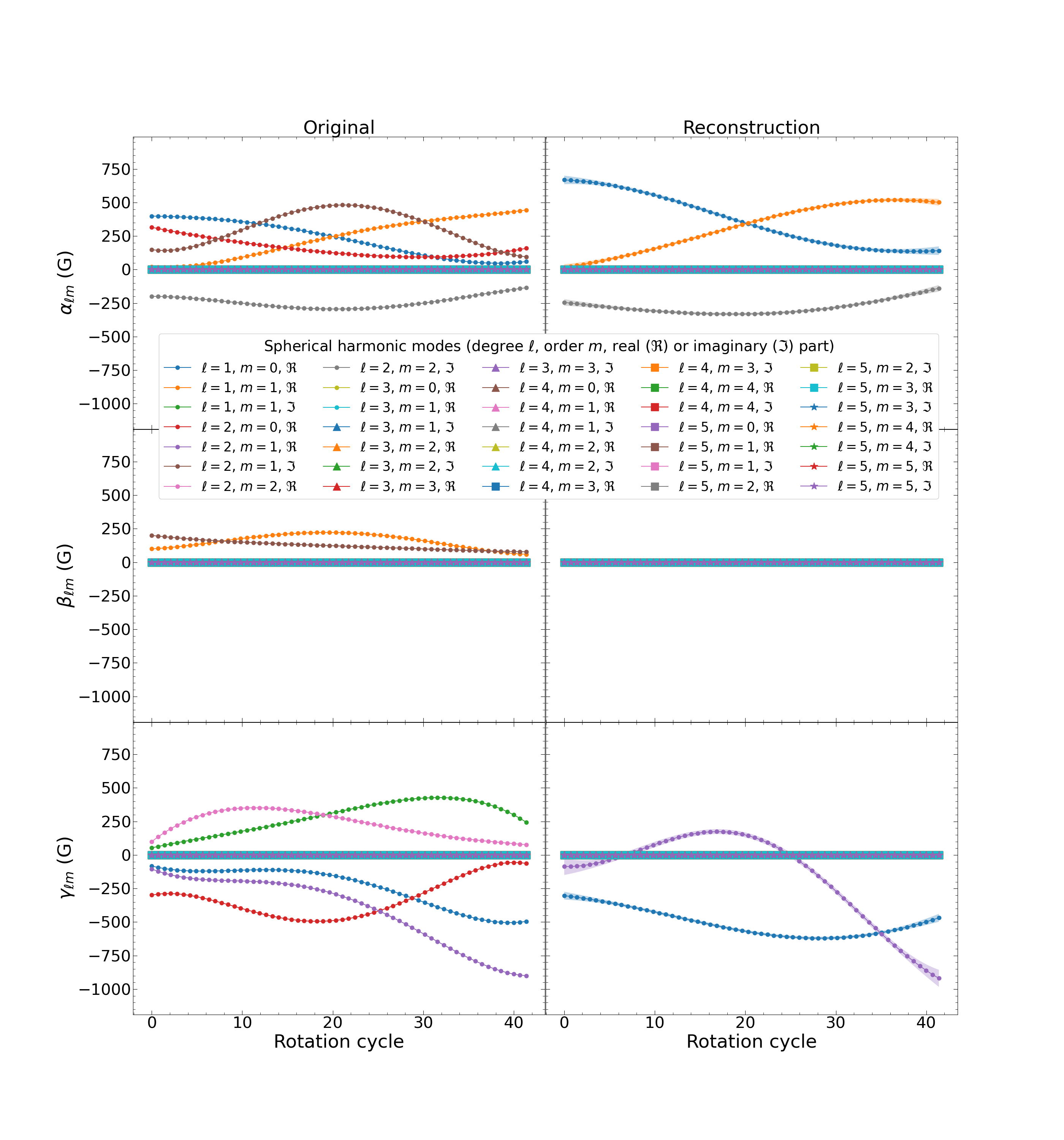}
    \includegraphics[scale=0.17,trim={0cm 0.5cm 0cm 0cm},clip]{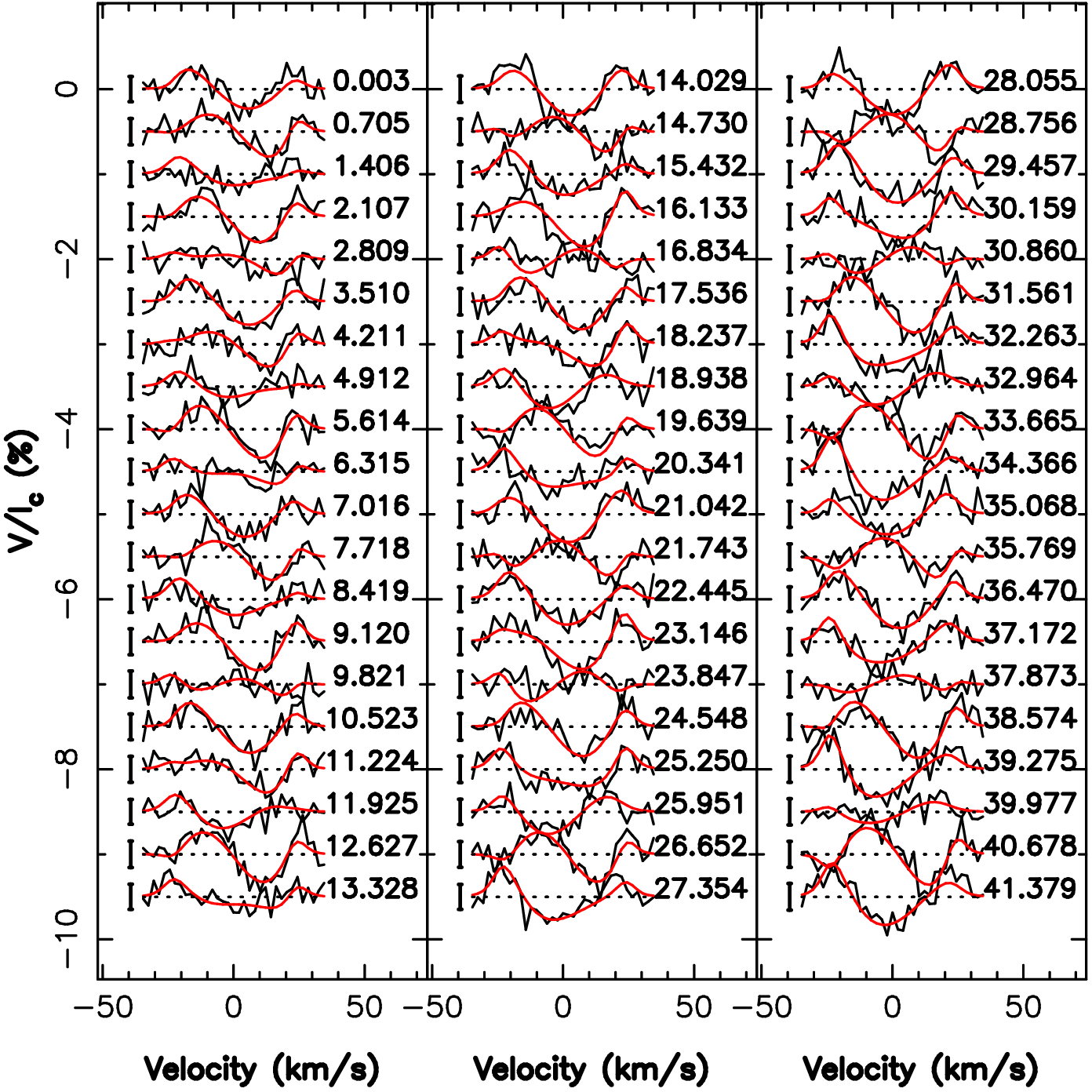}
    \caption{Same as Fig.~\ref{fig:inc30} for the same star as in the reference case (Sec.~\ref{sec:poloidal_toroidal}) except for the SNR of the Stokes~$V$ profiles decreased down to 1000.}
    \label{fig:snr1000}
\end{figure*}

\section{Impact of the penalization weights}

We present the results obtained when applying TIMeS on a purely toroidal field (Sec.~\ref{sec:toroidal}) and assuming penalization weights for all modes equal to $\ell / A_{\ell,m}$, where $A_{\ell,m}$ denotes the mean amplitude of the Stokes~$V$ profiles associated with the modes over a rotation cycle. Fig.~E1 (available as supplementary material) shows the reconstructed coefficients as well as the reconstructed magnetic maps.

\section{Impact of the sampling of the rotation cycle}

We present the reconstructed maps when considering the magnetic evolution of the reference case (Sec.~\ref{sec:poloidal_toroidal}), using $n=3$ (i.e. 1.4 rotation cycles) and $n=12$ (i.e. 7.7 rotation cycles) in Figs.~F1 and F2 (available as supplementary material), respectively. The Stokes~$V$ profiles are fitted down to $\chi^2_r = 5.18 $ and 1.07, respectively. These results demonstrate that considering too few profiles (poor sampling of the stellar rotation) or too many profiles (significant evolution of the field over the associated time interval) yields discrepant results.

\bsp	
\label{lastpage}
\end{document}



\appendix
\counterwithin{figure}{section}





\renewcommand{\thefigure}{B2}
\begin{figure*}
    \centering \hspace*{-0.5cm}
    \includegraphics[scale=0.09,trim={.5cm 20.cm 0cm 9cm},clip]{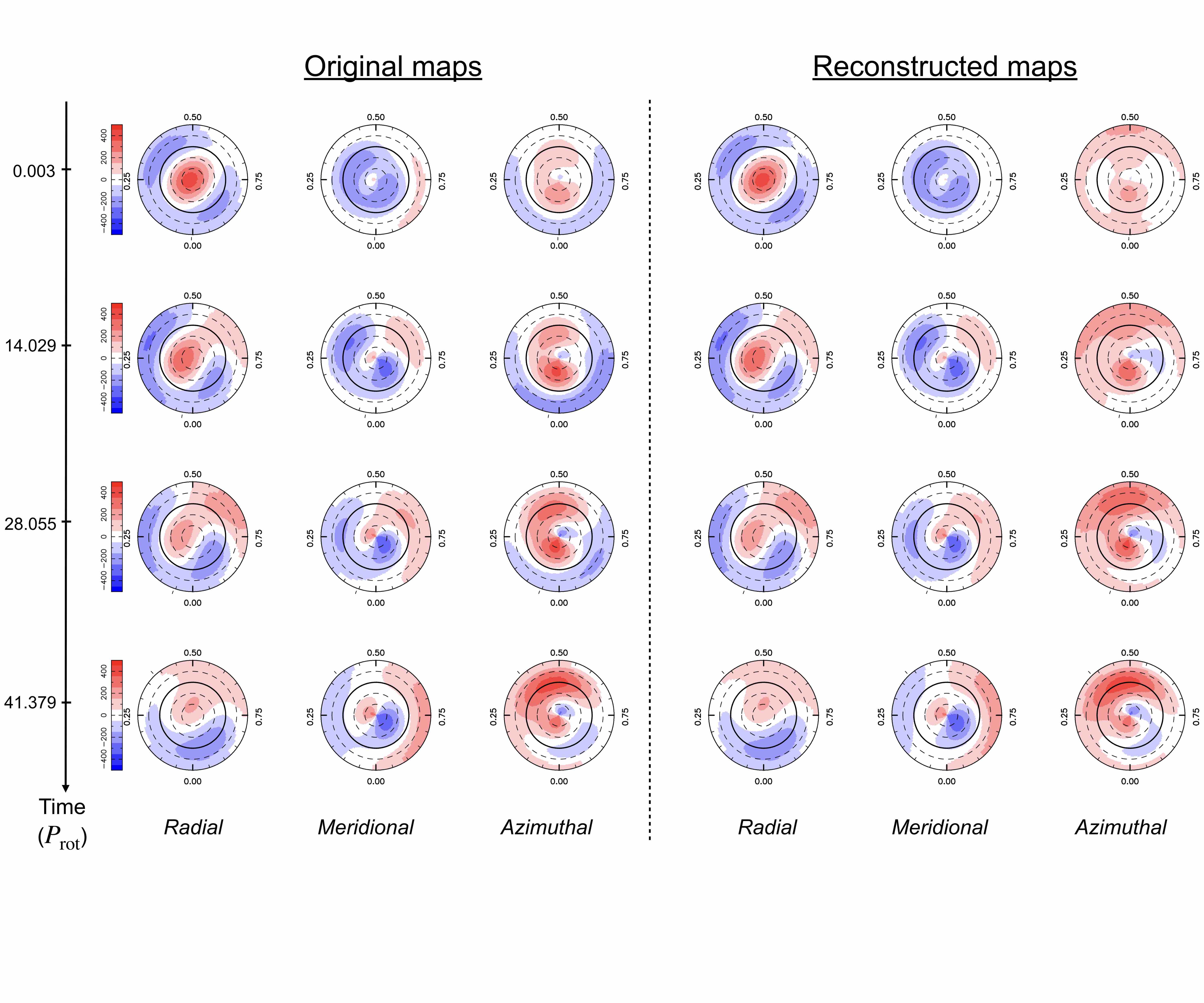} \hspace*{-0.8cm}
    \includegraphics[scale=0.16,trim={3cm 5.5cm 5cm 6cm},clip]{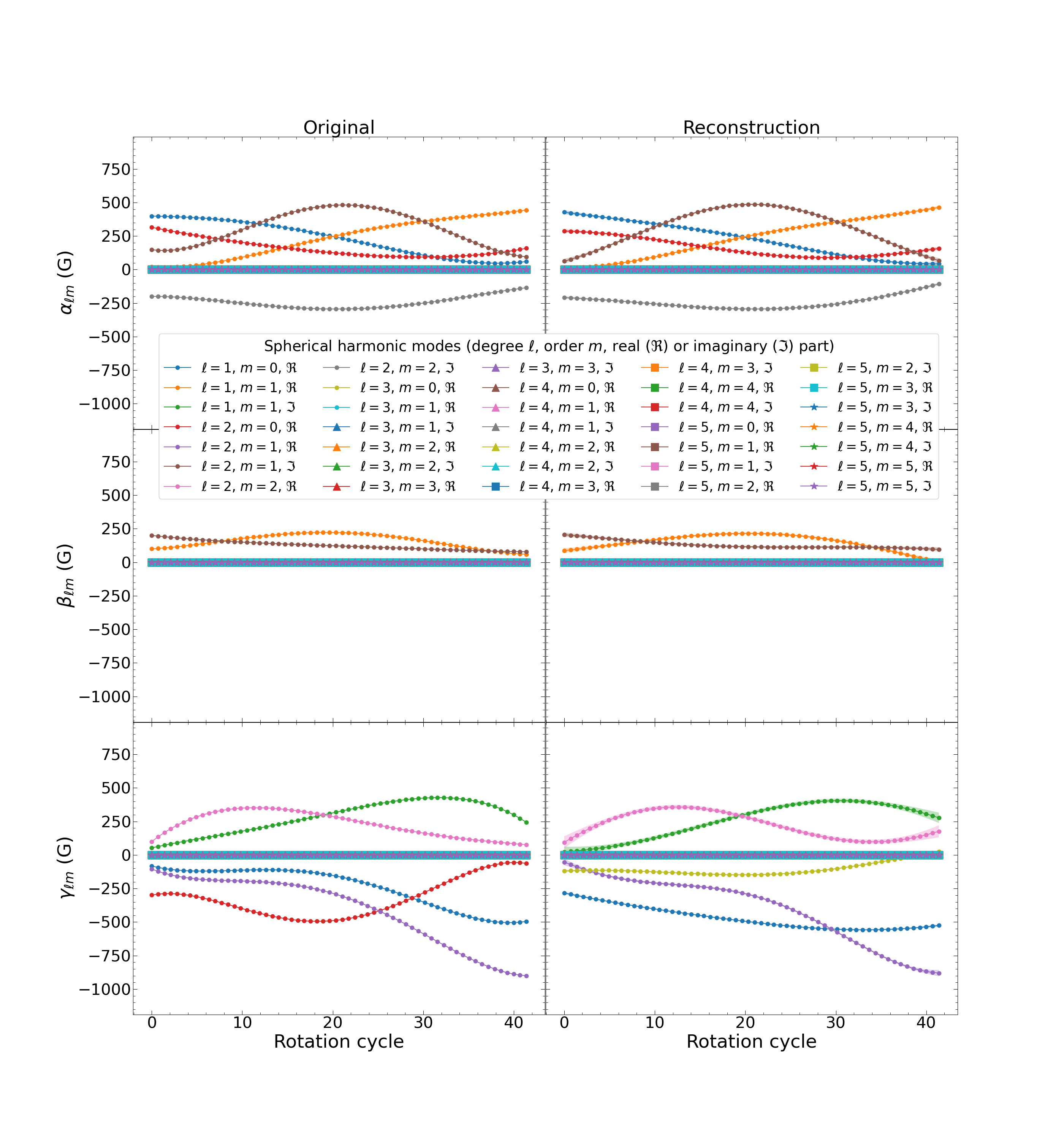}
    \includegraphics[scale=0.15,trim={0cm 0.5cm 0cm 0cm},clip]{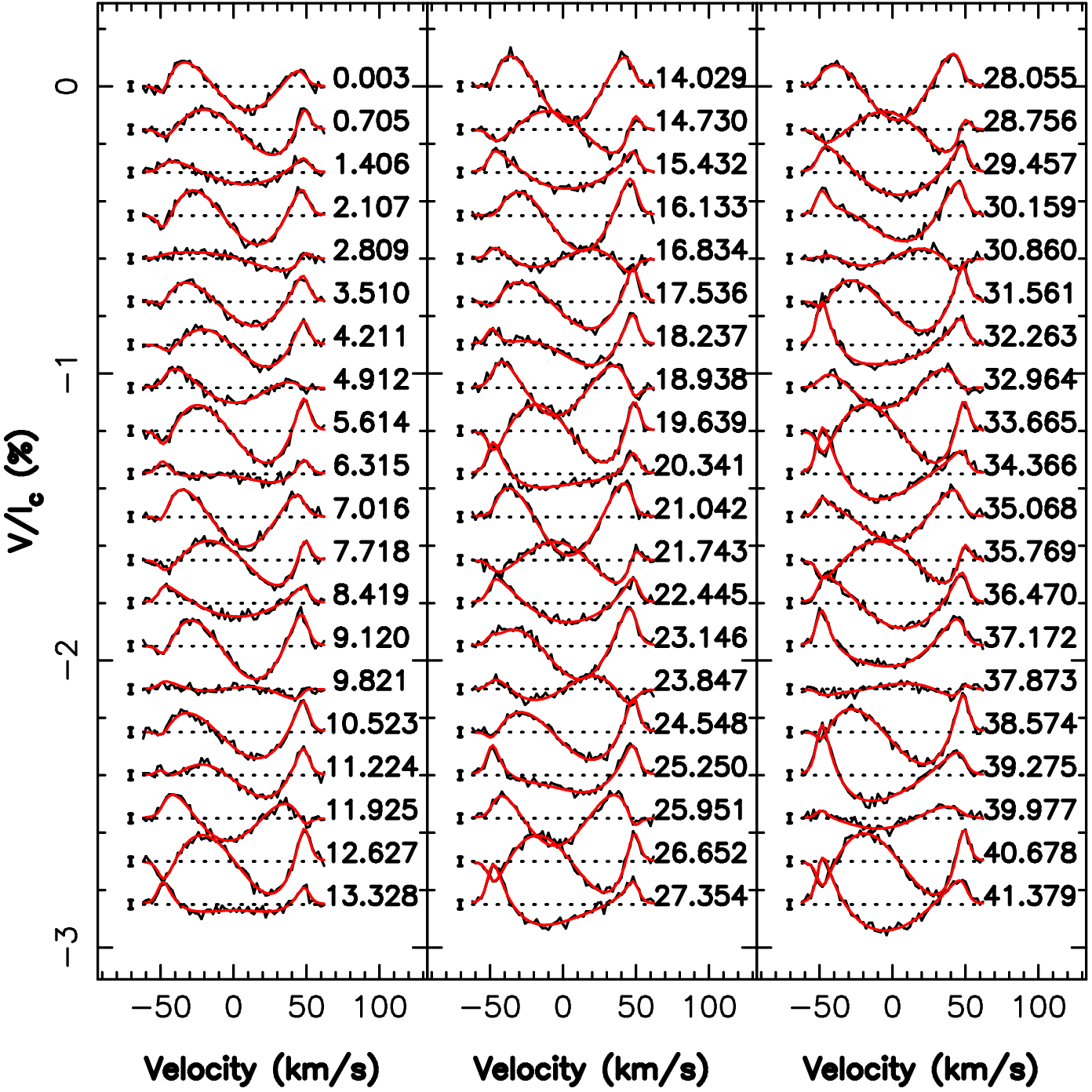}
   \caption{Same as Fig.~A1 for the same star as in the reference case (Sec.~3.3) except for the $v\sin{i}=50$~\kms.}
    \label{fig:coeff_poloidal_toroidal_vsini50}
\end{figure*}



\renewcommand{\thefigure}{C1}
\begin{figure*}
    \centering \hspace*{-0.5cm}
    \includegraphics[scale=0.09,trim={.5cm 27.cm 0cm 9cm},clip]{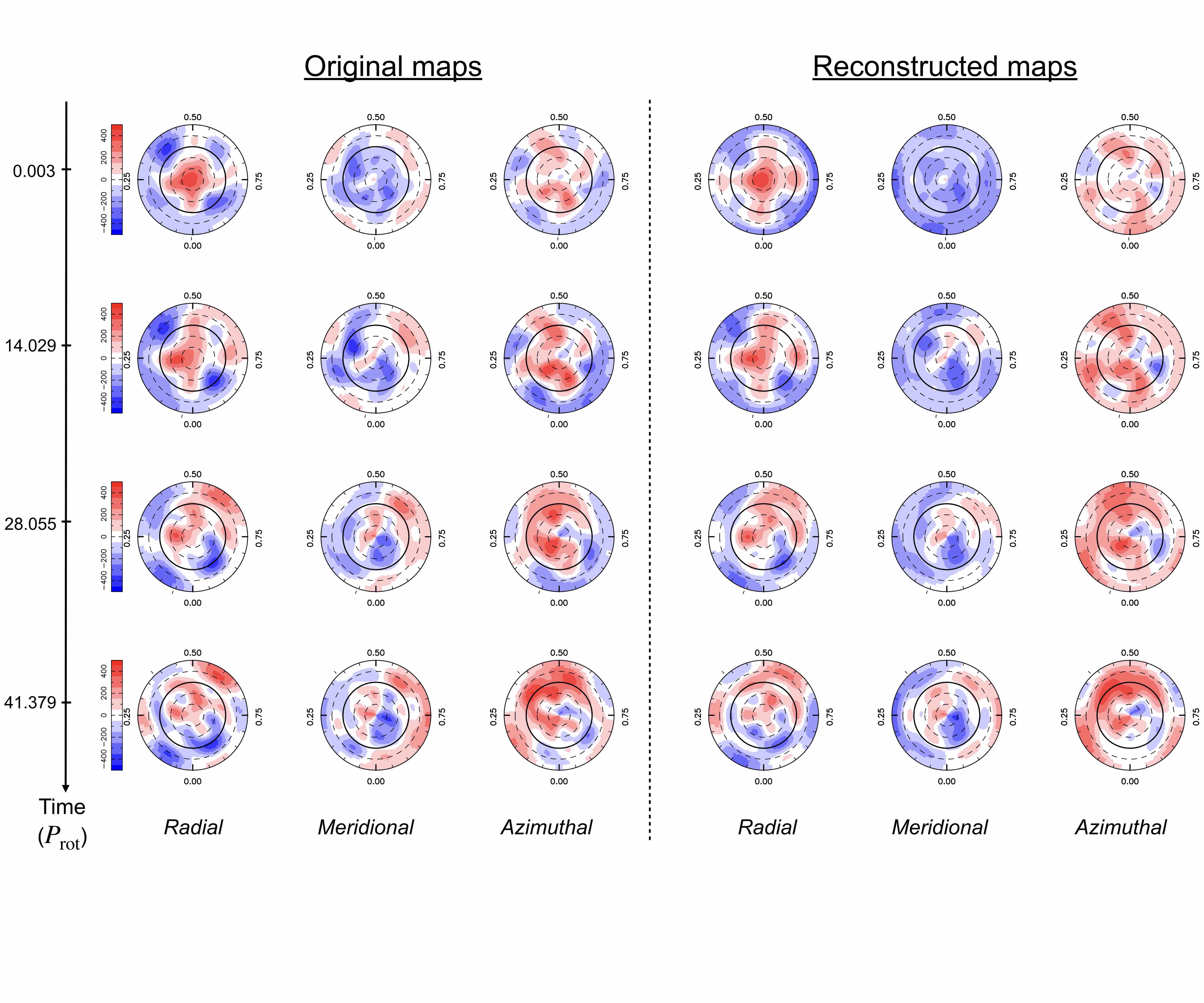} \hspace*{-0.8cm}
    \includegraphics[scale=0.12,trim={3cm 8cm 3cm 6cm},clip]{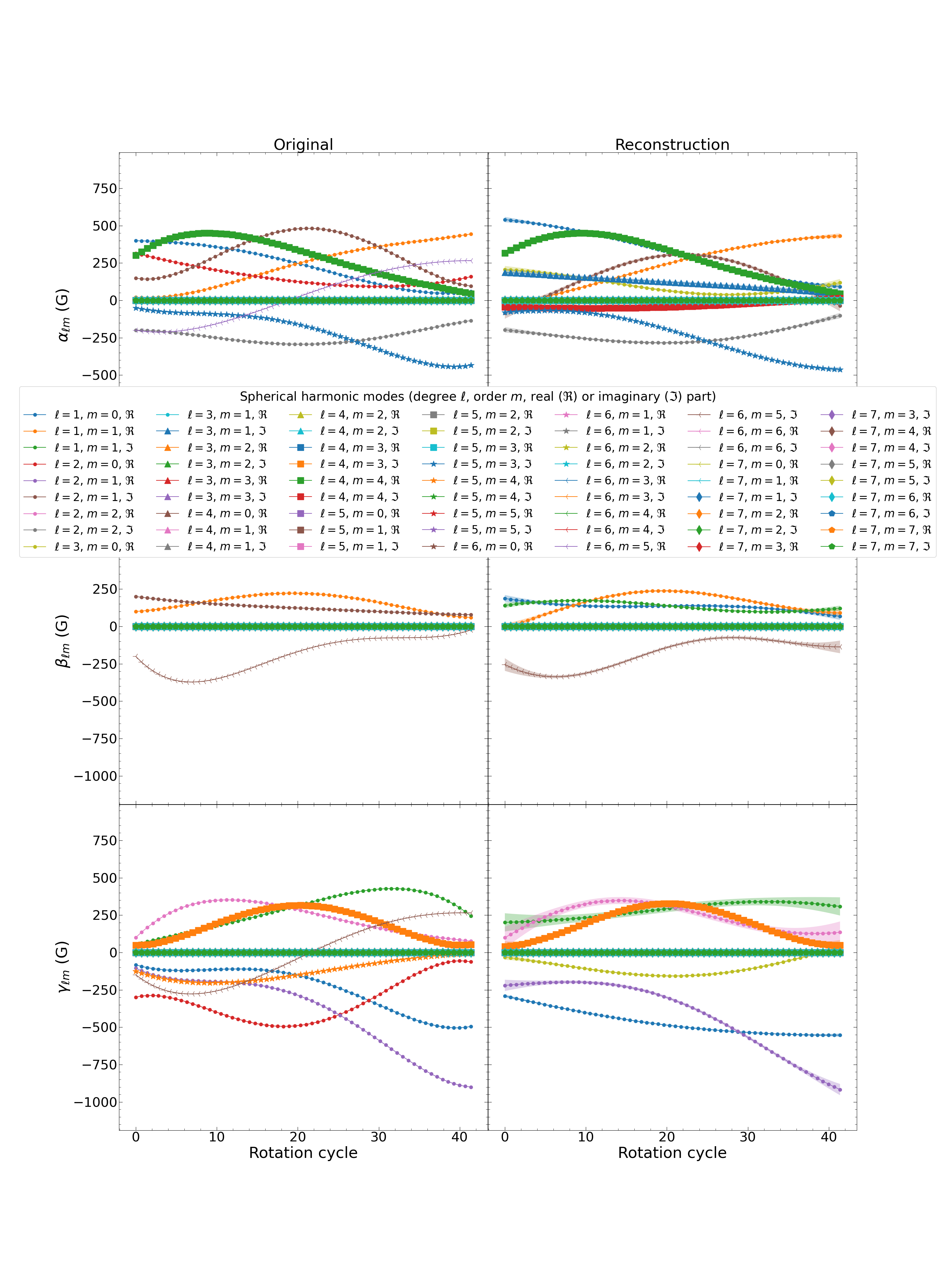}
    \includegraphics[scale=0.17,trim={0cm 0.5cm 0cm 0cm},clip]{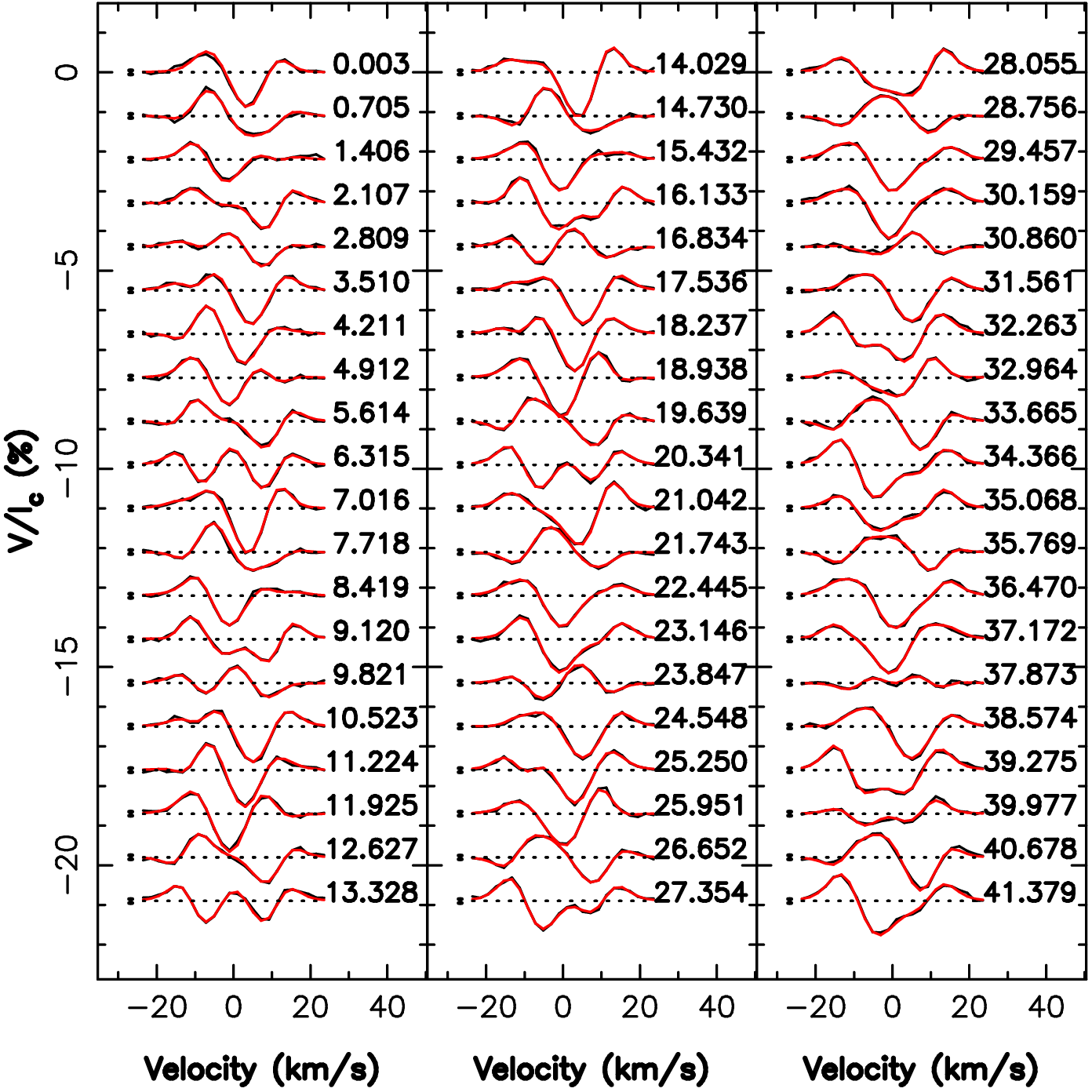}
    \caption{Reconstruction of a complex topology described by spherical harmonics modes up to $\ell=6$ for a star featuring $i=60^\circ$ and $v\sin{i}=15$~\kms. \textit{Top}: Reconstructed maps with the star shown in a flattened polar view. \textit{Bottom left}: Comparison between the input and reconstructed coefficients describing the field. \textit{Bottom right}: Observed (black) and reconstructed (red) Stokes~$V$ profiles ($\chi^2_r=1.18$). } 
    \label{fig:vsini15}
\end{figure*}

\renewcommand{\thefigure}{C2}
\begin{figure*}
    \centering \hspace*{-0.5cm}
    \includegraphics[scale=0.09,trim={.5cm 27.cm 0cm 9cm},clip]{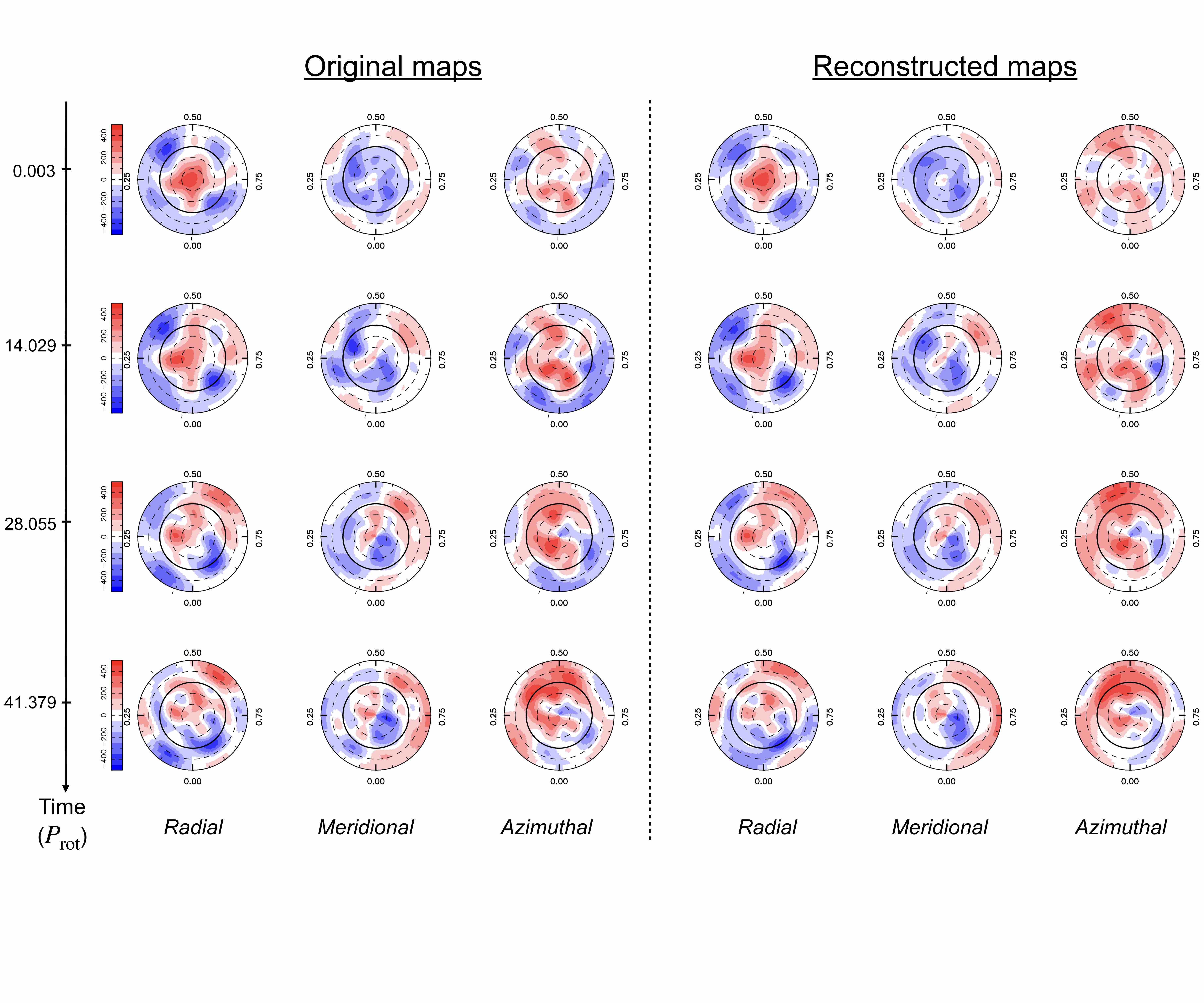} \hspace*{-0.8cm}
    \includegraphics[scale=0.12,trim={3cm 8cm 3cm 6cm},clip]{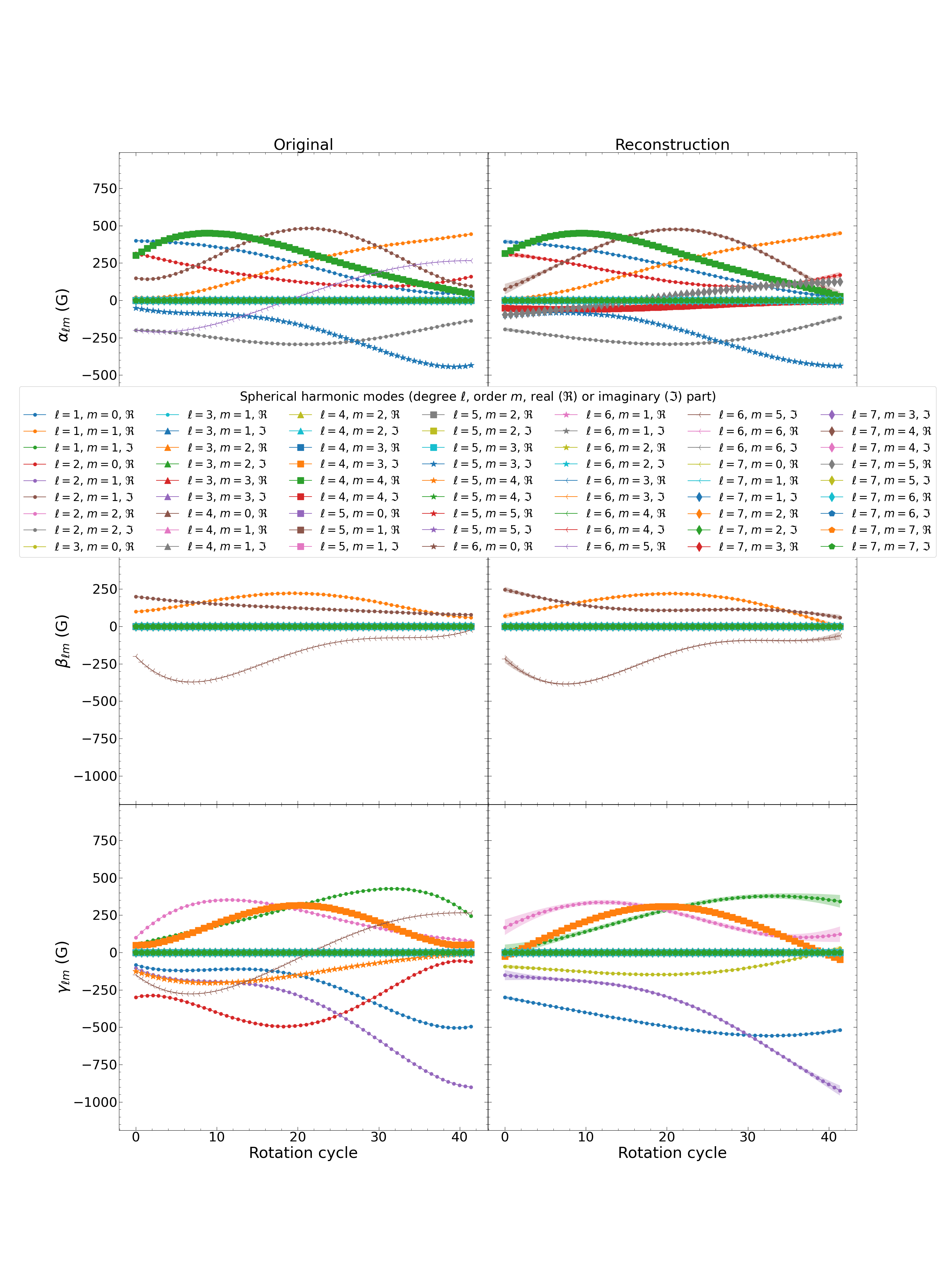}
    \includegraphics[scale=0.17,trim={0cm 0.5cm 0cm 0cm},clip]{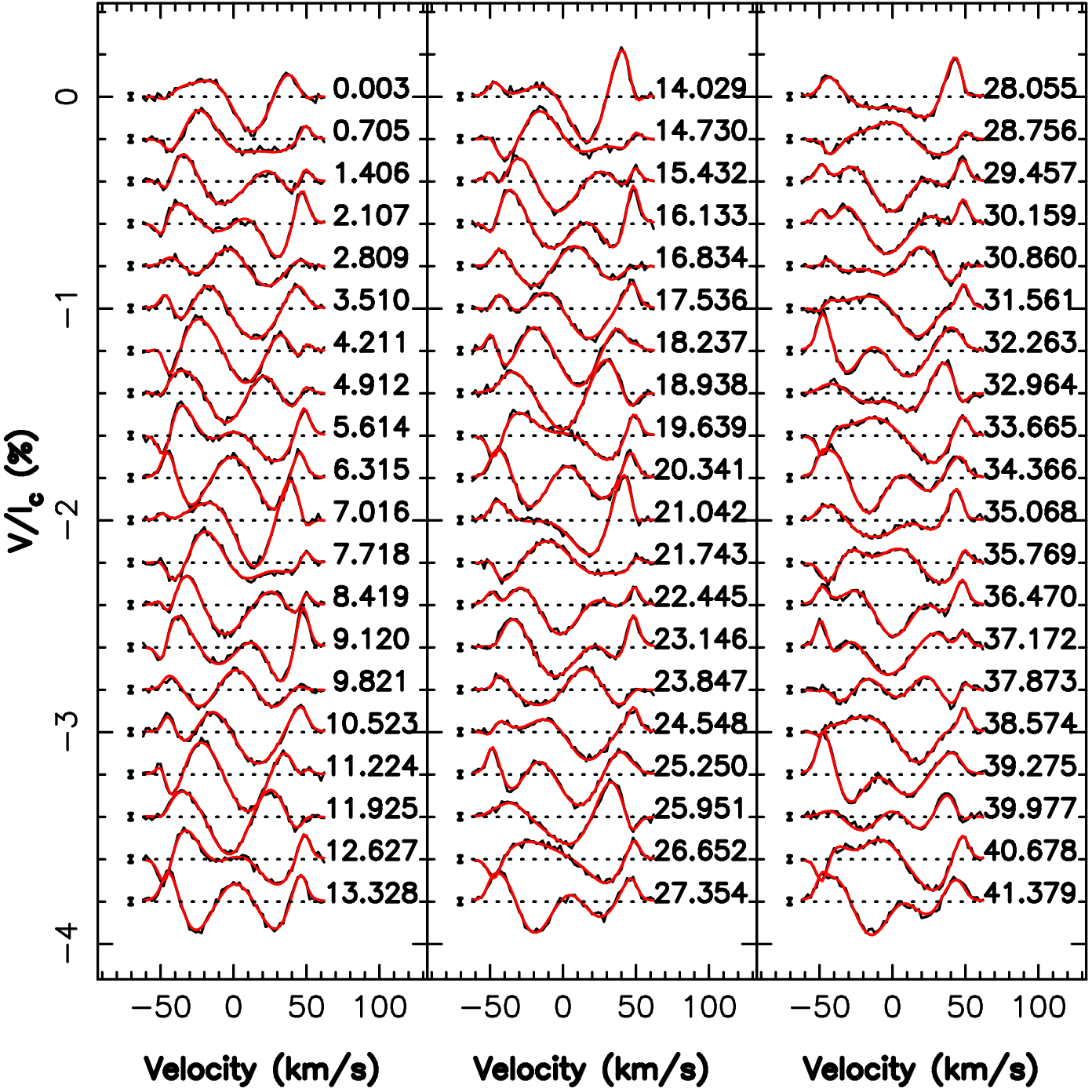}
     \caption{Same as Fig.~\ref{fig:vsini15} for $v\sin{i}=50$~\kms. The Stokes~$V$ profiles are fitted down to $\chi^2_r=1.10$.}
    \label{fig:vsini50}
\end{figure*}





\renewcommand{\thefigure}{E1}
\begin{figure*}
    \centering \hspace*{-0.5cm}
    \includegraphics[scale=0.09,trim={.5cm 27.cm 0cm 9cm},clip]{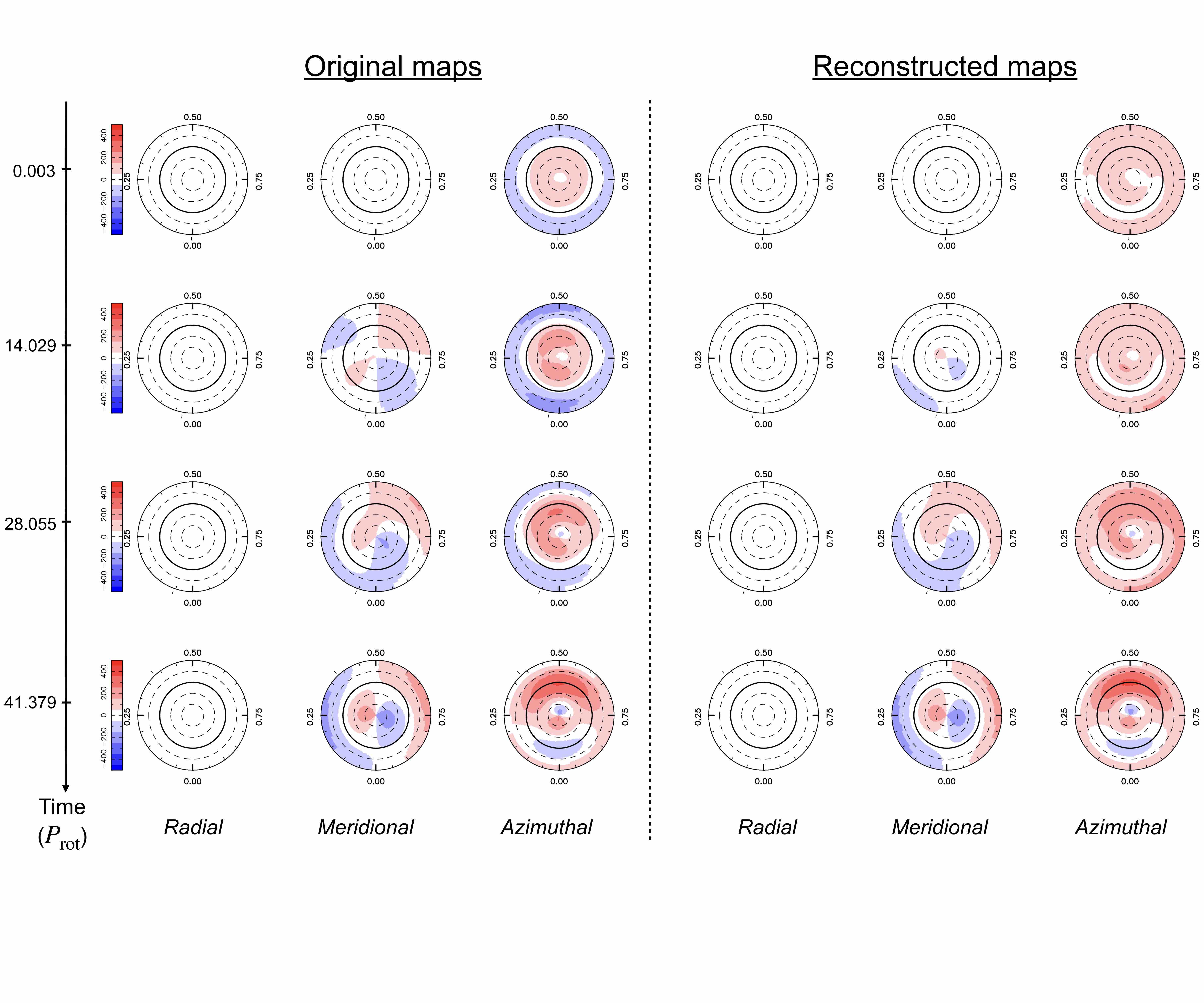} \hspace*{-0.8cm}
    \includegraphics[scale=0.16,trim={3cm 5.5cm 5cm 6cm},clip]{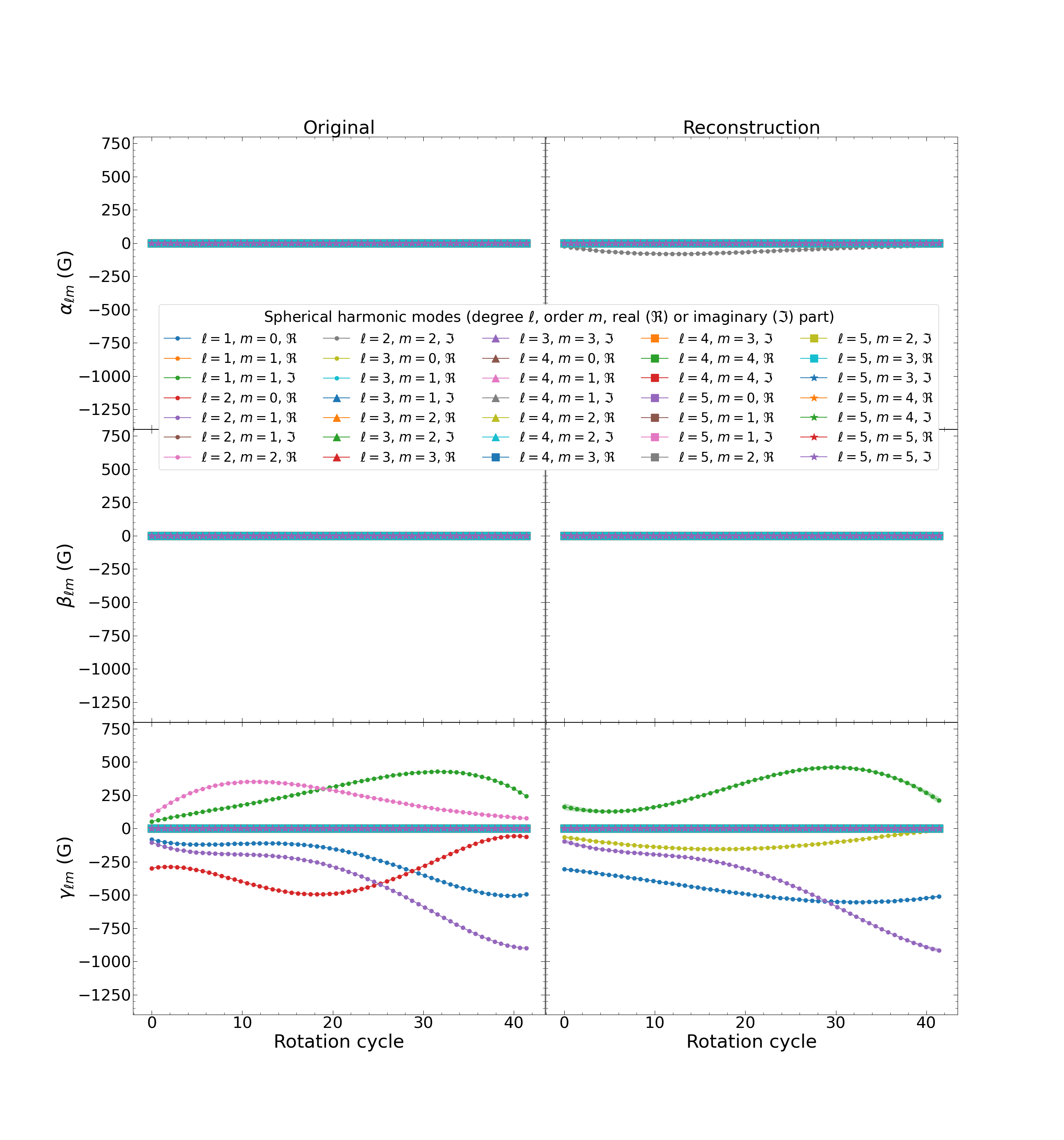}
    \includegraphics[scale=0.15,trim={0cm 0.5cm 0cm 0cm},clip]{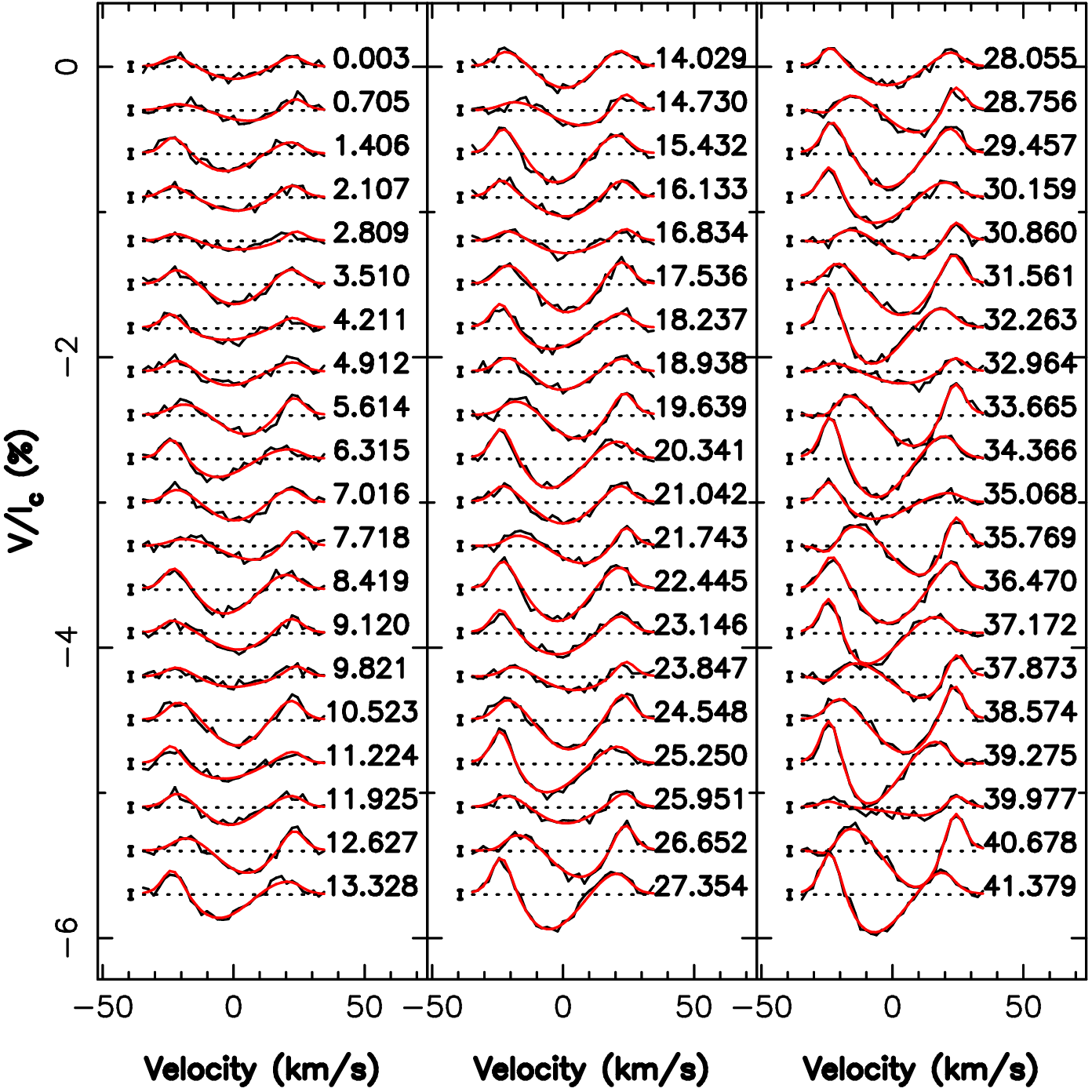}
    \caption{Same as Fig.~A1 when reconstructing the purely toroidal field (Sec.~3.2) with other penalization weights in the selection process. A weak spurious poloidal field is reconstructed, slightly altering the meridional and azimuthal components in the map. The observed Stokes~$V$ profiles are fitted down to $\chi^2_r=1.14$ instead of 1.02 (when using the weights presented in Sec.~2.4.1).}
    \label{fig:maps_other_weight}
\end{figure*}



\renewcommand{\thefigure}{F1}
\begin{figure*}
    \centering \hspace*{-0.5cm}
    \includegraphics[scale=0.09,trim={.5cm 25.cm 0cm 9cm},clip]{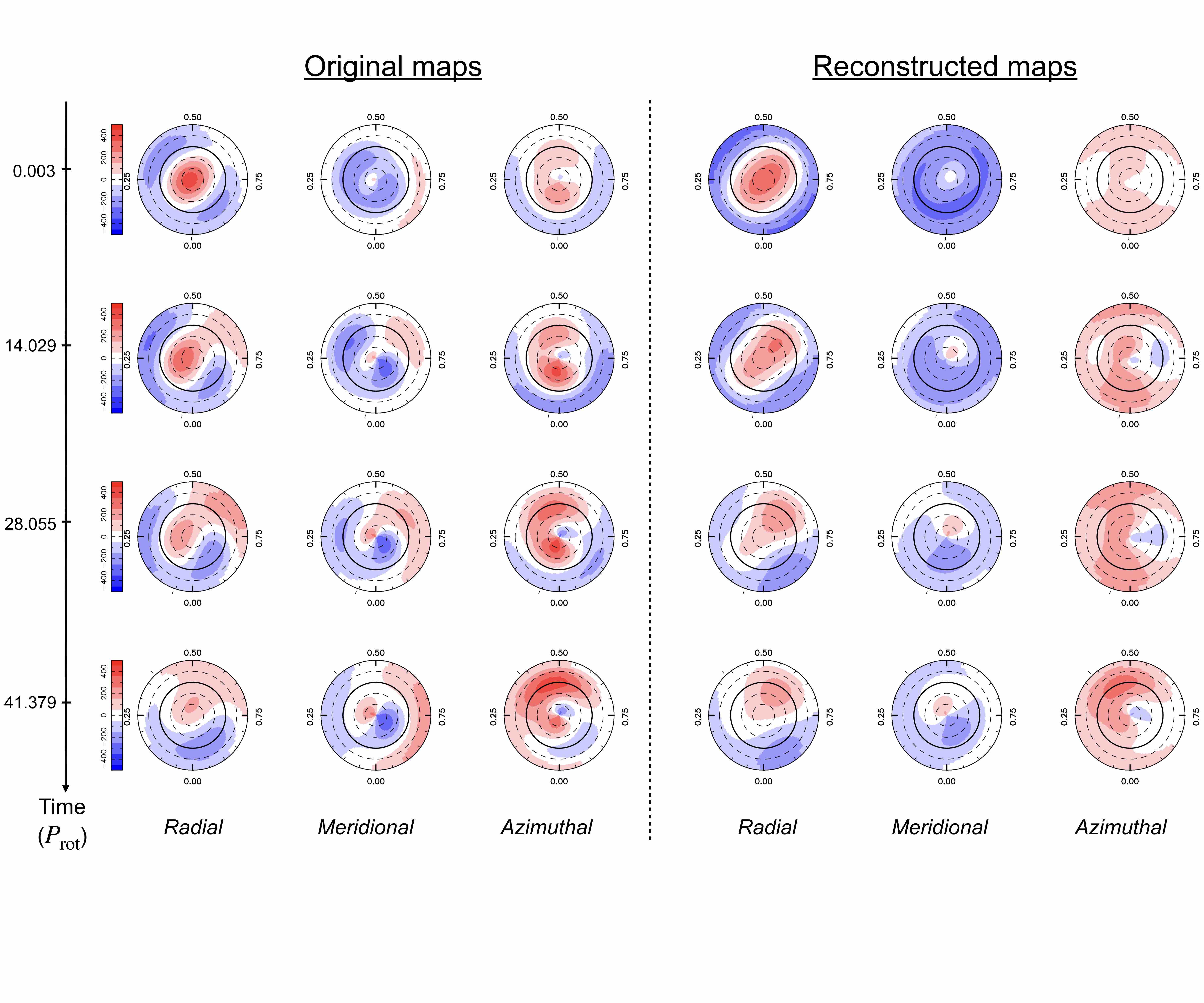} \hspace*{-0.8cm}
    \includegraphics[scale=0.16,trim={3cm 5.5cm 5cm 6cm},clip]{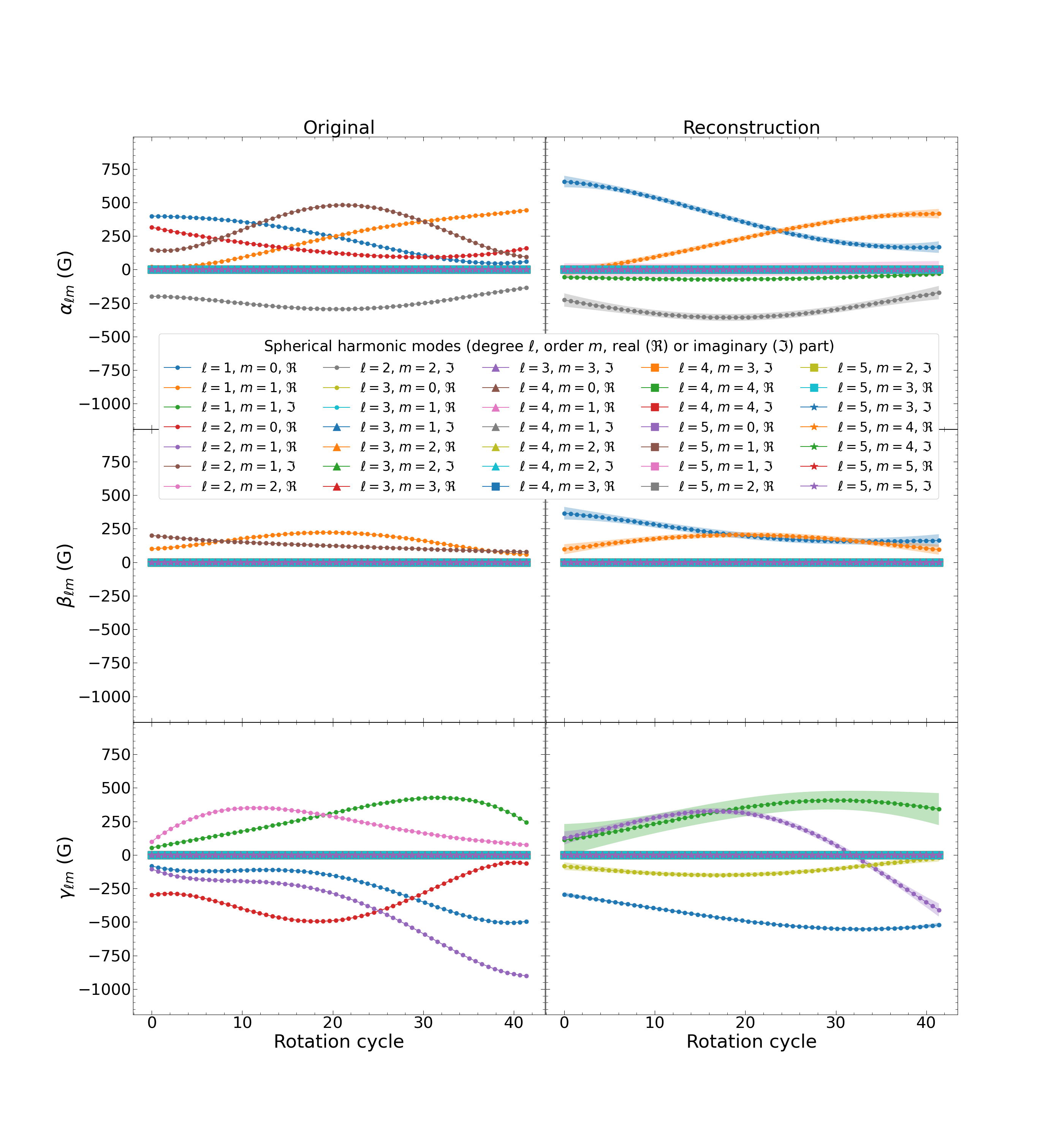}
    \includegraphics[scale=0.15,trim={0cm 0.5cm 0cm 0cm},clip]{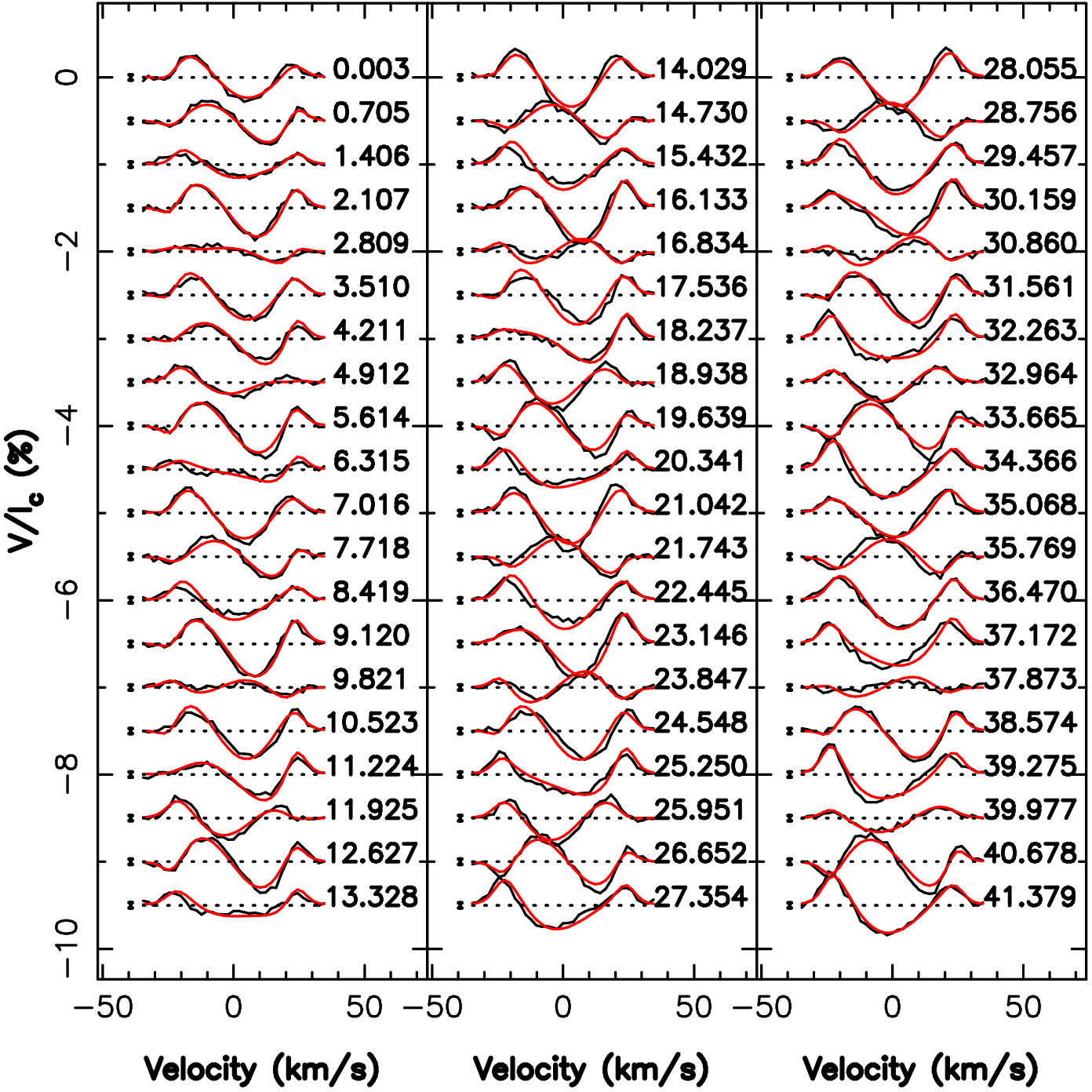}
    \caption{Same as Fig.~A1 when reconstructing the reference case (Sec.~3.3) using $n=3$ in the selection process, corresponding to a sampling of 1.4 rotation cycles (4.1~d).}
    \label{fig:maps_n3}
\end{figure*}

\renewcommand{\thefigure}{F2}
\begin{figure*}
    \centering \hspace*{-0.5cm}
    \includegraphics[scale=0.09,trim={.5cm 20.cm 0cm 9cm},clip]{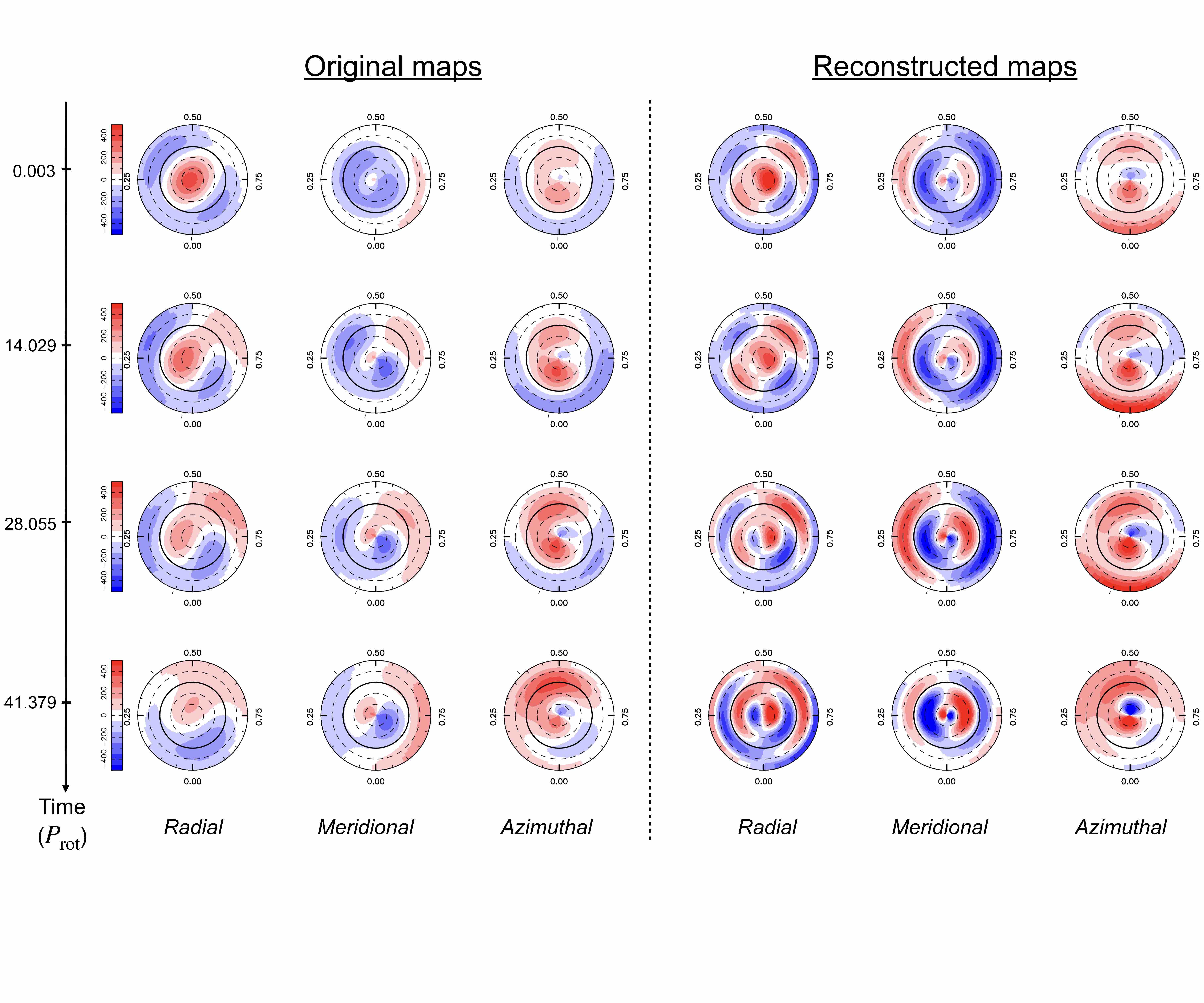} \hspace*{-0.8cm}
    \includegraphics[scale=0.15,trim={3cm 5.5cm 5cm 6cm},clip]{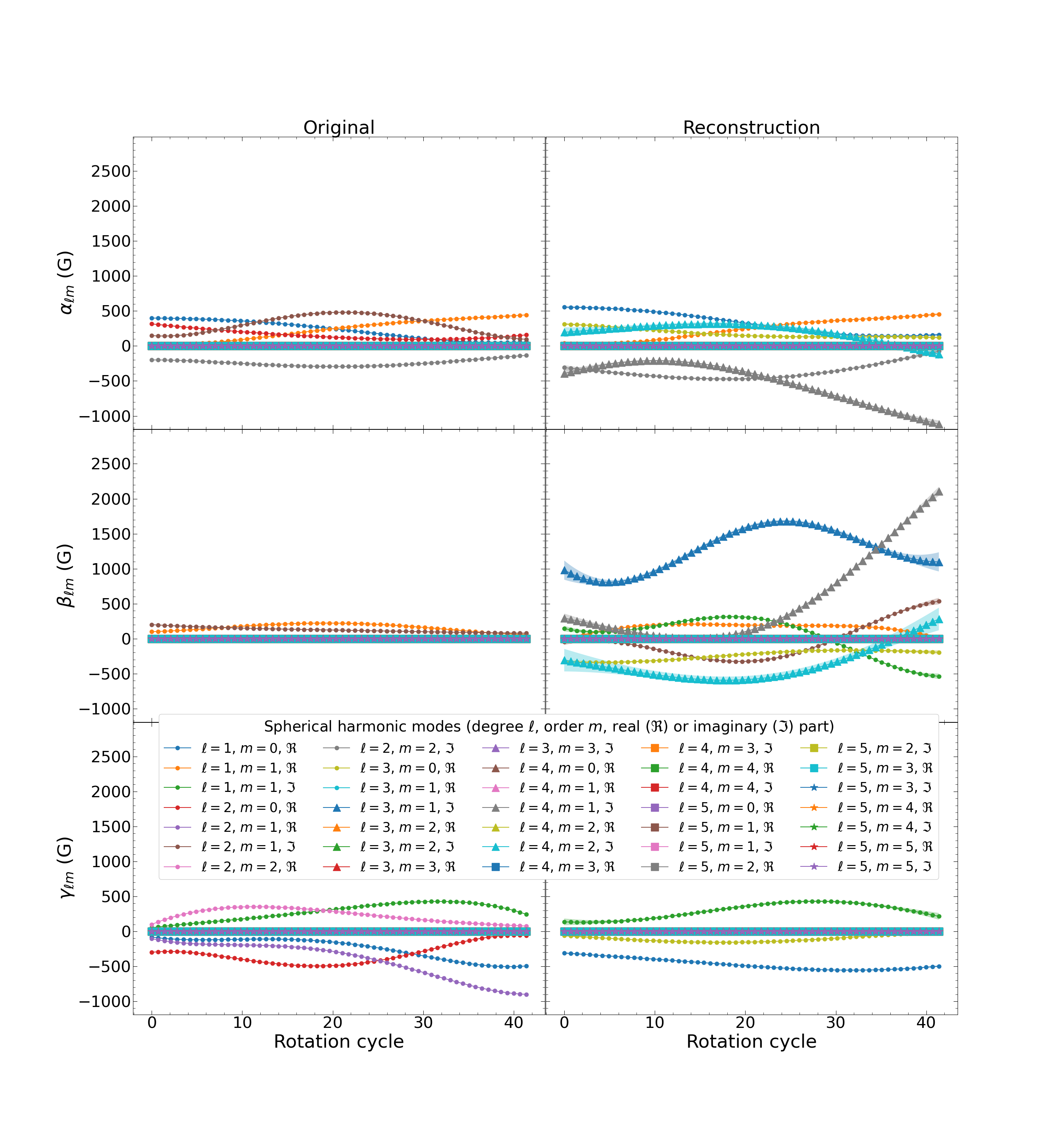}
    \includegraphics[scale=0.15,trim={0cm 0.5cm 0cm 0cm},clip]{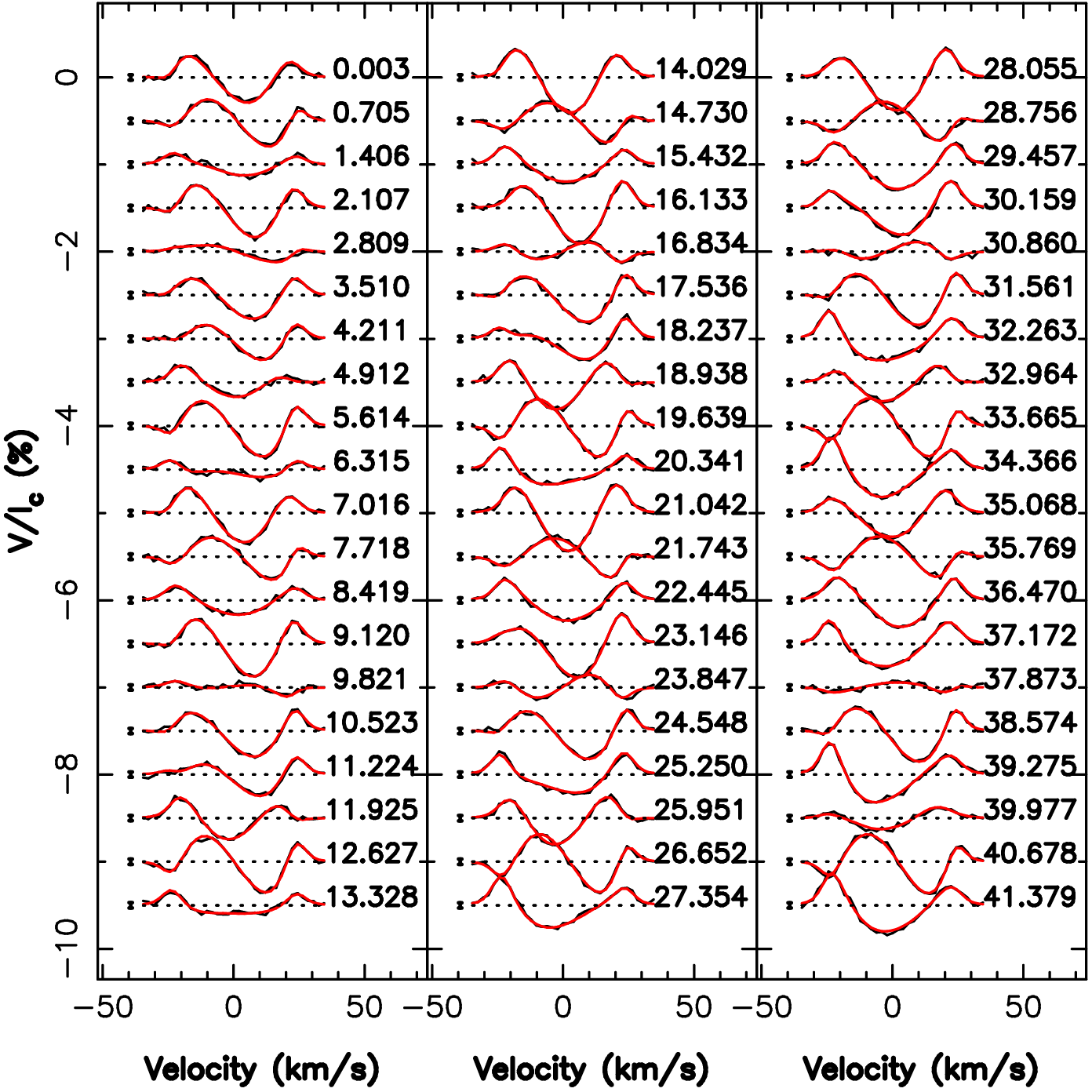}
    \caption{Same as Fig.~A1 when reconstructing the reference case (Sec.~3.3) using $n=12$ in the selection process, corresponding to a sampling of 7.7 rotation cycles (22.3~d).}
    \label{fig:maps_n12}
\end{figure*}

